\documentclass[11pt,reqno]{article}

\usepackage[bottom]{footmisc}
\usepackage{dirtytalk}
\usepackage{mathtools}
\usepackage{amsmath}
\usepackage{amsfonts}
\usepackage{amssymb}
\usepackage{amsxtra}
\usepackage{graphicx}
\usepackage{caption}
\usepackage{ushort}
\usepackage{color}
\usepackage{hyperref}
\usepackage[parsep]{collref}
\hypersetup{linktocpage=true}
\usepackage{enumitem}
\usepackage{titlesec}
\usepackage{cite}
\usepackage{float}
\usepackage{cleveref}
\usepackage{cancel}
\usepackage{tikz}

\DeclareMathOperator{\csch}{csch}
\DeclareMathOperator{\sech}{sech}
\crefformat{section}{\S#2#1#3}

 \def\bd{\begin{document}} \def\ed{\end{document}}
\def\ds{\documentstyle} \let\fr=\frac \let\bl=\bigl \let\br=\bigr
\let\Br=\Bigr \let\Bl=\Bigl
\let\bm=\bibitem
\let\na=\nabla
\let\pa=\partial \let\ov=\overline

\newcommand{\be}{\begin{equation}}
\newcommand{\ee}{\end{equation}}
\newcommand{\bea}{\begin{eqnarray}}
\newcommand{\eea}{\end{eqnarray}}
\newcommand{\ba}{\begin{array}}
\newcommand{\ea}{\end{array}}
\def\ie{{\it i.e.\ }}
\def\iec{{\it i.e.,\ }}
\def\eg{{\it e.g.\ }}
\def\egc{{\it e.g.,\ }}
\def\cf{{\it cf.\ }}

\newcommand\numberthis{\addtocounter{equation}{1}\tag{\theequation}}
\newcommand{\arctanh}{\text{arctanh}}
\newcommand{\bR}{\mathbb{R}}
\newcommand{\bZ}{\mathbb{Z}}
\newcommand{\cA}{\mathcal{A}}
\newcommand{\cE}{\mathcal{E}}
\newcommand{\cF}{\mathcal{F}}
\newcommand{\cI}{\mathcal{I}}
\newcommand{\cJ}{\mathcal{J}}
\newcommand{\cL}{\mathcal{L}}
\newcommand{\cM}{\mathcal{M}}
\newcommand{\cN}{\mathcal{N}}
\newcommand{\cO}{\mathcal{O}}
\newcommand{\cP}{\mathcal{P}}
\newcommand{\cQ}{\mathcal{Q}}
\newcommand{\cS}{\mathcal{S}}
\newcommand{\cV}{\mathcal{V}}
\newcommand{\fI}{\mathfrak{I}}
\newcommand{\fM}{\mathfrak{M}}
\newcommand{\Na}{\nabla}
\newcommand{\og}{\overline{g}}
\newcommand{\oh}{\overline{h}}
\newcommand{\oy}{\overline{y}}
\newcommand{\oz}{\overline{z}}
\newcommand{\ogamma}{\overline{\gamma}}
\newcommand{\oNa}{\overline{\nabla}}
\newcommand{\oR}{\overline{R}}
\newcommand{\orho}{\overline{\rho}}

\newcommand{\mf}{\mathfrak}
\newcommand{\mb}{\mathbb}
\newcommand{\mc}{\mathcal}
\newcommand{\wt}{\widetilde}
\newcommand{\ph}{\phantom}
\newcommand{\ut}{\underaccent{\tilde}}
\DeclareMathOperator{\im}{Im}
\DeclareMathOperator{\RP}{\mb{R}P}
\DeclareMathOperator{\Spin}{Spin}
\DeclareMathOperator{\SO}{SO}
\DeclareMathOperator{\GL}{GL}
\DeclareMathOperator{\SL}{SL}
\DeclareMathOperator{\U}{U}
\DeclareMathOperator{\SU}{SU}
\DeclareMathOperator{\Sp}{Sp}
\DeclareMathOperator{\vol}{vol}
\DeclareMathOperator{\Vol}{Vol}
\DeclareMathOperator{\Hol}{Hol}
\DeclareMathOperator{\tr}{tr}
\DeclareMathOperator{\diag}{diag}
\DeclareMathOperator{\Isom}{Isom}

\newcommand{\caltech}{\it Walter Burke Institute for Theoretical Physics, California Institute of Technology, Pasadena, CA 91125}

\newcommand{\brandeis}{\it Physics Department, Brandeis University, Waltham, MA 02454}

\newcommand{\imperial}{\it The Blackett Laboratory, Imperial College London\\
Prince Consort Road, London SW7 2AZ}

\newcommand{\auth}{
C. W. Erickson\,\footnote{\,christopher.erickson16@imperial.ac.uk},
Rahim Leung\,\footnote{\,rahim.leung14@imperial.ac.uk},
and K. S. Stelle\,\footnote{\,k.stelle@imperial.ac.uk}} 

\let\oldabstract\abstract
\let\oldendabstract\endabstract
\makeatletter
\renewenvironment{abstract}
{\renewenvironment{quotation}%
               {\list{}{\addtolength{\leftmargin}{1em} 
                        \listparindent 1.5em%
                        \itemindent    \listparindent%
                        \rightmargin   \leftmargin%
                        \parsep        \z@ \@plus\p@}%
                \item\relax}%
               {\endlist}%
\oldabstract}
{\oldendabstract}
\makeatother

\numberwithin{equation}{subsection}

\let\oldsubsection\subsection
\renewcommand{\subsection}{\renewcommand{\theequation}{\thesubsection.\arabic{equation}}\oldsubsection}

\titlespacing*{\section}
{0pt}{5.5ex plus 1ex minus .2ex}{4.3ex plus .2ex}
\titlespacing*{\subsection}
{0pt}{5.5ex plus 1ex minus .2ex}{4.3ex plus .2ex}
\begin{document}
\setcounter{page}{0}
\thispagestyle{empty}
\begin{flushright}
\hfill{
Imperial/TP/21/KS/01}\\
\end{flushright} 
\vspace{10pt}

\begin{center}  

{\Large {\bf Taxonomy of Brane Gravity Localisations}}   

\vspace{20pt}

\auth
\setcounter{footnote}{0}

\vspace{7pt}
\imperial

\end{center} 
\vspace{15pt}
\begin{abstract}
Generating an effective theory of lower-dimensional gravity on a submanifold within an original higher-dimensional theory can be achieved even if the reduction space is non-compact. Localisation of gravity on such a lower-dimensional worldvolume can be interpreted in a number of ways. The first scenario, Type I, requires a mathematically consistent Kaluza--Klein style truncation down to a theory in the lower dimension, in which case solutions purely within that reduced theory exist. However, that situation is not a genuine localisation of gravity because all such solutions have higher-dimensional source extensions according to the Kaluza--Klein ansatz. Also, there is no meaningful notion of Newton's constant for such Type I constructions. 

Types II and III admit coupling to genuinely localised sources in the higher-dimensional theory, with corresponding solutions involving full sets of higher-dimensional modes. Type II puts no specific boundary conditions near the worldvolume aside from regularity away from sources. In a case where the wave equation separated in the non-compact space transverse to the worldvolume admits a normalisable zero mode, the Type III scenario requires boundary conditions near the worldvolume that permit the inclusion of that zero mode in mode expansions for gravitational wave fluctuations or potentials. In such a case, an effective theory of lower-dimensional gravity can emerge at sufficiently large worldvolume distance scales. 

This taxonomy of brane gravity localisations is developed in detail for linearised perturbations about a background incorporating the vacuum solution of Salam--Sezgin theory when embedded into ten-dimensional supergravity with a hyperbolic non-compact transverse space. Interpretations of the Newton constant for the corresponding Type III localisation are then analysed.

\end{abstract}
\newpage
\tableofcontents
\newpage
\section{Introduction}
A standout feature of string and supergravity theories is the fundamental r\^ole of higher-dimensional spacetimes in the most mathematically natural formulations of these theories. Relating these theories to realistic physics requires some method of producing a four-dimensional effective theory. A standard method (\cf \eg \cite{Duff:1986hr}) is Kaluza--Klein dimensional reduction, making an ansatz in which the space transverse to the four-dimensional spacetime is compact, thus producing a naturally discrete spectrum of transverse-space wavefunctions. The discrete eigenvalues of such transverse wavefunctions correspond to discrete $(\hbox{mass})^2$ values for the excitations of the four-dimensional effective theory. In order for such a model to describe four-dimensional gravity, there must exist massless graviton states, \ie the transverse-space problem must admit a zero eigenvalue. The existence of a gap in the effective-theory mass spectrum then allows for a low-energy regime in which physical phenomena are effectively four-dimensional. A mathematically attractive situation occurs when the coupling between the four-dimensional massless sector of the theory and the massive sector allows, at least at the classical level, for the massive modes to be set to zero consistently with the original higher-dimensional field equations. In that case, one has a {\em consistent truncation}, a topic which has been much studied in the Kaluza--Klein literature. However, many physically interesting constructions derived from higher-dimensional theories are not endowed with such massive-mode consistent truncations. The existence of a mass gap and consequent lower-dimensional behaviour in a low-energy \say{grace zone} with only small massive-mode-induced corrections is more essential.

Other ways of obtaining effectively lower-dimensional physics from a higher-dimensional theory have also been explored.  A theme in the development of supergravity theories which has been explored but not widely applied is the existence of supergravity models with non-compact gauge symmetries (\cf \eg \cite{Hull:1988jw}). Such lower-dimensional models can be obtained from a variant of the Kaluza--Klein idea, in which the transverse space is taken to be non-compact, and with correspondingly non-compact symmetries. In such cases where there is also a consistent truncation to the lower-dimensional massless sector, one may consider that one has embedded a kind of lower-dimensional physics with non-compact symmetries into the higher-dimensional theory. A trivial example of such a non-compact reduction would be from higher-dimensional Minkowski space, with the consistent reduction ansatz requiring translational invariance in all dimensions higher than four. That trivial example, however, illustrates a frequent problem with non-compact reductions: the massive-sector spectrum is continuous right down to zero mass. Any small excitation of the massive modes causes the theory to behave as originally formulated in the higher dimension, and there is no low-energy grace zone. This can prevent the effective localisation of the effective-theory gravity in the lower dimension, unless somehow a mass gap can be arranged below the edge of the continuous spectrum.

In analysing transverse wave equation eigenvalue spectra, it is often convenient to make a change of variables so as to remove first-derivative terms from the wave equation, leaving just second-derivative terms and potential terms without derivatives. The resulting transverse wave equation then has the form of a time-independent Schr\"odinger equation. An approach to the localisation of gravity on a four-dimensional subsurface of a higher-dimensional spacetime somewhat along such lines was given by Randall and Sundrum in \cite{Randall:1999vf}, reflecting segments of $\hbox{AdS}_5$ with the reflection surface corresponding to an extended delta-function source for the Einstein equations. This gives rise to a normalisable bound state in the corresponding effective-theory Schr\"odinger problem and to the existence of a low-energy grace zone. One criticism of that construction, however, has been that it is still a kind of compactification of the transverse space, because the location of the reflection surface near the $\hbox{AdS}_5$ horizon in Poincar\'e patch coordinates leads to a finite normalisation integral for the corresponding zero mode: upon reflecting, an infinite volume of $\hbox{AdS}_5$ spacetime has been excluded. The nature of the implied delta-function source also remained unmotivated.

A different construction possessing a normalisable transverse-space zero mode which is based on physical excitations around a genuinely non-compact solution of ten-dimensional Type IIA supergravity was explored in Reference \cite{Crampton:2014hia}. The underlying supergravity solution is based on the $\mb{R}^{1,3}\times S^2$ ground-state solution of the 1984 six-dimensional supergravity found by Salam and Sezgin \cite{Salam:1984cj}, later lifted into Type IIA supergravity by Cveti\v{c}, Gibbons and Pope in Reference \cite{Cvetic:2003xr}. The lift of Salam--Sezgin supergravity into Type IIA supergravity is made on the non-compact hyperbolic space $H(2,2)$ times a circle, and the Salam--Sezgin ground state has a flat ${\rm Minkowski}_4\!$ \say{worldvolume} lifted to $D=10$ via a transverse space incorporating four-dimensional Eguchi--Hansen space. We shall refer to this lifted ground state solution as the SS--CGP background. Gravitational excitations on the $D=4$ worldvolume then have a transverse-space field equation whose Schr\"odinger reformulation has a potential $V(\rho)=2- (\coth(2\rho))^2$ giving an integrable system of P\"oschl--Teller type \cite{Crampton:2014hia}. This transverse system possesses a single normalisable zero mode, then a gap, above which is the expected continuum of nonzero eigenvalue states. Accordingly, the $D=4$ effective field theory has a massless graviton separated from a continuum of massive states by a mass gap, so the system is endowed with a low-energy grace zone in which gravity-wave physics is effectively four-dimensional.

In the present paper, we shall continue further an exploration of the hyperbolic spacetime construction of Reference \cite{Crampton:2014hia} by considering the consequences of including a massive point source situated on the worldvolume but fully localised in the higher-dimensional spacetime. Our discussion will mostly be carried out at linearised level in perturbations about the SS--CGP background and for simplicity we mostly will work in the context of a five-dimensional reduction of the $D=10$ theory. Since we will maintain throughout the transverse-space symmetries of the SS--CGP solution as well as spherical symmetry on the worldvolume, the discussion could be lifted back up into a full $D=10$ context. Accordingly, the key coordinates for our analysis will mostly be just the non-compact transverse-space coordinate $\rho$ and the $D=4$ worldvolume radius $r$. 

The key question that we shall address is whether the response to the inclusion of a massive point source produces a genuinely four-dimensional massless gravitational field response at large $r$ distances or not. We shall find that there are three different situations. We begin in Section \ref{se: blackspokes} firstly by reviewing the well-known construction of sourced gravitational solutions just in the dimensionally reduced $D=4$ theory, following References \cite{Brecher:1999xf,Chamblin:1999by,Lu:2000xc}. Such solutions do not have a genuinely localised massive point source from the point of view of the higher-dimensional theory, however, since they are based upon a dimensional reduction ansatz in which the full SS--CGP transverse space structure remains unchanged. In this situation, a $D=4$ point source is really an extended source from the higher-dimensional point of view. In Reference \cite{Chamblin:1999by}, such sources were called \say{black strings}, but in order to emphasise the radially extended nature of such sources in the $D=10$ higher-dimensional theory, we shall prefer to call them \say{black spokes}. We shall call this situation Type I structure. 

The rest of the paper will be concerned with genuinely localised sources from a higher-dimensional perspective. In Section \ref{se: Newton's Constant}, we shall discuss how to define an effective Newton constant in the $D=4$ subspace by the study of geodesics perturbed by the effect of a massive point source. In Section \ref{se: PertH00} we shall simplify the linearised analysis of the gravitational perturbation about the SS--CGP background, showing how one may reduce the study of the Newtonian gravitational potential to that just of a scalar Green function on that background.

The main results of the paper then come in Section \ref{se: GFforCPS} and Section \ref{ldm}. In Section \ref{se: GFforCPS}, we shall first study what might be considered the most natural context of simply putting a massive point source on the worldvolume of the SS--CGP background and finding its behaviour at large worldvolume radius $r$, while imposing Neumann boundary conditions at $\rho=0$ in the transverse coordinate. This does not turn out to reproduce an effectively $D=4$ Newtonian gravitational potential at large $r$, however. Instead, one finds a $D=5$ Yukawa-type gravitational potential, which we will call Type II structure. This is quite different from what one might have expected from the massless gravity-wave analysis of Reference \cite{Crampton:2014hia}. The reason for this hangs upon the type of transverse-space boundary conditions that are imposed at the $\rho=0$ worldvolume surface. The P\"oschl--Teller zero mode of the transverse-space Schr\"odinger problem has a logarithmic structure as $\rho\to0$, and in order to incorporate such modes, one needs to impose a different type of boundary condition at $\rho=0\,$: a generalised Robin boundary condition. When this boundary condition is imposed at $\rho=0$, the anticipated four-dimensional $1/r$ behaviour of the gravitational potential at large $r$ worldvolume radius makes its appearance, with, however, a $\log(\rho)$ prefactor. This we will call Type III structure. In Section \ref{ldm}, the transition between Type II and Type III structures will be explored using a technique due to Bender \cite{benderprivate} which we call \say{long distance mirrors}. This transition turns out to be analogous to the detailed analysis of the Randall--Sundrum system given in Reference \cite{Giddings:2000mu}, which also hinged on consideration of boundary conditions at the reflection surface.

In Section \ref{se: Newton effective}, we show how an effective-theory Newton constant may be obtained from a variety of local averagings in position near the $\rho=0$ worldvolume, one of which precisely reproduces the effective-theory value found in Reference  \cite{Crampton:2014hia}.
\section{Type I: Radial Black Spokes}\label{se: blackspokes}

\subsection{Geometry of the Salam--Sezgin lift}

 The uplift of the Salam--Sezgin $\mb{R}^{1,3}\times S^2$ solution into Type IIA supergravity was obtained in Reference \cite{Cvetic:2003xr}. In Einstein frame, this uplifted solution, the SS--CGP solution, is 
\begin{equation}\label{CGPSS}
\begin{split}
    &ds^2_{10} = {\cal H}^{-1/4}\left(\eta_{\mu\nu}dx^\mu dx^\nu + dy^2 + \frac{1}{4g^2}\big(d\psi + \sech2\rho\,(d\chi+\cos\theta\,d\varphi)\big)^2\right) + \frac{1}{g^2}{\cal H}^{3/4}ds^2_{EH} \,, \\
    &B_{(2)} = \frac{1}{4g^2}\big(d\chi+\sech2\rho\,d\psi\big)\wedge\big(d\chi+\cos\theta\,d\varphi\big) \,,\quad e^{2\phi} = {\cal H} \,,\quad {\cal H} = \sech2\rho \,,
\end{split}
\end{equation}
where $g$ is a constant, and $ds^2_{EH}$ is the metric on the four-dimensional Eguchi--Hanson space, 
\begin{equation}
    ds^2_{EH} = \cosh2\rho\Big(d\rho^2 + \frac{1}{4}(\tanh2\rho)^2(d\chi+\cos\theta\,d\varphi)^2 + \frac{1}{4}(d\theta^2+\sin^2\theta\,d\varphi^2)\Big) \,.
\end{equation} 
The coordinates take values in ranges $x^\mu \in \mb{R}^{1,3}$, $y\in[0,l_y)$, $\chi,\varphi \in [0,2\pi)$, $\psi \in [0,4\pi)$, $\theta \in [0,\pi]$ and $\rho \in[0,\infty)$. The fact that $\chi$ has a period of $2\pi$ as opposed to $4\pi$ means that the boundary of the Eguchi--Hanson space at infinity is given by $\RP^3 \cong S^3/\mb{Z}_2$, where the $S^3$ is realised as a Hopf fibration over $S^2$, with $\chi$ being the fibre coordinate. Near $\rho=0$, the geometry of the Eguchi--Hanson space is $\mb{R}^2\times S^2$ for constant $(\theta,\varphi)$, with $(\rho,\chi)$ acting as plane polar coordinates on $\mb{R}^2$. 

 As explained in \cite{Crampton:2014hia}, the SS--CGP solution preserves 8 supercharges, and has the form of an NS5-brane wrapped on $(y,\psi)\in T^2$ with an effective worldvolume $\mb{R}^{1,3}$ that has a singularity which is resolved by transgression. The function ${\cal H}$, which is usually a harmonic function on the transverse space (Eguchi--Hanson in our case), is now a particular solution to the sourced Laplacian
\begin{equation}\label{transgeqn}
    \Delta_{EH}{\cal H} = -\frac{g^2}{2}(\mc{F}_{(2)})^2 \,,
\end{equation}
where $\Delta_{EH}$ is the Laplacian on Eguchi--Hanson space, and $\mc{F}_{(2)}$ is the field strength of the 1-form 
\begin{equation}
    \mc{A}_{(1)} = \sech2\rho\,(d\chi+\cos\theta\,d\varphi) \,,
\end{equation} 
and is the unique, anti-self-dual 2-form on Eguchi--Hanson space. Geometrically, this transgression is realised as a $U(1)$ fibration of the worldvolume over Eguchi--Hanson space with fibre coordinate $\psi$ and connection $\mc{A}_{(1)}$. We will call this $U(1)$ bundle the transgression bundle. For generic values of $\rho$, this bundle is non-trivial with second Chern character
\begin{equation}
    \int \text{ch}_2(\mc{F}_{(2)}) = \frac{1}{2(2\pi)^2}\int \mc{F}_{(2)}\wedge\mc{F}_{(2)} = 1 \,,
\end{equation} 
where the integral is over the Eguchi--Hanson space. There is a special limit of the SS--CGP solution which makes the connection to NS5-branes even more manifest. As $\rho\to\infty$, the field strength $\mc{F}_{(2)}$ vanishes, so the transgression bundle trivialises. In this limit, the solution asymptotes to the linear dilaton solution, which is the near-horizon limit of the NS5-brane. Consequently, there is also an enhancement of supersymmetry to 16 supercharges in this limit. 

 It is worth mentioning that it is possible \cite{Crampton:2014hia} to include an additional NS5-brane into the SS--CGP solution without breaking any more supersymmetry by adding to ${\cal H}$ a homogeneous solution to \eqref{transgeqn}. Explicitly, one has
\begin{equation}
    {\cal H} = \sech2\rho - k\log\tanh\rho\,,
\end{equation}
where $k$ is a positive constant that is proportional to the tension of the NS5-brane. The logarithmic behaviour of ${\cal H}$ for small $\rho$ is indicative of the fact that the topology of the Eguchi--Hanson space is $\mb{R}^2\times S^2$ in this neighbourhood. In order for this to remain a solution, the NSNS 2-form is also modified to be 
\begin{equation}
    B_{(2)} = \frac{1}{4g^2}\big((1+k)d\chi+\sech2\rho\,d\psi\big)\wedge\big(d\chi+\cos\theta\,d\varphi\big) \,.
\end{equation}
We will not be studying this solution further in the present paper, but more information about it can be found in Reference \cite{Crampton:2014hia}. 
\subsection{Ricci-flat Branes and Radial Black Spokes}

 For every supersymmetric brane solution, resolved or unresolved, it is possible to replace the flat worldvolume (or effective worldvolume in the resolved case) by an arbitrary Ricci-flat manifold. The same is also true for its transverse space. As such, the metrics of these doubly-Ricci-flat branes are given by
\begin{equation}\label{drfbranes}
    ds^2 = {\cal H}^ag_{\mu\nu}(x)dx^\mu dx^\nu + {\cal H}^b G_{ij}(y)dy^idy^j \,,
\end{equation}
where $a, b$ are appropriate constants, $g_{\mu\nu}$ and $G_{ij}$ are the Ricci-flat metrics on the effective worldvolume and transverse space respectively, and ${\cal H}$ is a harmonic function on the transverse space. We note that for a flat transverse space, the above is an example of a brane with Ricci-flat worldvolume as first explored in \cite{Brecher:1999xf}, which is also a special case of a \say{branes on branes} construction, where one considers a consistent truncation to a supergravity theory on the lower-dimensional worldvolume \cite{Lu:2000xc}. It is not difficult to show that the solution \eqref{drfbranes}, along with its appropriate scalars and fluxes, is supersymmetric provided that $g_{\mu\nu}$ and $G_{ij}$ admit covariantly constant spinors with an appropriate projection condition. In Appendix A, we will provide an explicit example of this solution realised in a supergravity model. From  \cite{Figueroa-OFarrill:1999waz}, there are no static, irreducible Ricci-flat, Lorentzian manifolds other than Minkowski space that admit covariantly constant spinors. For the transverse space, on the other hand, there are many options other than Euclidean space. In particular, depending on dimension, one can select Calabi--Yau, hyper-K\"{a}hler, $G_{(2)}$, or $Spin(7)$ manifolds. The resulting number of preserved supercharges is then determined by the number of singlets in the decomposition of the representation of the $Spin(m)$ spinor with respect to the holonomy group of the transverse space, where $m$ is the dimension of the transverse space. For a more detailed account of special holonomy manifolds in relation to supersymmetry, we refer the reader to the review in Reference \cite{Gauntlett:2003di}. 

 A compact Riemannian manifold without boundary only admits a constant solution to the Laplace equation, so it is necessary for the transverse space to be non-compact in order for ${\cal H}$ to be non-trivial. Due to the non-compactness of the transverse space, ${\cal H}$ will generically have a singularity. For example, if we take the transverse space to be a conical Calabi--Yau space with metric
\begin{equation}
    ds^2(CY) = dR^2 + R^2ds^2(SE_{m-1}) \,,
\end{equation}
where $SE_{m-1}$ is a $(m-1)$-dimensional Sasaki-Einstein manifold with $m$ even,\footnote{Taking $SE_{m-1}=S^{m-1}$ gives the flat metric on $\mb{R}^m$.} and let ${\cal H}$ only be $R$-dependent, then 
\begin{equation}
    {\cal H} = 1 + \frac{k}{R^{m-2}} \,.
\end{equation}
This has a singularity at $R=0$, which for generic dimensions translates into a curvature singularity of the solution. This is not the case for the M2 and M5-branes, however, where the $R=0$ singularity is a horizon; see \cite{Duff:1990xz, Gueven:1992hh, Acharya:1998db} for further details. 

 Leaving supersymmetry aside, one enticing aspect of these doubly-Ricci-flat branes is that they seemingly allow for localised gravitational physics on the brane worldvolume. As an example, the Ricci-flat worldvolume can be chosen to be the Schwarzschild black hole in isotropic coordinates,
\begin{equation}\label{schwarzschild}
ds^2= -\left(\frac{1-\frac{M}{r^{n-3}}}{1+\frac{M}{r^{n-3}}}\right)^2dt^2 + \left(1+\frac{M}{r^{n-3}}\right)^{\frac{4}{n-3}}\left(dr^2+r^2ds^2(S^{n-2})\right)\,,
\end{equation}
where $n$ is the worldvolume dimension. The singularity structure of the solution, ignoring the contribution from the harmonic function ${\cal H}$, will now be at $r=0$. However, this $r=0$ singularity is located everywhere in the transverse space. More precisely, in a perturbative picture, a doubly-Ricci-flat brane with worldvolume given by \eqref{schwarzschild} can be written as a perturbation of a doubly-Ricci-flat brane with a $\mb{R}^{1,n-1}$ worldvolume,
\begin{equation}\label{perturbeddrfbrane}
ds^2 = {\cal H}^a\left(\eta_{\mu\nu}dx^\mu dx^\nu + Mh_{\mu\nu}dx^\mu dx^\nu\right) + {\cal H}^b G_{ij}(y)dy^idy^j + \mc{O}(M^2) \,,
\end{equation}
with
\begin{equation}
    h_{00} = \frac{4}{r^{n-3}}\,,\quad h_{mn} = \frac{4}{(n-3)r^{n-3}}\delta_{mn} \,,
\end{equation}
where $\delta_{mn}$ is the flat metric on the $\mb{R}^{n-1}$ slice of $\mb{R}^{1,n-1}$. This perturbation is not traceless, and obeys the de Donder gauge
\begin{equation}
    \partial^\mu h_{\mu\nu} - \frac{1}{2}\partial_\nu h^{\rho}_\rho = 0 \,.
\end{equation}
The stress tensor that sources this perturbed solution \eqref{perturbeddrfbrane} has the form
\begin{equation}
    T_{MN} = M\delta_{M0}\delta_{N0}f(y)\frac{\delta(r)}{r^{n-2}} \,,
\end{equation}
where $f(y)$ is a smooth function on the transverse space. From the delta function structure of the stress tensor, we observe that the source of this solution is not localised in the higher dimension, but is spread out radially like a spoke. 

For the rest of this paper, we will be interested in braneworld localisation on the SS--CGP background with a source that has a genuinely higher-dimensional origin, so that the curvature singularity is located at a single point in the higher-dimensional space. More precisely, we require a source that is localised at a single point in $(r,\rho)$ space, where $r$ is the isotropic, spatial radius on the effective $\mb{R}^{1,3}$ worldvolume. We will be employing a perturbative treatment as in \eqref{perturbeddrfbrane}, but since the source for the perturbation will be higher-dimensional, the perturbation itself will also depend importantly on the transverse radius $\rho$. For the perturbation in \eqref{perturbeddrfbrane}, we will not assume tracelessness. 

\section{Inferring Newton's Constant; Geodesics}\label{se: Newton's Constant}

 Owing to the non-trivial nature of the SS--CGP background, it is difficult to solve for all components of a gravitational perturbation that is sourced at the $(r,\rho) = (0,0)$ origin. However, our goal is to understand whether brane-gravity localisation is possible, and, to this end, we only need to compute the effective gravitational potential associated with the perturbation and from that infer the lower-dimensional Newton constant. The problem of defining a lower-dimensional Newton constant can be interpreted at the level of the field theory action by reading off the coupling of matter fields to the metric. However, one alternate method by which one might determine the Newton constant, and also determine the dimension to which the effective gravity corresponds,  would be to measure the response of a test particle to a known source mass and so to infer the corresponding Newton constant by the way geodesics in spacetime are distorted by the gravitational perturbation.

 If one considers a weak-field limit in the neighbourhood of a source mass for a perturbation caused by the source in a Minkowski spacetime, only the details of the time-time component\footnote{This statement makes use of an implicit gauge. We give full details of our gauge below.} of the perturbation need be known. We will show that this is also true in the SS--CGP background.
\subsection{The SS--CGP Background}

 In this section, we'll consider timelike geodesics on the SS--CGP background \eqref{CGPSS}. The affinely parametrised geodesic equation for a path $\gamma$ is given by
\begin{equation}
    \frac{d^2Z^M}{d\tau^2} + \Gamma^{M}_{\ph{M}KL}(Z)\frac{dZ^K}{d\tau}\frac{dZ^L}{d\tau} = 0 \,,
\end{equation}
where $Z^M = (X^\mu, Y, P, \Theta, \Phi, \Sigma, \Psi)$ are the coordinates of the path $\gamma$, and $\tau$ is the proper time. We use capital letters here in order to avoid confusion with the global coordinates in \eqref{CGPSS}.\footnote{$(X^\mu, Y, P, \Theta, \Phi, \Sigma, \Psi)$ correspond to $(x^\mu, y, \rho,\theta,\varphi,\chi,\psi)$.} This equation of motion gives extrema for the Lagrangian\footnote{We use the form \eqref{eq:lag} of the particle Lagrangian in this discussion for simplicity, instead of the worldline reparametrisation invariant proper-time action $\int \frac{d\tau(p)}{dp}\,dp$ involving a square root. The Lagrangian \eqref{eq:lag} can of course be obtained from the \say{einbein} form \cite{Brink:1976uf} $L_{\rm BdVH} = \frac12(e^{-1}g_{MN}(Z)\frac{dZ^M}{d\tau}\frac{dZ^N}{d\tau}+m_{\rm particle}^2e)$ by choosing the reparametrisation gauge $e=\frac12$.}
\begin{equation}\label{eq:lag}
    L = g_{MN}(Z)\frac{dZ^M}{d\tau}\frac{dZ^N}{d\tau}\,. 
\end{equation}

 The isometry group of \eqref{CGPSS} is given by
\begin{equation}
    \text{Isom}_{10} = ISO(1,3)\times U(1)^3 \times SO(3)^2 \,,
\end{equation}
where the $U(1)^3$ corresponds to the 3-torus parametrised by $(y, \psi, \chi)$, and the $SO(3)^2$ is the isometry of the $S^2$ parametrised by $(\theta,\varphi)$. Using these isometries, we find that a solution to the geodesic equation is
\begin{equation}\label{eq:isomsol}
    Y= 0 \,,\quad \Theta = \pi\,,\quad \Phi=0\,,\quad \Sigma = 0\,,\quad \Psi = 0\,.
\end{equation}
Simplifying the Lagrangian \eqref{eq:lag} for a solution that obeys conditions \eqref{eq:isomsol}, we find
\begin{equation}
    L = (\cosh2P)^{1/4}\Big[\eta_{\mu\nu}\frac{dX^\mu}{d\tau}\frac{dX^\nu}{d\tau} + \frac{1}{g^2}\Big(\frac{dP}{d\tau}\Big)^2\Big]\,.
\end{equation}
The equation of motion for $P(\tau)$ is then
\begin{equation}
    \frac{d^2P}{d\tau^2} + \frac{1}{4}(\tanh2P)\Big(\frac{dP}{d\tau}\Big)^2 - \frac{g^2}{4}(\tanh2P)\eta_{\mu\nu}\frac{dX^\mu}{d\tau}\frac{dX^\nu}{d\tau} = 0 \,.
\end{equation}
This admits the solution
\begin{equation}
    P(\tau) = 0 \,.
\end{equation}
The remaining equations for $X^\mu$ are the usual geodesic equations on $\mb{R}^{1,3}$. Remembering that we're looking for a timelike geodesic, the appropriate solution is
\begin{equation}
X^0 = \tau \,,\quad X^i = 0\,.
\end{equation}

 To summarise, we find that the SS--CGP geometry admits the stable timelike geodesic
\begin{equation}
    X^0=\tau\,,\quad X^i = 0 \,,\quad Y = 0 \,,\quad P = 0\,,\quad \Theta = \pi\,,\quad \Phi=0\,,\quad \Sigma = 0\,,\quad \Psi = 0\,. \label{eq:geo1}
\end{equation}
This will be the starting point for the next section, where we look for a timelike geodesic on a perturbed SS--CGP geometry.
\subsection{Perturbed SS--CGP Geodesics}

 We consider the perturbed geometry described by a metric $\hat g$, with
\begin{equation}
    \hat g_{MN} = (\cosh2\rho)^{1/4}(\bar g_{MN} + H_{MN})
\end{equation}
where $\bar g$ is the string-frame metric on the SS--CGP background,\footnote{The string-frame metric is related to the Einstein-frame metric by $ds^2_{\text{str}}=e^{\phi/2}ds^2_{\text{Ein}}$.} and $H_{MN}$ is a perturbation. The perturbation that we are interested in is independent of the time coordinate $x^0$, and has components only along the $x^\mu$ and $\rho$ directions, with $H_{0i}=0$. As such, the $U(1)^3\times SO(3)^2$ isometry of the SS--CGP background is unbroken, and we have it that
\begin{equation}
    Y= 0 \,,\quad \Theta = \pi\,,\quad \Phi=0\,,\quad \Sigma = 0\,,\quad \Psi = 0
\end{equation}
solves the perturbed geodesic equations. With this choice, the perturbed particle Lagrangian is
\begin{equation}\begin{split}
L &= \hat g_{MN}(Z)\frac{dZ^M}{d\tau}\frac{dZ^N}{d\tau} \\
&= (\cosh2P)^{1/4}\Big[(\eta_{\mu\nu}+H_{\mu\nu})\frac{dX^\mu}{d\tau}\frac{dX^\nu}{d\tau} + 2A_{\mu}\frac{dX^\mu}{d\tau}\frac{dP}{d\tau} + \frac{1}{g^2}(1 + B)\Big(\frac{dP}{d\tau}\Big)^2\Big] \,,
\end{split}\end{equation}
where we have defined
\begin{equation}
    H_{\mu\rho}(X^i,P) \equiv A_\mu(X^i,P)\,,\quad H_{\rho\rho}(X^i,P) = \frac{1}{g^2}B(X^i,P) \,.
\end{equation}
It is important to note that $H_{\mu\nu}$, $A_\mu$ and $B$ are functions of $X^i$ and $P$ only. 

 The resulting equations for $X^\mu$ and $P$ are given by
\begin{equation}\begin{split}
    &(\delta^\mu_\nu + H^\mu_{\ph{\mu}\nu})\frac{d^2 X^\nu}{d\tau^2} + \frac{1}{2}\eta^{\mu\nu}\big(\partial_\sigma H_{\lambda\nu} + \partial_\lambda H_{\sigma \nu} - \partial_\nu H_{\sigma\lambda}\big)\frac{dX^\sigma}{d\tau}\frac{dX^\lambda}{d\tau}  \\
    &+ A^\mu\Big(\frac{d^2P}{d\tau^2} + \frac{1}{2}(\tanh2P)\Big(\frac{dP}{d\tau}\Big)^2\Big) +\Big(\partial_PA^\mu- \frac{1}{2g^2}\partial^\mu B\Big)\Big(\frac{dP}{d\tau}\Big)^2  \\
    &+ \Big(F_\nu^{\ph{\nu}\mu} + \partial_PH^\mu_{\ph{\mu}\nu}+\frac{1}{2}(\tanh2P)(\delta^\mu_\nu + H^\mu_{\ph{\mu}\nu})\Big)\frac{dX^\nu}{d\tau}\frac{dP}{d\tau} = 0 \,,
\end{split}\end{equation}
and
\begin{equation}\begin{split}
    &\frac{1}{g^2}(1+B)\Big(\frac{d^2P}{d\tau^2} + \frac{1}{4}(\tanh2P)\Big(\frac{dP}{d\tau}\Big)^2\Big) + \frac{1}{2g^2}\partial_PB\Big(\frac{dP}{d\tau}\Big)^2 + A_\mu\frac{d^2X^\mu}{d\tau^2} \\
    &+\frac{1}{g^2}\partial_\mu B \frac{dX^\mu}{d\tau}\frac{dP}{d\tau} + \Big(\partial_{(\mu}A_{\nu)}-\frac{1}{2}\partial_PH_{\mu\nu} -\frac{1}{4}\tanh2P\,\big(\eta_{\mu\nu}+H_{\mu\nu}\big)\Big)\frac{dX^\mu}{d\tau}\frac{dX^\nu}{d\tau} = 0\,,
\end{split}\end{equation}
where 
\begin{equation}
    \partial_\mu \equiv \frac{\partial}{\partial X^\mu} \,,\quad \partial_P \equiv \frac{\partial}{\partial P} \,,\quad F_{\mu\nu} \equiv 2\partial_{[\mu}A_{\nu]}
\end{equation}
and the $\mu,\nu$ indices are raised by $\eta^{\mu\nu}$. 

 We now consider a deviation of the original timelike geodesic on the SS--CGP background as given in \eqref{eq:geo1}. We write
\begin{equation}
    X^0 = \tau + \delta X^0 \,,\quad X^i = \delta X^i \,,\quad P = \delta P \,.
\end{equation}
We'll treat $\delta P$ and the $\tau$-derivatives of these deviations as small (the Newtonian limit), and will only consider terms of order 1 in perturbations. Here, the considered perturbations include $H_{\mu\nu}$, $A_\mu$, and $B$, the $\tau$-derivatives of $\delta X^\mu$ and $\delta P$, as well as $\delta P$ itself; so we'll neglect, for example, terms of the form
\begin{equation}
    B\frac{d^2\delta P}{d\tau^2} = \mc{O}(\text{pert}^2) \,.
\end{equation}
The resulting linearised equations for $X^\mu$ are then
\begin{equation}
    \frac{d^2\delta t}{d\tau^2} = 0\,,\quad \frac{d^2\delta X^i}{d\tau^2} = \frac{1}{2}\frac{\partial}{\partial\delta X^i} H_{00} \,,
\end{equation}
where we have used that $H_{\mu\nu}$ is independent of $X^0$. The first equation allows us to set $\delta t =0$, and so $X^0 = \tau$ at least in this linearised regime. Thus, we are allowed to interpret $\tau$ as the underlying manifold's time, which we'll write as $t$. The $\delta X^i$ equation is then Newton's equation, with a gravitational potential
\begin{equation}
    V_N(\delta X^i,\delta P) = -2m_{\text{particle}}H_{00}(\delta X^i,\delta P) \,,\label{eq:newt}
\end{equation}
where $m_{\text{particle}}$ is the small mass of the test particle following the geodesic. 

 Finally, we also have the $\delta P$ equation, which in our approximation reads
\begin{equation}
    \frac{d^2\delta P}{dt^2} + \frac{g^2}{2}\delta P - \frac{g^2}{2}\frac{\partial}{\partial\delta P}  H_{00} = 0 \,.
\end{equation}
where we have used that $A_\mu$ is independent of $X^0$. Using the Newtonian potential defined in \eqref{eq:newt}, and removing the $\delta$'s from $\delta X^i$ and $\delta P$ for convenience, we can rewrite their equations as
\begin{align}\label{eq:system}
    &m_{\text{particle}}\frac{d^2X^i}{dt^2} = -\frac{\partial}{\partial X^i} V_N \,,\\
    &m_{\text{particle}}\Big(\frac{d^2P}{dt^2} + \frac{g^2}{2}P \Big) = -g^2\frac{\partial}{\partial P} V_N \,.\label{eq:system2}
\end{align}

 In conclusion, the leading effect of perturbations about the SS--CGP background is through $V_N\propto H_{00}$.
\section{Perturbations and the Scalar Green Function}\label{se: PertH00}

 One method of finding solutions to the perturbation problem about the SS--CGP background is to find solutions to the perturbation problem of the 5-dimensional theory\footnote{Our convention for the Hodge dual is $\ast(dx^{m_1}\wedge\cdots\wedge dx^{m_p}) = \frac{1}{q!}\sqrt{|g|}\epsilon_{\ph{m_1\cdots m_p}n_1\cdots n_q}^{m_1\cdots m_p} dx^{n_1}\wedge\cdots\wedge dx^{n_q}$, where $\epsilon_{n_1\cdots n_D}$ with lowered indices is numerical, and $q = D-p$.}
\begin{equation}
    \mc{L}_5 = R{\ast} 1 - \frac{1}{2}d\Phi_i\wedge{\ast} d\Phi_i - \frac{1}{2}e^{\sqrt{2}\Phi_1}d\sigma\wedge{\ast} d\sigma - V{\ast} 1\,,
\end{equation}
obtained from type I supergravity reduced on $T^3\times S^2$, the details of which are presented in Appendix \ref{se:consistenttrunc}. The scalar potential $V$ is
\begin{equation}
    V = 2g^2e^{\sqrt{\frac{2}{5}}\Phi_2-\frac{8}{\sqrt{15}}\Phi_3}\Big(e^{-\sqrt{2}\Phi_1}+\sigma^2 + \frac{1}{4}e^{\sqrt{2}\Phi_1}(\sigma^2-2)^2 - 4e^{-\sqrt{\frac{2}{5}}\Phi_2+\sqrt{\frac{3}{5}}\Phi_3}\Big) \,.
\end{equation}
The Salam--Sezgin solution (which lifts to the SS--CGP background) of type I supergravity in $D=5$ is
\begin{equation}\label{eq:ss5}\begin{gathered}
    ds^2_5 = (\sinh2\rho)^{\frac{2}{3}}\Big(\eta_{\mu\nu}dx^\mu dx^\nu + \frac{1}{g^2}d\rho^2\Big) \,, \qquad e^{-\sqrt{2}\Phi_1} = (\tanh2\rho)^2 \,, \\
    e^{\sqrt{10}\Phi_2} = e^{\sqrt{15}\Phi_3} = (\sinh2\rho)^2\,,\qquad \sigma = \sqrt{2}\sech2\rho \,. 
\end{gathered}\end{equation}

Now consider a perturbation about the background \eqref{eq:ss5},
\begin{equation}\label{eq:pertmet}
    g_{MN} = (\sinh2\rho)^{\frac{2}{3}}\big(\ov g_{MN} + H_{MN}\big) \,,\quad \Phi_i = \ov\Phi_i + \phi_i \,,\quad \sigma = \ov\sigma + \Sigma \,, 
\end{equation}
where
\begin{equation}
    d\ov s^2_5 = \ov g_{MN} dX^MdX^N = \eta_{\mu\nu}dx^\mu dx^\nu + \frac{1}{g^2}d\rho^2 \,,
\end{equation}
and $\ov\Phi_i$ and $\ov\sigma$ are the background values of the scalars. Here, we have used $X^M=(x^\mu,\rho)$. For notational convenience, we will define a function $A(\rho)$ by
\begin{equation}
    e^{2A(\rho)} = (\sinh2\rho)^{\frac{2}{3}} \,, \label{eq:a}
\end{equation}
and also make the coordinate rescaling $\rho\to z(\rho) = \rho/g$. In the $X^M=(x^\mu, z)$ coordinate system, $\ov g_{MN} = \eta_{MN}$, and the linearised Ricci tensor of \eqref{eq:pertmet} is given by
\begin{align*}
    R^{(1)}_{MN} &= \frac{1}{2}\big(\partial^P\partial_MH_{NP} + \partial^P\partial_NH_{MP} - \Box_{5} H_{MN} - \partial_M\partial_NH\big) +\frac{3}{2}A'\big(\partial_MH_{Nz}+\partial_NH_{Mz} - \partial_zH_{MN}\big) \\
    &+ \Big(A'\Big(\partial^MH_{Mz} - \frac{1}{2}\partial_zH\Big) + \big(A''+3(A')^2\big)H_{zz}\Big)\eta_{MN} - \big(A''+3(A')^2\big)H_{MN} \numberthis \,,
\end{align*}
where $\Box_5 =\eta^{MN}\partial_M\partial_N$, and $H = \eta^{MN}H_{MN}$. Using the definition of $A(z)$ given in \eqref{eq:a}, we find
\begin{align*}
    R^{(1)}_{MN} &= \frac{1}{2}\big(\partial^P\partial_MH_{NP} + \partial^P\partial_NH_{MP} - \Delta_{5} H_{MN} - \partial_M\partial_NH\big) + g\coth(2gz)\big(\partial_MH_{Nz}+\partial_NH_{Mz}\big) \\
    &+ \Big(\frac{2}{3}g\coth(2gz)\Big(\partial^MH_{Mz} - \frac{1}{2}\partial_zH\Big) + \frac{4g^2}{3}H_{zz}\Big)\eta_{MN} - \frac{4g^2}{3}H_{MN} \,.\numberthis \label{eq:ricci}
\end{align*}
The operator $\Delta_5$ in \eqref{eq:ricci} is defined as
\begin{equation}\label{eq:cpsop}
    \Delta_5 = \Box_5 + 2g\coth(2gz)\partial_z = \eta^{\mu\nu}\partial_\mu\partial_\nu + g^2\left(\partial^2_\rho + 2\coth(2\rho)\partial_\rho\right) \,.
\end{equation} 
\subsection{Scalar Equations}

 Before looking at $H_{MN}$'s field equations, we will first consider the equations for $\phi_i$ and $\Sigma$. These come from the 5-dimensional field equations,
\begin{equation}
    \Box_{5}\Phi_{2,3} = \frac{\partial V}{\partial\Phi_{2,3}} \,,\quad \Box_{5}\Phi_1 = \frac{1}{\sqrt{2}}e^{\sqrt{2}\Phi_1}\big(\partial\sigma\big)^2 + \frac{\partial V}{\partial \Phi_1} \,,\quad \nabla_M\big(e^{\sqrt{2}\Phi_1}\partial^M\sigma\big) = \frac{\partial V}{\partial \sigma} \,.
\end{equation}
We use that
\begin{equation}
\sqrt{-g} = e^{5A}\Big(1 + \frac{1}{2}H+\cO(H^2)\Big)
\end{equation}
and find after some algebra
\begin{align}
    \Box_{5}\Phi_i =\;& e^{-2A}\Delta_5\ov\Phi_i + e^{-2A}\Big(\Delta_5\phi_i - \Big(\partial^MH_{Mz} - \frac{1}{2}\partial_zH\Big)\ov\Phi_i' - H_{zz}\Delta_5\ov\Phi_i\Big) \,, \\
    \nabla_M\big(e^{\sqrt{2}\Phi_1}\partial^M\sigma\big) =\;& e^{\sqrt{2}\,\ov\Phi_1-2A}\wt{\Delta}_5\ov\sigma + e^{\sqrt{2}\,\ov\Phi_1-2A}\Big(\wt{\Delta}_5\Sigma + \big(\sqrt{2}\phi_1 - H_{zz}\big)\wt{\Delta}_5\ov\sigma \nonumber \\
    &- \Big(\partial^MH_{Mz} - \frac{1}{2}\partial_zH\Big)\ov\sigma' + \sqrt{2}\ov\sigma'\phi_1'\Big) \,,
\end{align}
where the operator $\wt{\Delta}_5$ is defined as
\begin{equation}
    \wt{\Delta}_5 = \Delta_5 + \sqrt{2}\,\ov\Phi_1'\partial_z = \Delta_5 - 8g\csch(4gz)\partial_z \,.
\end{equation}
For the right-hand-side of the scalar equations, we have, to first-order in perturbations,
\begin{equation}
    \frac{\partial V}{\partial S_\alpha} = \frac{\partial V}{\partial S_\alpha}\Bigg\rvert_{\ov S} + \frac{\partial^2 V}{\partial S_\beta\partial S_\alpha}\Bigg\rvert_{\ov S}\delta S_\beta \,,
\end{equation}
where $S_\alpha =\{\Phi_i,\sigma\}$, $\delta S_\alpha = \{\phi_i,\Sigma\}$, and $\ov S$ denote the scalars, scalar perturbations, and background scalars respectively. Note that there is no $\frac12$ prefactor on the second derivative of the potential.

 We also have
\begin{equation}
e^{\sqrt{2}\Phi_1}\big(\partial\sigma\big)^2 = e^{\sqrt{2}\,\ov\Phi_1-2A}\big((\ov\sigma')^2 + (\sqrt{2}\phi_1-H_{zz})(\ov\sigma')^2 + 2\ov\sigma'\Sigma'\big) \,.
\end{equation}
Now a straightforward calculation shows that
\begin{equation}
    \frac{\partial^2V}{\partial\Phi_{2,3}\partial\Phi_1}\Bigg\rvert_{\ov S} = \frac{\partial^2V}{\partial\Phi_{2,3}\partial\sigma}\Bigg\rvert_{\ov S} = 0 \,.
\end{equation}
Consequently, $\{\phi_1, \Sigma\}$ and $\{\phi_2,\phi_3\}$ are decoupled from each other at this order in perturbations. Explicitly, the scalar equations are
\begin{align}
    \phi_1: \quad &\big(\Delta_5-8g^2\big)\phi_1 = -4g\csch(4gz)\Big(\sqrt{2}\mc{G}_z + 2\cosh(2gz)\big(\partial_z\Sigma + 2g\tanh(2gz)\Sigma\big)\big) \,, \\
    \Sigma: \quad &\big(\wt{\Delta}_5 - 8g^2(\sech(2gz))^2\big)\Sigma = -2g\sech(2gz)\tanh(2gz)\big(\sqrt{2}\mc{G}_z \nonumber \\
    &\qquad\qquad\qquad\qquad\qquad\qquad- 2\big(\partial_z\phi_1 - 2g\tanh(2gz)\phi_1\big)\big) \,, \\
    \phi_2: \quad &\Delta_5\phi_2 - \frac{8g^2}{5}\phi_2 + \frac{32}{5}\sqrt{\frac{2}{3}}g^2\phi_3 = 2\sqrt{\frac{2}{5}}g\big(\coth(2gz)\mc{G}_z +2gH_{zz}\big) \,, \\
    \phi_3: \quad &\Delta_5\phi_3 - \frac{56g^2}{15}\phi_3 + \frac{32}{5}\sqrt{\frac{2}{3}}g^2\phi_2 = \frac{4g}{\sqrt{15}}\big(\coth(2gz)\mc{G}_z +2gH_{zz}\big) \,, 
\end{align}
where for brevity, we define
\begin{equation}
\mc{G}_z = \partial^MH_{Mz} - \frac{1}{2}\partial_zH \,.
\end{equation}
We can solve one of the $\{\phi_1,\Sigma\}$ equations and one of the $\{\phi_2,\phi_3\}$ equations by requiring
\begin{equation}
    \Sigma = \sinh(2gz)\tanh(2gz)\phi_1 \,,\quad \phi_3 = \sqrt{\frac{2}{3}}\phi_2 \,.
\end{equation}
The resulting equations are
\begin{align}\label{eq:5dscalareom}
    \phi_1: \quad &\big(\Box_5 + 2g\csch(4gz)(3\cosh(4gz)-1)\partial_z + 8g^2\big)\phi_1 = -4\sqrt{2}g\csch(4gz)\mc{G}_z\,, \\
    \phi_2: \quad &\Delta_5\phi_2 + \frac{8g^2}{3}\phi_2 - 4\sqrt{\frac{2}{5}}g^2H_{zz} = 2\sqrt{\frac{2}{5}}g\coth(2gz)\mc{G}_z \,.
\end{align}

 The right-hand-side of the remaining scalar equations are proportional to $\mc{G}_z$. We recognise this to be the $z$-component of the de Donder combination. Since the supergravity equations are invariant under linearised diffeomorphisms
\begin{equation}
    H_{MN} \mapsto H_{MN} + \partial_{(M}\xi_{N)} +2A'\xi_z\eta_{MN} \,,\quad \xi_M\coloneqq \eta_{MN}\xi^N\,, \label{eq:diff} 
\end{equation}
with similar expressions for the transformations of $\phi_i$ and $\Sigma$, we can set $\mc{G}_z = 0$ as a gauge condition.

 In this gauge, $\phi_1$ decouples from the gravity sector. So, for simplicity, we will set $\phi_1=0$. The same is not true for $\phi_2$, as it couples to $H_{zz}$. For completeness, the equation for $\phi_2$ in this gauge is
\begin{equation} 
    \Delta_5\phi_2 + \frac{8g^2}{3}\phi_2 - 4\sqrt{\frac{2}{5}}g^2H_{zz} = 0 \,. \label{eq:p2}
\end{equation} 
\subsection{Einstein Equations}

 Now, let us analyse the equations of motion for $H_{MN}$. The linearised (trace-reversed) stress tensor $\theta^{(1)}_{MN}$, where $R^{(1)}_{MN} = \theta^{(1)}_{MN}$, is given by
\begin{align}
    \theta^{(1)}_{MN} &= \partial_{(M}\ov\Phi_2\partial_{N)}\delta\Phi_2 + \partial_{(M}\ov\Phi_3\partial_{N)}\delta\Phi_3 + \frac{e^{2A}}{3}\Big(\frac{\partial V}{\partial\Phi_2}\Big\rvert_{\ov S}\delta\Phi_2+\frac{\partial V}{\partial\Phi_3}\Big\rvert_{\ov S}\delta\Phi_3\Big)\eta_{MN} + \frac{e^{2A}}{3} V\Big\rvert_{\ov S}H_{MN} \nonumber \\
    &= \frac{\sqrt{10}}{3}g\coth(2gz)\big(\delta_{M z}\partial_N\phi_2 + \delta_{Nz}\partial_M\phi_2\big) + \frac{4\sqrt{10}}{9}g^2\phi_2\,\eta_{MN} - \frac{4g^2}{3}H_{MN} \,, \label{eq:linst1}
\end{align}
where we have used $\phi_1 = \Sigma =0$ and the $\mc{G}_z = 0$ gauge. 

 For simplicity, we now use our remaining diffeomorphism invariance to set the full de Donder gauge
\begin{equation}
    \partial^MH_{MN} - \frac{1}{2}\partial_NH = 0 \,.
\end{equation}
In this gauge, the linearised Ricci tensor given in \eqref{eq:ricci} becomes
\begin{equation}
    R^{(1)}_{MN} = -\frac{1}{2}\Delta_{5} H_{MN}+ g\coth(2gz)\big(\partial_MH_{Nz}+\partial_NH_{Mz}\big) + \frac{4g^2}{3}H_{zz}\eta_{MN} - \frac{4g^2}{3}H_{MN} \,,
\end{equation}
and the Einstein equations now simplify to
\begin{equation}\label{eq:einstein}
    \Delta_{5} H_{MN}-4g\coth(2gz)\partial_{(M}H_{N)z} - \frac{8g^2}{3}H_{zz}\eta_{MN} = -\frac{4\sqrt{10}}{3}g\coth(2gz)\delta_{z(M}\partial_{N)}\phi_2 - \frac{8\sqrt{10}}{9}g^2\phi_2\,\eta_{MN} \,.
\end{equation}

Firstly, we examine the $zz$ component of \eqref{eq:einstein}. It reads
\begin{equation}
    \Delta_5H_{zz} - \frac{8g^2}{3}H_{zz} + \frac{8\sqrt{10}}{9}g^2\phi_2 = 4g\coth(2gz)\Big(\partial_zH_{zz} - \frac{\sqrt{10}}{3}\partial_z\phi_2\Big) \,. \label{eq:zz}
\end{equation}
Recall that $\phi_2$ obeys \eqref{eq:p2}. Performing the field redefinitions
\begin{equation}
    H_{zz} = \frac{1}{\sqrt{2}}\phi + \varphi \,,\quad \phi_2 = \frac{3}{2\sqrt{5}}\phi \,,
\end{equation}
we find that \eqref{eq:p2} and \eqref{eq:zz} become
\begin{align}
    \phi_2: \quad &\Delta_5\phi = \frac{8\sqrt{2}}{3}g^2 \varphi \,, \label{eq:peqn} \\
    \varphi: \quad &\ov{\Delta}_5\varphi = 0 \,, \label{eq:vpeqn}
\end{align}
where
\begin{equation}\label{eq:cpsop2}
\ov\Delta_5 = \Box_5 - 2g\coth(2gz)\partial_z \,.
\end{equation}
Next, we have the $\mu z$ components of \eqref{eq:einstein}, which read
\begin{equation}\label{eq:muz}
    \Box_5H_{\mu z} = 2g\coth(2gz)\Big(\partial_\mu\varphi- \frac{1}{\sqrt{2}}\partial_\mu\phi\Big) \,. 
\end{equation}
Since $\phi$ and $\varphi$ are fixed, if the operator $\Box_5$ is invertible (which it is in the case of time-independent solutions), the solution to $H_{\mu z}$ is symbolically,
\begin{equation}\label{hmuzeqn}
    H_{\mu z} = 2g \frac{1}{\Box_5}\coth(2gz)\Big(\partial_\mu\varphi- \frac{1}{\sqrt{2}}\partial_\mu\phi\Big) \,.
\end{equation}
Finally, the $\mu \nu$ components of \eqref{eq:einstein} are
\begin{equation}
    \Delta_5H_{\mu\nu} = 4g\coth(2gz)\partial_{(\mu}H_{\nu)z} + \frac{8g^2}{3}\varphi\eta_{\mu\nu} \,.
\end{equation}
Since $\Delta_5$ is a linear operator, we can split $H_{\mu\nu}$ into three parts 
\begin{equation}
    H_{\mu\nu} = \mc{H}_{\mu\nu} + K_{\mu\nu} + J\eta_{\mu\nu} \,,
\end{equation}
where
\begin{align}
    &\Delta_5\mc{H}_{\mu\nu} = 0 \,,\label{eq:munu1} \\
    &\Delta_5K_{\mu\nu} = 4g\coth(2gz)\partial_{(\mu}H_{\nu)z}  \,, \label{eq:munu2} \\
    &\Delta_5J = \frac{8g^2}{3}\varphi \,. \label{eq:munu3} 
\end{align}
As with the $\mu z$ equation, all of the quantities on the right-hand-sides are known. In fact, \eqref{eq:munu3} is equivalent to \eqref{eq:peqn} with the choice $J = \phi/\sqrt{2}$. Thus, provided that appropriate boundary conditions are imposed, $\Delta_5$ can be inverted to solve \eqref{eq:munu1}-\eqref{eq:munu3}. 
\subsection{Time Independent \texorpdfstring{$H_{00}$}{H00}}

 For time-independent solutions, it is clear from \eqref{hmuzeqn} that $H_{0z} = 0$ is a solution, and consequently, we have $K_{00} = 0$ as a solution. Then, $H_{00} = \mc{H}_{00} -  J = \mc{H}_{00} - \phi/\sqrt{2}$, where, for completeness, $\mc{H}_{00}$ and $\phi$ satisfy
\begin{equation} \label{eq:final}
    \Delta_5\mc{H}_{00} = 0 \,, \quad \Delta_5\phi  = \frac{8\sqrt{2}}{3}g^2 \varphi \,, \quad \ov\Delta_5\varphi=0\,.
\end{equation}
with operators $\Delta_5$ and $\ov\Delta_5$ as defined in \eqref{eq:cpsop} and \eqref{eq:cpsop2} respectively. For solutions that are also radially symmetric in $\mb{R}^{1,3}$, we have, recalling that $z=\rho/g$,
\begin{equation}
    \Delta_5 = {\partial_r}^2 + \frac{2}{r}\partial_r + g^2\big({\partial_\rho}^2 + 2\coth2\rho\,\partial_\rho\big) \,,\quad \ov\Delta_5 = {\partial_r}^2 + \frac{2}{r}\partial_r + g^2\big({\partial_\rho}^2 - 2\coth2\rho\,\partial_\rho\big) \,,
\end{equation}
with $r$ the isotropic, spatial radius in $\mb{R}^{1,3}$. For simplicity, we will consider the case $\varphi=0$.

 In summary, we find that, as for Minkowski spacetime, the leading component of any perturbative solution for $H_{00}$ is given by a Green function associated with $\Delta_5$, the CPS operator \cite{Crampton:2014hia}. 
\section{Types II and III: Green Functions for the CPS Operator}\label{se: GFforCPS}
From the previous sections, the key to understanding the effective Newton potential is understanding the behaviour of $H_{00}$, which is given by a Green function of the CPS operator $\Delta_5$. Since we are interested in computing Newton's constant, which arises from the interaction of a small test particle orbiting a massive source, we consider the sourced equation,
\begin{equation}\label{diffeq}
    \Delta_5 G(r,\rho) = \frac{g\hat\kappa^2M\delta(r)\delta(\rho)}{4 \pi r^2 \mu(\rho)}= \frac{g\hat\kappa^2M\delta(r)\delta(\rho)}{4 \pi r^2 \sinh2\rho}\;,
\end{equation}
where $\hat\kappa^2$ is the five-dimensional Newton constant, $M$ is the mass of the source, $\mu(\rho) =\sinh2\rho$ is the appropriate measure for integrating over $\rho$, as seen from consideration of the $H^2_{\mu\nu}$ terms in the perturbative action, and $V_N(r,\rho) = -2m_{\text{particle}}G(r,\rho)$ is the Newtonian potential. Eigenfunctions of this operator have previously been studied in \cite{Crampton:2014hia}. There, time dependent solutions were found that localise gravity to four dimensions via a non-constant, normalisable zero mode $\xi_0$ of the $\rho$-dependent part of $\Delta_5$:
\begin{equation}\label{zerocps}
\left(\partial^2_\rho + 2\coth2\rho\,\partial_\rho\right)\xi_0 = 0 \,.
\end{equation} 
For convenience, we will call this $\rho$-dependent part the transverse operator $\Delta$. The solution to \eqref{zerocps} that is normalisable (and normalised) with respect to the measure $\mu(\rho)$ is
\begin{equation}\label{zeromode}
\xi_0 = \pm\frac{2\sqrt{3}}{\pi}\log\tanh\rho \,.
\end{equation}
The existence of this normalisable zero mode is special. In many examples of non-compact, transverse geometries realised in supergravity, such as BPS branes, the zero modes of the associated transverse operator are non-normalisable, and the coupling of the lower dimensional massless gravitational sector to all other modes in the effective field theory consequently vanishes. This is a consequence of the extended nature of the source in the higher dimension, as is the case with black spokes as discussed in Section \ref{se: blackspokes}.

In this section, we will first inspect asymptotic solutions to \eqref{diffeq} in order to understand the general behaviour of the Green functions. Then, following \cite{Crampton:2014hia}, we will solve for the Green functions by expanding in a basis of eigenfunctions of the transverse operator $\Delta$. There are two bases of eigenfunctions of interest which are distinguished by their boundary conditions in $\rho$. We will start with a mode decomposition where the Green function $G(r,\rho)$ vanishes at infinity and is continuous everywhere away from the source at the $(r,\rho) = (0,0)$ origin. We will find that this solution does not become effectively lower dimensional (\ie 4D) for a massless field, but instead becomes exponentially suppressed in the worldvolume radius $r$. We will secondly consider a mode decomposition that includes the zero mode \eqref{zeromode} as found in \cite{Crampton:2014hia}, and will find that the corresponding solution then does effectively become lower-dimensional, but that it also has logarithmic structure as $\rho\to0$. The relationship between these two cases will be explained in more detail in Section \ref{ldm}.
\subsection{Asymptotic Solutions}

 There are two main regimes where the $\Delta_5$ operator simplifies greatly. The first is when $\rho \ll 1$, and the second is when $\rho \gg 1$. The relevant asymptotic expansions of the operator are
\begin{equation}\label{rhosmallasymp}
    \Delta_5 = {\partial^2_r} + \frac{2}{r}\partial_r + g^2\left({\partial^2_\rho} + \frac{1}{\rho}\partial_\rho + \frac{4}{3}\rho\;\partial_\rho+\cO(\rho^3)\right)
\end{equation}
when $\rho$ is small, and
\begin{equation}\label{rholargeasymp}
    \Delta_5 = {\partial^2_r} + \frac{2}{r}\partial_r + g^2\left({\partial^2_\rho} + 2\;\partial_\rho + 4 \exp(- 2 \rho)\partial_\rho+ \cO(\exp(-4\rho))\right)
\end{equation}
when $\rho$ is large. Since we are interested in sources at $r=0$ near the $\rho=0$ submanifold, we should inspect the Green function in that limit. Specifically, by substituting the coordinate redefinition
\begin{equation}
    R^2 = g^2 r^2 + \rho^2 \;,\qquad \theta = \arctan\left(\frac{ \rho}{g r}\right)\;,
\end{equation}
\eqref{rhosmallasymp} becomes
\begin{equation}
    \Delta_5 = g^2 \left({\partial^2_R}+ \frac{4}{R} \partial_R + \frac{1}{R^2}\left({\partial^2_\theta}+\left(\cot(\theta)-2 \tan(\theta)\right)\partial_\theta\right)\right)+\cO(\rho)\;.
\end{equation}
For $\theta$-independent functions, this is just the Laplacian on $\mb{R}^5$. The precise normalisation of the radially symmetric Green function on $\mb{R}^5$ is given by
\begin{equation}\label{r5laplacian}
    \left({\partial^2_P}+\frac{4}{P}\partial_P\right)\frac{1}{2 \pi^2 P^3} = -\delta^5\left(X^M\right)\;,
\end{equation}
where $X \in \mb{R}^5$ and $P^2 = X\cdot X$. Defining $r^2=(X^1)^2+(X^2)^2+(X^3)^2$ and $\rho^2=(X^4)^2+(X^5)^2$, we may integrate over the angular dimensions in \eqref{r5laplacian} to find
\begin{equation}
    \left({\partial^2_P}+\frac{4}{P}\partial_P\right)\frac{1}{2 \pi^2 P^3}= -\frac{\delta(r)\delta(\rho)}{8 \pi^2 r^2\rho}\;.
\end{equation}
Now, the right-hand-side of \eqref{diffeq} in the $\rho\to0$ limit reads 
\begin{equation}
 \frac{g\hat\kappa^2M\delta(r)\delta(\rho)}{4 \pi r^2 \sinh2\rho} \sim \frac{g\hat\kappa^2M\delta(r)\delta(\rho)}{8 \pi r^2 \rho} +\cO(\rho^2)\;.
\end{equation}
Consequently, we expect the leading component of the Green function in the $R\to0$ limit to be
\begin{equation}\label{eq:Rtom3}
    G(r,\rho) = -\frac{g^4\hat\kappa^2M}{2 \pi\left(g^2 r^2+\rho^2\right)^\frac{3}{2}}+\cO\left(\frac{1}{R^2}\right)\;.
\end{equation}
There are two more regimes of interest. The first is when $r \gg 1$ and $\rho \ll 1$. For $r\gg1$, the differential operator takes the same form as in \eqref{rhosmallasymp}. However, we are interested in solutions expanded as a Laurent series about $r=\infty$. As such, we may use separability to find the leading term, which can be expanded in inverse integer powers of $r$. We have
\begin{equation}
    \Delta_5 f(r,\rho) = 0 \qquad\Rightarrow\qquad f(r,\rho) = \frac{A}{r}+\frac{B \log(\rho)}{r} + \cO\left(\frac{1}{r^2}\right)\;.
\end{equation}

The second regime is when $\rho \gg 1$. In this regime, the transverse operator $\Delta$ can be manipulated into the form of the Helmholtz operator in leading order by writing
\begin{equation}\label{helmholzeq}
    \Delta_5 \left(\frac{\exp(- \rho)}{r}f( g r,\rho)\right) = 0 \qquad \Rightarrow\qquad \left({\partial^2_x}+{\partial^2_\rho}-1\right) f(x,\rho) = 0\;,
\end{equation}
for $x=gr$. In this regime, we will be interested in the $f= \exp(-\rho)$ solution to \eqref{helmholzeq}, since this is the leading component of the $\tfrac{\xi_0(\rho)}{r}$ solution found in \cite{Crampton:2014hia}:
\begin{equation}
    \frac{\xi_0(\rho)}{r}\propto\frac{\log\tanh\rho}{r} = - \frac{2}{r} \exp(- 2 \rho)+\cO\left(\exp(-4 \rho)\right)\;.
\end{equation}

 Knowing the leading components of a Green function in asymptotic regimes, however, does not tell us how the solution for a given source near $(r,\rho) = (0,0)$ evolves as it approaches infinity in various directions. We are left with the question: does the solution with leading behaviour \eqref{eq:Rtom3} asymptote to the $\tfrac{\xi_0(\rho)}{r}$ solution at large $r$? And if so, what is the coefficient of this term?
\subsection{Type II: Higher-dimensional Black Holes; Neumann Boundary Conditions}

 One method of computing the relationship between the small $R=\sqrt{g^2r^2+\rho^2}$ and large $r$ asymptotic limits of a Green function is to appeal to a specific basis expansion. To specify a basis expansion, boundary conditions at $\rho=0$ and $\rho=\infty$ need to be imposed. Physically, we expect Green functions to vanish as $\rho\rightarrow \infty$, so we select Dirichlet conditions there.

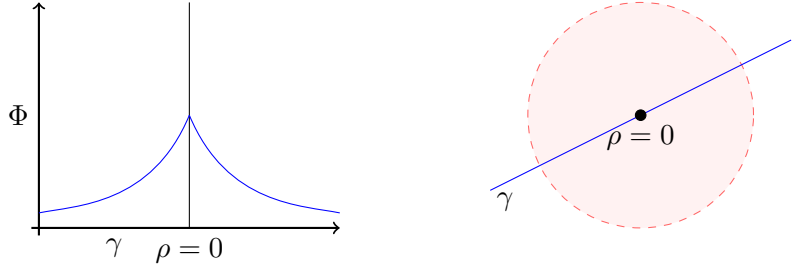
\begin{figure}[ht]
\begin{center}
\begin{tikzpicture}
    \filldraw[black] (-5,0) circle (0pt) node[anchor=east] {$\Phi$};
    \draw[thick,->] (-5,-1.6) -- (-5,1.5);
    \filldraw[black] (-4,-1.5) circle (0pt) node[anchor=north] {$\gamma$};
    \draw[thick,->] (-5.1,-1.5) -- (-1,-1.5);
    \filldraw[black] (-3,-1.5) circle (0pt) node[anchor=north] {$\rho=0$};
    \draw (-3,-1.5) -- (-3,1.5);
    \draw[blue] (-5,-1.3) .. controls (-4.5,-1.2) and (-3.5,-1.2) .. (-3,0);
    \draw[blue] (-1,-1.3) .. controls (-1.5,-1.2) and (-2.5,-1.2) .. (-3,0);
    
    \filldraw[color=red!60, fill=red!5, dashed](3,0) circle (1.5);
    \filldraw[black] (1.2,-.9) circle (0pt) node[anchor=north] {$\gamma$};
    \draw[blue] (1,-1) -- (5,1);
    \filldraw[black] (3,0) circle (2pt) node[anchor=north] {$\rho=0$};
\end{tikzpicture}
\caption{The value of a scalar function along a path through the origin}\label{fig:scalarfunctionpath}
\end{center}
\end{figure}

 For $\rho=0$, there is more freedom of choice. Since $\rho$ is a radial coordinate, any solution that is symmetric about the related angular coordinate -- $\chi$ in ten dimensions -- will be Neumann at $\rho=0$ for finite $r\ne0$, at which the solution will firstly be taken to be regular. That is, if we were to follow the value of a circularly symmetric scalar function along a path at fixed $\chi$ through $\rho=0$, we would see the value of that function mirrored around $\rho=0$. If the scalar function had a non-vanishing derivative at that point, this would appear as a cusp and our function would not be differentiable along such a path, as illustrated in Figure \eqref{fig:scalarfunctionpath}.\\
Consequently, we are interested in Green functions which obey special Neumann-Dirichlet conditions. That is $G^N$ which obey
\begin{equation}
    \partial_\rho G^N(r,\rho) \Big|_{\rho=0}=0\;,\qquad G^N(r,\rho)\Big|_{\rho\rightarrow\infty} \rightarrow 0\;.
\end{equation}
\indent We suppose $G^N$ may be written in terms of a superposition of separated solutions
\begin{equation}
    G^N(r,\rho) = \int_\mathcal{I} f^\omega(r) \zeta_\omega(\rho) d\omega\,,
\end{equation}
where $\mathcal{I}$ is the spectrum of the transverse operator, which will be discussed shortly, and, away from $r=\rho=0$ the functions are eigenfunctions of the worldvolume or transverse differential operators,\footnote{The sourcing of the Green function will also be handled shortly.}
\begin{align}
    \left({\partial^2_r}+\frac{2}{r}\partial_r \right) f^\omega(r) &= g^2\omega^2 f^\omega(r)\;,\\
    \left({\partial^2_\rho} + 2 \coth(2 \rho) \partial_\rho\right) \zeta_\omega(\rho) &= - \omega^2 \zeta_\omega(\rho)\;.\label{eq:transverseeigenvalueequation}
\end{align}
Given this separation of variables we may now restate the boundary conditions as conditions on the eigenfunctions
\begin{align}
    \partial_\rho G^N(r,\rho)\Big|_{\rho=0} = \int_\mathcal{I} f^\omega(r)\partial_\rho \zeta_\omega(\rho)\Big|_{\rho=0} d \omega = 0 \qquad &\Rightarrow\qquad \partial_\rho \zeta_\omega(\rho)\Big|_{\rho=0} = 0\;,\label{eq:lowerNeumann}\\
    G^N(r,\rho)\Big|_{\rho\rightarrow\infty}=\int_\mathcal{I} f^\omega(r) \zeta_\omega(\rho)\Big|_{\rho\rightarrow\infty} d \omega = 0\;\qquad&\Rightarrow\qquad \zeta_\omega(\rho)\Big|_{\rho\rightarrow\infty} = 0\;.\label{eq:upperDirichlet}
\end{align}
\indent Now, in order to write a separation of variables in the first place, we require orthonormalisation of the transverse basis eigenfunctions. That is, we require that any bound states $\zeta_i$ (with eigenvalue $\omega_i$) be Kronecker delta orthonormalised with respect to the transverse space inner product,
\begin{equation}
    \int_0^\infty\sinh(2 \rho)\zeta_i(\rho)\zeta_j(\rho) d\rho =\delta_{i,j}\;,
\end{equation}
and that any scattering states $\zeta_\omega$ be Dirac delta distribution orthonormalised,
\begin{equation}
    \int_0^\infty \sinh(2 \rho)\zeta_\omega(\rho) \zeta_\tau(\rho) d\rho =\delta(\omega - \tau)\;.
\end{equation}
To find the orthonormalised $\{\zeta_\omega\}$, we first note that they are required to be in a self-adjoint domain of $\Delta$, and we recall that two functions $(f,g)$ are in a self-adjoint domain of $\Delta$ if
\begin{equation}\begin{split}\label{selfadjoint}
    0=\int_0^\infty \sinh(2 \rho) \left(f\Delta g - g\Delta f\right) d \rho=  \sinh(2 \rho)\left( f \partial_\rho g - g \partial_\rho f\right)\Big|_{\rho=0}^{\rho\rightarrow\infty}\;.
\end{split}\end{equation}
From this we find the overlap integral of two solutions to the eigenvalue equation may be simplified
\begin{equation}
    \int_0^\infty \sinh(2 \rho)\zeta_\omega(\rho) \zeta_\tau(\rho) d\rho = \frac{\sinh(2 \rho)}{\omega^2-\tau^2}\left(\zeta_\omega \partial_\rho \zeta_\tau - \zeta_\tau \partial_\rho \zeta_\omega\right)\Big|_{\rho=0}^{\rho\rightarrow\infty}\;.
\end{equation}
We now pick boundary conditions so that these boundary terms vanish. The standard Neumann condition at $\rho=0$ is traded for a weaker generalised Neumann condition,
\begin{equation}\label{eq:lowergeneralNeumann}
    \sinh(2 \rho) \partial_\rho \zeta_\omega(\rho) \Big|_{\rho=0} = 0\;.
\end{equation}
For bound states we require a stronger generalised Dirichlet condition as $\rho\rightarrow\infty$
\begin{equation}\label{eq:upperboundgeneralDirichlet}
    \sqrt{\sinh(2 \rho)} \zeta_i(\rho) \Big|_{\rho\rightarrow\infty} =0\;,
\end{equation}
and for scattering states, we require a different, and again stronger generalised Dirichlet condition
\begin{equation}\label{eq:upperscatteringgeneralDirichlet}
    \sqrt{\sinh(2 \rho)} \zeta_\omega(\rho) \Big|_{\rho\rightarrow\infty} < \infty\;.
\end{equation}
The difference between equality and inequality in equations \eqref{eq:upperboundgeneralDirichlet} and \eqref{eq:upperscatteringgeneralDirichlet} is due to the requirement of bound states obeying $L_2((0,\infty),\sinh(2\rho)d\rho)$ normalisability but scattering states only being normalisable in the distributional sense.\\
\indent The detailed derivation of eigenfunctions which obey equations \eqref{eq:transverseeigenvalueequation}, \eqref{eq:lowergeneralNeumann}, and \eqref{eq:upperboundgeneralDirichlet} or \eqref{eq:upperscatteringgeneralDirichlet} requires a careful study of the limiting forms of special functions, and we have included this in Appendix \eqref{se:spegre}. To summarise the results in Appendix \eqref{se:spegre}: given Neumann-Dirichlet conditions on the transverse space, there are no bound states, the scattering states have eigenvalues $\omega>1$, and are given by
\begin{equation}\label{neumannbasis}
    \zeta_\omega(\rho) = \cN_\omega\; \cP_{-\frac{1}{2}+\frac{\sqrt{1-\omega^2}}{2}}\left(\cosh\left(2 \rho\right)\right)\;.
\end{equation}
Here $\cP_\nu(z)$ is a Legendre function of the first type defined with its branch cuts extending from $z=1$ to $z=-1$ and from $z=-1$ to $z\rightarrow-\infty$, while $\cN_\omega$ is a normalisation constant. Further details, including the value of $\cN_\omega$, are given in Appendix \eqref{se:spegre}.\\
\indent From \eqref{neumannbasis}, we see that any eigenfunction $\zeta_\omega$ which obeys our weaker general Neumann condition, \eqref{eq:lowergeneralNeumann}, also obeys the stronger Neumann condition \eqref{eq:lowerNeumann}. Similarly, if it obeys our weaker Dirichlet condition \eqref{eq:upperDirichlet}, it also obeys the stronger general Dirichlet conditions  \eqref{eq:upperboundgeneralDirichlet} or \eqref{eq:upperscatteringgeneralDirichlet}. This is a generic property of solutions to ordinary differential equations. For any boundary condition we write, there are entire families of boundary conditions we could write which do not change the set of solutions. If we consider solutions to the worldvolume eigenvalue problem that vanish as $r\rightarrow\infty$ we find that all such solutions vanish at least linearly fast, since all functions that vanish more slowly simply do not solve the differential equation.\\
\indent We may now use these basis functions to give a resolution of the identity following the analysis of Appendix \eqref{se:psgf}:
\begin{equation}\label{resolution}
    \int_1^\infty \zeta_{\omega}(\rho) \zeta_\omega(\eta) d \omega = \frac{\delta(\rho-\eta)}{\sinh(2\rho)}\;.
\end{equation}
To construct the full Green function, we recall that the solution to
\begin{equation}
\left(\partial^2_r + \frac{2}{r}\partial_r - g^2\omega^2\right)f_\omega = \frac{g\hat\kappa^2M}{4\pi r^2}\delta(r) \,,
\end{equation}
is 
\begin{equation}\label{wvmassive}
f_\omega = -\frac{g\hat\kappa^2M}{4\pi}\frac{\exp(-g\omega r)}{r} \,.
\end{equation}
Using \eqref{resolution} and \eqref{wvmassive}, the Green function is given by
\begin{equation}\label{greenfneq}
G^N(r,\rho-\eta) = -\int_1^\infty \frac{g\hat\kappa^2M\exp( -g\omega r)}{4 \pi r} \zeta_\omega(\rho)\zeta_\omega(\eta)d \omega   \;.
\end{equation}

 The closed-form expression of this integral is unknown to the authors. Using approximations, however, we can find its leading behaviour in various limits. Most importantly, we can see that the leading behaviour when $\eta\ll1$ and $R\ll1$ is
\begin{equation}
    -\int_1^\infty \frac{g\hat\kappa^2M\exp( -g \omega r)}{4 \pi r} \zeta_\omega(\rho)\zeta_\omega(\eta) d \omega=-\frac{g^4\hat\kappa^2M}{2 \pi\left(g^2 r^2+(\rho-\eta)^2\right)^\frac{3}{2}}+\cO\left(\frac{1}{R^2}\right)\;,
\end{equation}
which is exactly what is expected from analysis of the asymptotic operator given in \eqref{eq:Rtom3}.

 A similar analysis using approximate forms of special functions will give us the leading behaviour of the Green function $G(r,\rho)$ for $r \gg 1$ and $\rho \ll 1$. However, before stating the results in detail, we can find bounds on the leading behaviour by studying the form of the integrand in \eqref{greenfneq}. Specifically, if we consider the large $r$ limit, factorise the $\omega$ independent part of the integral, and change variables using $\tau = g \omega$, we find
\begin{equation}
    G^N(r,\rho-\eta) = -\frac{\hat\kappa^2M}{4 \pi r}\int_g^\infty \exp( - \tau r) \zeta_{\frac{\tau}{g}}(\rho) \zeta_{\frac{\tau}{g}}(\eta)d \tau\;.
\end{equation}
We recognise this to be the Laplace transform in $\tau$ with respect to frequency $r$ of some quantity $\theta(\tau - g)\zeta_{\frac{\tau}{g}}(\rho)\zeta_{\frac{\tau}{g}}(\eta)$, where $\theta$ is the Heaviside theta function. Since the integral starts from $\tau = g$, the Green function will be exponentially suppressed unless the Laplace transform of $\zeta_{\frac{\tau}{g}+g}(\rho)\zeta_{\frac{\tau}{g}+g}(\eta)$ grows exponentially in $r$. However, exponentially growing functions lie outside the region of convergence for the inverse Laplace transform. Therefore, we expect the Green function $G(r,\rho-\eta)$ to be exponentially suppressed when $r\gg1$. As seen in Appendix \eqref{se:spegre}, we in fact find 
\begin{equation}\label{expsup}
    G^N(r,\rho-\eta) = \exp(- g r)\left(-\frac{X}{r^2} +\cO\left(\frac{1}{r^3}\right)\right)\;,
\end{equation}
where $X$ is a $\rho$-dependent function. The $\tfrac{1}{r^2}\exp(- g r)$ behaviour at large $r$ in this Type II case is more characteristic of a massive theory in five spacetime dimensions than of a massless theory in four spacetime dimensions.

\subsection{Type III: Higher-dimensional Black Holes; Robin Boundary Conditions}

 The exponential suppression seen in \eqref{expsup} might seem surprising, as one might expect that the theory reduces to a massless gravitational theory in four dimensions. However, one should note that the $\zeta_\omega$ basis expansion used above did not include the zero mode found in \cite{Crampton:2014hia}, $\xi_0 \propto \log\tanh\rho$. Perhaps the choice of basis was at fault? We can repeat the analysis of the previous subsection except for now requiring that the boundary condition on the transverse modes now be the boundary condition that defines a self-adjoint domain for the transverse operator $\Delta$ including $\xi_0$. Using \eqref{selfadjoint}, this gives a generalised Robin boundary condition at $\rho=0$ and a Dirichlet condition at $\rho=\infty$:
\begin{equation}\label{robinBCS}
    \left(\sinh(2 \rho) \log\tanh\rho\,\partial_\rho - 2 \right) \xi_\omega(\rho) \Big|_{\rho= 0}=0\;,\qquad\sqrt{\sinh(2 \rho)}\;\xi_\omega(\rho)\Big|_{\rho\rightarrow \infty}<\infty\;.
\end{equation}

 Given these conditions, we find that the zero mode \eqref{zeromode} is the unique bound state with zero eigenvalue. The scattering states are separated from the $\omega=0$ bound state by a mass gap, with eigenvalues $\omega>1$, and can be written as
\begin{equation}
    \xi_\omega(\rho) = \cM_\omega \cQ_{-\frac{1}{2}+\frac{\sqrt{1-\omega^2}}{2}}\left(\cosh(2\rho)\right) + c.c.\;,
\end{equation}
where $\cQ_\nu$ is a Legendre function of the second type with branch cuts from $z=1$ to $z=-1$ and from $z=-1$ to $z\rightarrow-\infty$, while $\cM_\omega$ is a normalisation constant. Once again, we find the Green function for this system, now Type III, by invoking a resolution of the identity. It is given by
\begin{equation}\label{greensRD}
  G^R(r,\rho-\eta)=  -\frac{g\hat\kappa^2M}{4 \pi r} \xi_0(\rho)\xi_0(\eta)-\int_1^\infty \frac{g\hat\kappa^2M\exp( -g \omega r)}{4 \pi r} \xi_\omega(\rho)\xi_\omega(\eta) d \omega \;,
\end{equation}
where we put a superscript $R$ to distinguish it from $G^N$, the Green function obeying a Neumann boundary condition at $\rho=0$.  

Note that for any point except $\rho =0$, any function in the Robin basis $\{\xi_\omega(\rho)\}$ can be re-expanded as a superposition of the Neumann basis $\{\zeta_\omega(\rho)\}$ introduced in the previous subsection. However, such a re-expansion does not converge at the $\rho=0$ boundary. The key point about the $\{\xi_\omega(\rho)\}$ Robin basis is that the $G^R$ expansion in it {\em will} converge at the $\rho=0$ boundary, with $G^R$ consequently inheriting the Robin boundary condition \eqref{robinBCS} at $\rho=0$:
\begin{equation}
    \left(\sinh(2 \rho)\log\tanh\rho\,\partial_\rho-2\right) G^R(r,\rho-\eta)\Big|_{\rho=0}=0 \,,\quad G^R(r,\rho-\eta)\Big|_{\rho\to\infty}=0 \,.
\end{equation}

Looking at \eqref{greensRD}, the leading behaviour in the large $r$ regime is seen to be
\begin{equation}\label{typeIIIlargedist}
    G^R(r,\rho-\eta) = -\frac{g\hat\kappa^2M}{4 \pi r} \xi_0(\rho)\xi_0(\eta)+\cO\left(\exp(- g r)\right)\;.
\end{equation}
Consequently, we find that the $G^R$ Green function is effectively lower-dimensional at large $r$, corresponding to three spatial dimensions, but also that it diverges logarithmically when either $\rho$ or $\eta$ tends to zero.

The existence of a single normalisable bound state separated from a continuum of delta function normalisable states is a consequence of the P\"oschl--Teller integrable structure of the Schr\"odinger reformulation of the transverse wavefunction problem \cite{Crampton:2014hia}.

\section{Type III from Type II; Long Distance Mirrors}\label{ldm}

We have seen a variety of effective gravitational behaviours on the lower-dimensional worldvolume.  Is there a way to relate them? First, let's recall the Green functions $G^N$ and $G^R$ that respectively obey the Neumann and Robin boundary conditions at $\rho=0$: 
\begin{equation}
    G^N(r,\rho-\eta)=-\int_1^\infty \frac{g\hat\kappa^2M\exp( -g \omega r)}{4 \pi r} \zeta_\omega(\rho)\zeta_\omega(\eta) d \omega\;,
\end{equation}
and
\begin{equation}
    G^R(r,\rho-\eta) = -\frac{g\hat\kappa^2M}{4 \pi r} \xi_0(\rho)\xi_0(\eta)-\int_1^\infty \frac{g\hat\kappa^2M\exp( -g \omega r)}{4 \pi r} \xi_\omega(\rho)\xi_\omega(\eta) d \omega\,.
\end{equation}
Both $G^N$ and $G^R$ solve the same sourced equation \eqref{diffeq}. The only thing that distinguishes them is the boundary condition imposed at $\rho = 0$. Since they solve the same sourced equation, their difference solves the homogeneous equation, 
\begin{equation}
\Delta_5F = 0 \,,\quad F=G^R- G^N \,.
\end{equation}
In particular, the interpolating function $F$ must be regular everywhere in the $(r,\rho)$ plane even though $G^R$ and $G^N$ are divergent at certain points.  Due to the exponential suppression of $G^N$ as $r\to\infty$, it is clear that the leading behaviour of the homogeneous solution $F$ at large $r$ is
\begin{equation}\label{homogeneouslarger}
F(r,\rho-\eta) \sim -\frac{g\hat\kappa^2M}{4 \pi r} \xi_0(\rho)\xi_0(\eta) \,.
\end{equation}
Consequently, we can interpret this homogeneous solution as the localising mechanism that takes $G^N$, which has the expected higher-dimensional behaviour near the origin, and \say{flattens} it to give $G^R$, which exhibits lower-dimensional behaviour at large $r$ radius. The homogeneous solution $F$ has lower-dimensional structure at large $r$ radius, but remains regular near the $r$ origin. One may view this as akin to how a point source of light in four dimensions sandwiched between two three-dimensional mirrors will effectively give a three-dimensional intensity field at large distance. Another analogy might be to a sound source at a large distance down a corridor between two plane walls -- a \say{whispering corridor}.

In the following, we will outline a method of constructing the asymptotics of the homogeneous solution $F$ using the corresponding asymptotics of $G^N$ and $G^R$, and showing explicitly the regularity of $F$ near the origin. We will first demonstrate this method using a simpler example where the transverse space is a line interval $I=[-l,l]$, where the two boundaries correspond to the mirrors.  We will then apply this technique to the case at hand.  From the analogy with light sandwiched by mirrors, we will call this method \say{long distance mirrors}.\footnote{We learned this technique from private discussions with Carl Bender on an analogous treatment of the heat equation with a nontrivial set of boundary conditions \cite{benderprivate}.}

\subsection{Long Distance Mirrors and a Transverse Interval}

Let's consider the Green function on $\mb{R}^3\times[-l,l]$, which solves
\begin{equation}\label{greens4d}
    \left({\partial^2_r}+\frac{2}{r}\partial_r + {\partial^2_z}\right) G(r,z) = \frac{\delta(r)\delta(z)}{4 \pi r^2}\;,
\end{equation}
where $z\in[-l,l]$. If the transverse interval is infinitely large $(l\to\infty)$, then the exact solution, which is Dirichlet in all directions at infinity, is given by
\begin{equation}
    G^{(4)}(r,z)=-\frac{1}{4 \pi^2\left(r^2+z^2\right)}\;.
\end{equation}
On a transverse interval, however, we require different boundary conditions. One possible set of boundary conditions which trivialises this problem is 
\begin{equation}
    G(r,\pm l) =- \frac{1}{4 \pi^2\left(r^2+l^2\right)}\;.
\end{equation}
We are, however, more interested in \textit{special Neumann} conditions. That is,
\begin{equation}\label{specialneumann}
    \partial_z G(r,z)\Big|_{z=\pm l} = 0\;.
\end{equation}
We may then choose to write the Green function as
\begin{equation}\label{decomposition4d}
    G(r,z) =G^{(4)}(r,z)+F(r,z)= -\frac{1}{4\pi^2\left(r^2+z^2\right)}+F(r,z)\;.
\end{equation}
Since both $G$ and $G^{(4)}$ solve the same equation \eqref{greens4d}, the interpolating function $F$ must solve the unsourced Laplace equation,
\begin{equation}\label{homogeneous4d}
    \left({\partial^2_r}+\frac{2}{r}\partial_r + {\partial^2_z}\right) F(r,z) = 0\;.
\end{equation}
Furthermore, the boundary condition  \eqref{specialneumann} on $G$ combined with the boundary behaviour of $G^{(4)}$ allows us to deduce the following \textit{general Neumann} condition on $F(r,z)$,
\begin{equation}\label{gbc4d}
\left(\partial_z F(r,z)+\frac{2z}{4 \pi^2\left(r^2+z^2\right)^2}\right)\Bigg|_{z=\pm l} = 0\;. 
\end{equation}
We now invoke a basis decomposition of $F$ with respect to a complete set of orthonormalised eigenmodes of $\partial^2_z$, writing
\begin{equation}
F(r,z) = \frac{1}{\sqrt{2l}}F^0(r) + \frac{1}{\sqrt{l}}\sum_{n\neq0}\sin\left(\frac{\pi n}{2 l} z +\frac{\pi}{4}\left(1+(-1)^n\right)\right)F^n(r) \,.
\end{equation}
Equivalently, the constituent modes can be obtained from $F$ by the following projections:
\begin{equation}\label{zeroproj4d}
    F^0(r) = \int_{-l}^l \frac{1}{\sqrt{2 l}} F(r,z) dz 
\end{equation}
and 
\begin{equation}
    F^n(r) = \int_{-l}^l \frac{1}{\sqrt{l}}\sin\left(\frac{\pi n}{2 l} z +\frac{\pi}{4}\left(1+(-1)^n\right)\right) F(r,z) dz\;.
\end{equation}
The equations of motion obeyed by $F^0$ and $F^n$ can be obtained by projecting \eqref{homogeneous4d} over the eigenmodes, giving
\begin{equation}\label{zero4d}
F^0:\quad \int_{-l}^l \frac{1}{\sqrt{2 l}} \left({\partial^2_r}+\frac{2}{r}\partial_r + {\partial^2_z}\right) F(r,z) dz = 0\;,
\end{equation}
and
\begin{equation}\label{massive4d}
F^n:\quad \int_{-l}^l \frac{1}{\sqrt{l}}\sin\left(\frac{\pi n}{2 l} z +\frac{\pi}{4}\left(1+(-1)^n\right)\right)\left({\partial^2_r}+\frac{2}{r}\partial_r + {\partial^2_z}\right) F(r,z) dz = 0\;.
\end{equation}

The key steps of the long distance mirrors technique involve integrating by parts in the mode-expanded $(r,z)$ partial differential equation so as to produce ordinary differential equations for the various modes, in which the PDE boundary conditions give rise to ODEs with inhomogeneous terms. To see how this works, let's first focus on \eqref{zero4d}. We can integrate by parts to give
\begin{equation}
    \left({\partial^2_r}+\frac{2}{r}\partial_r\right)\int_{-l}^l \frac{1}{\sqrt{2 l}}\; F(r,z) dz +\left( \frac{1}{\sqrt{2 l}} \partial_z F(r,z) - F(r,z)\partial_z \frac{1}{\sqrt{2 l}}\right)\Big|_{-l}^l = 0\;.
\end{equation}
The first integral on the left simply returns $F^0$ as seen in \eqref{zeroproj4d}, and the boundary term is evaluated using \eqref{gbc4d}. The $F^0$ zero mode equation is then
\begin{equation}\label{zero4deqn}
    \left({\partial^2_r}+\frac{2}{r}\partial_r\right) F^0(r) =\sqrt{\frac{l}{2}}\frac{1}{\pi^2\left(r^2+l^2\right)^2}\;.
\end{equation}
The technique for deriving equations for the other modes is the same. We integrate \eqref{massive4d} by parts, giving
\begin{equation}
\left({\partial^2_r}+\frac{2}{r}\partial_r-\frac{\pi^2 n^2}{4 l^2}\right) \int_{-l}^l \xi_n(z) F(r,z) dz +\left(\xi_n(z) \partial_z F(r,z) - F(r,z)\partial_z  \xi_n(z)\right)\Big|_{-l}^l = 0\;,
\end{equation}
where we have defined $ \xi_n(z) =\tfrac{1}{\sqrt{l}}\sin\left(\tfrac{\pi n}{2 l} z +\tfrac{\pi}{4}\left(1+(-1)^n\right)\right)$. For the odd $n$ modes, this is
\begin{equation}\label{nodd4d}
    \left({\partial^2_r} + \frac{2}{r}\partial_r - \frac{\pi^2n^2}{4 l^2}\right) F^n(r) = 0\;,\quad \text{$n$ odd} \,,
\end{equation}
and for the even $n$ modes, this is
\begin{equation}\label{neven4d}
    \left({\partial^2_r} + \frac{2}{r}\partial_r - \frac{\pi^2n^2}{4 l^2}\right) F^n(r) = -\frac{1}{\sqrt{l}}\frac{4l}{4 \pi^2\left(r^2+l^2\right)^2}\;,\quad \text{$n$ even} \,.
\end{equation}
For regularity at $r=0$ and for vanishing of $F$ at infinity, we impose Neumann-Dirichlet boundary conditions on the modes,
\begin{equation}\label{nd4d}
    \partial_r F^{i}(r)\Big|_{r=0} = 0\;,\qquad F^{i}(r)\Big|_{r\rightarrow\infty} = 0\;,\quad i \in \mb{Z} \,.
\end{equation}
Of greatest interest is the zero mode. The solution to \eqref{zero4deqn} obeying the above boundary conditions is
\begin{equation}
    F^0(r)=-\sqrt{\frac{2}{l}}\frac{\tan^{-1}(\frac{r}{l})}{4 \pi^2r} \,.
\end{equation}
From this, we observe that $F^0$ encodes a lower-dimensional behaviour. When $r\to\infty$, one has asymptotically
\begin{equation}
    F^0(r) = -\sqrt{\frac{2}{l}}\frac{1 }{4 \pi r}+ \cO\left(\frac{1}{r^2}\right) \;.
\end{equation}
As $r\to\infty$, the modes $F^n$ with $n\neq0$ are exponentially suppressed, as is evident from the mass term that appears in their equations of motion \eqref{nodd4d} and \eqref{neven4d}. Therefore, in the large $r$ regime, the zero mode $F^0$ encodes the leading behaviour of the full solution $F(r,z)$. Recalling the relation \eqref{decomposition4d} between the homogeneous solution $F(r,z)$ and the full Green function $G(r,z)$, we find that for large $r$,
\begin{equation}\label{4dasymp}
    G(r,z) = \frac{1}{\sqrt{2 l}} F^0(r) + \cO\left(\frac{1}{r^2}\right) = -\frac{1}{4 \pi l}\frac{1}{r} + \cO\left(\frac{1}{r^2}\right)\;.
\end{equation}
We can now compare this to the Green function obtained by taking a superposition of higher-dimensional fundamental solutions, which is another way to obtain the Green function for the light source between mirrors:
\begin{equation}\begin{split}
    G(r,z) =&- \sum_{k=-\infty}^\infty \frac{1}{4 \pi^2\left(r^2+\left(z-l k\right)^2\right)}\\
    =&-\frac{1}{4\pi l}\frac{1}{r}\frac{\sinh\left(\frac{2\pi r}{l}\right)}{\cosh\left(\frac{2\pi r}{l}\right)-\cos\left(\frac{2\pi z}{l}\right)} \;.
\end{split}\end{equation}
For large $r$, we have
\begin{equation}
G(r,z) = -\frac{1}{4\pi l}\frac{1}{r} + \cO\left(\frac{1}{r^2}\right)\;,
\end{equation}
which agrees with the asymptotic behaviour \eqref{4dasymp} found by the long-distance mirrors technique, showing complete agreement between the two methods.
\subsection{Long Distance Mirrors and SS--CGP}
As we have learned from the above example with a transverse interval, the first thing we have to write down are the boundary conditions for $G^N$ and $G^R$. As found in Section \ref{se: GFforCPS}, these are
\begin{equation}\begin{gathered}\label{neumannforG}
    \partial_\rho G^N(r,\rho-\eta) \Big|_{\rho=0}\;,\qquad \partial_r G^N(r,\rho-\eta)\Big|_{r=0\;\text{and}\;\rho\neq\eta}=0\;,\\
    G^N(r,\rho-\eta)\Big|_{r\rightarrow\infty}= 0\;, \qquad G^N(r,\rho-\eta)\Big|_{\rho\rightarrow\infty}= 0\;,
\end{gathered}\end{equation}
and
\begin{equation}\begin{gathered}\label{robinbcforG}
    \left(\sinh(2 \rho)\log\tanh\rho\,\partial_\rho-2\right) G^R(r,\rho-\eta)\Big|_{\rho=0}=0\;,\qquad \partial_r G^R(r,\rho-\eta)\Big|_{r=0\text{ and }\rho\neq \eta}=0\;,\\ G^R(r,\rho-\eta)\Big|_{r\rightarrow\infty}=0\;,\qquad G^R(r,\rho-\eta)\Big|_{\rho\rightarrow\infty}=0\;.
\end{gathered}\end{equation}
We wish to underline that \eqref{neumannforG} are the boundary conditions for the Type II case, now expressed in terms of the full Type II Green function $G^N$, and \eqref{robinbcforG} are the boundary conditions appropriate for the Type III case, now expressed in terms of the full Type III Green function $G^R$. It is the unusual choice of boundary conditions in  \eqref{robinbcforG} that allows for the localisation of gravity on the brane worldvolume. 

Carrying on, we now decompose $G^R$ into a sourced and an unsourced part
\begin{equation}
    G^R(r,\rho-\eta) = G^N(r,\rho-\eta) + F(r,\rho-\eta)\;,
\end{equation}
where the interpolating function $F$ solves the unsourced Laplace equation,
\begin{equation}\label{homogeneous} 
\left({\partial^2_r}+\frac{2}{r}\partial_r + {\partial^2_\rho}+2 \coth(2 \rho) \partial_\rho\right) F(r,\rho-\eta) = 0 \,.
\end{equation}
From the boundary condition for $G^R$ we find the following general\footnote{That is, the right-hand-side does not vanish, as it does for special.} generalised\footnote{That is, there is $\rho$ functional dependence in the coefficients of the condition.} Robin condition for $F$, 
\begin{equation}
\left(\sinh(2 \rho)\log\tanh\rho\,\partial_\rho-2\right) F(r,\rho)\Big|_{\rho = 0} = -\left(\sinh(2 \rho)\log\tanh\rho\,\partial_\rho-2\right)G^N(r,\rho)\Big|_{\rho = 0}\;.
\end{equation}
When $r\ll1$, this reads
\begin{equation}\begin{split}
    &-\left(\sinh(2 \rho)\log\tanh\rho\,\partial_\rho-2\right)G^N(r,\rho)\Big|_{\rho = 0}\\
    &\qquad=\frac{2\hat\kappa^2M}{\left(r ^2+\eta^2\right)^{3/2}}-\left(\frac{6\hat\kappa^2M (\eta  \log (\rho )+\eta )}{\left(r ^2+\eta^2\right)^{5/2}}\rho+\cO\left(\rho^2\right)+\cO\left(\frac{1}{R}\right)\right)\Big|_{\rho = 0}\\
    &\qquad=\frac{2\hat\kappa^2M}{\left(r ^2+\eta^2\right)^{3/2}}+\cO\left(\frac{1}{R}\right)\;.
\end{split}\end{equation}
The interpolating function $F$ can be expanded in either  of the $\{\zeta_\omega\}$ or $\{\xi_\omega\}$ bases. Choosing the $\{\xi_\omega\}$ basis for convenience, the Laplace equation \eqref{homogeneous} gives
\begin{equation}
    \left({\partial^2_r}+\frac{2}{r}\partial_r -\omega^2\right) F^\omega(r) = -\mu(\rho)\left(\xi_\omega(\rho) \partial_\rho F(r,\rho-\eta)-\left(\partial_\rho \xi_\omega\right)F(r,\rho-\eta)\right)\Big|_{\rho=0}^{\rho\rightarrow\infty}\;.
\end{equation}
Here, we have projected \eqref{homogeneous} into the $\{\xi_\omega\}$ basis and have integrated by parts as in the previous subsection. For the $\omega = 0$ zero mode, this simplifies, in the $r\to 0$ limit, to
\begin{equation}\label{sscgpsmallr}
    \left({\partial^2_r}+\frac{2}{r}\partial_r\right) F^0(r) =  \pm \frac{2\sqrt{3}}{\pi} \frac{2\hat\kappa^2M}{\left(\eta^2+r^2\right)^\frac{3}{2}}+\cO\left(\frac{1}{R}\right)\;.
\end{equation}
We shall momentarily disregard the $\cO\left(\frac{1}{R}\right)$ corrections. The solution to \eqref{sscgpsmallr} is then
\begin{equation}\label{inside}
    F^0(r) = \pm\frac{2\sqrt{3}\hat\kappa^2M}{\pi}\left(\frac{1}{\eta} - \frac{\sinh^{-1}\left(\frac{r}{\eta}\right)}{r}\right)+\frac{c_1}{r} + k_1\;,
\end{equation}
where $c_1$ and $k_1$ are integration constants. Since $F^0$ must be regular at $r=0$, we must have $c_1 = 0$. Now let's consider the $\cO(\frac{1}{R})$ corrections. Since the explicit form of $G^N$ is not known, these cannot be written in closed-form. However, we do know that $G^N$ must vanish as $r\rightarrow\infty$, at least exponentially fast. So, as $r\to\infty$, the zero mode of $F$ must solve
\begin{equation}
    \left({\partial^2_r}+\frac{2}{r}\partial_r\right)F^0(r) = \pm \frac{A \exp(-r)}{r^2}\;,
\end{equation}
where $A$ is some unspecified constant given by the asymptotic form of $G^N$. Assuming that $F^0$ vanishes when $r\rightarrow\infty$, the solution to this is
\begin{equation}
    F^0(r) = \pm A\left(\frac{\exp(-r)}{r} + \text{Ei}(-r)\right) + \frac{c_2}{r} \;,
\end{equation}
where $\text{Ei}(-r)$ is the exponential integral function, and $c_2$ is a constant. We now define
\begin{equation}
F^{\rm in} = \pm\frac{2\sqrt{3}\hat\kappa^2M}{\pi}\left(\frac{1}{\eta} - \frac{\sinh^{-1}\left(\frac{r}{\eta}\right)}{r}\right)+ k_1\,,\quad F^{\rm out} = \pm A\left(\frac{\exp(-r)}{r} + \text{Ei}(-r)\right) + \frac{c_2}{r}\,,
\end{equation}
for inside and outside solutions respectively. We assume some crossover point $r=l$ where $\tfrac{\exp(-r)}{r^2}$ becomes a better estimate of $G^N$ than $\tfrac{1}{R^3}$ and require continuity of the functions, 
\begin{equation}
F^{\rm in}(r)\Big|_{r=l} = F^{\rm out}(r)\Big|_{r=l}\;,\quad \partial_r F^{\rm in}(r)\Big|_{r=l} = \partial_r F^{\rm out}(r)\Big|_{r=l}\;.
\end{equation}
Solving these junction conditions fixes the remaining constants $k_1$ and $c_2$\,:
\begin{equation}
\begin{split}
&k_1 = \pm\left(\frac{2\sqrt{3}\hat\kappa^2M}{\pi\sqrt{l^2+\eta^2}} - \frac{2\sqrt{3}\hat\kappa^2M}{\pi\eta} + A\,\text{Ei}(-l)\right) \,,\\
&c_2 = \pm\left(\frac{2\sqrt{3}\hat\kappa^2Ml}{\pi\sqrt{l^2+\eta^2}} - \frac{2\sqrt{3}\hat\kappa^2M\sinh^{-1}\left(\frac{l}{\eta}\right)}{\pi}-Ae^{-l}\right) \,.
\end{split}
\end{equation}
The constant $k_1$ is irrelevant to the large $r$ behaviour, but $c_2$ gives the lower-dimensional behaviour at large $r$. Specifically, when $\eta\ll1$, one has
\begin{equation}
c_2 = \pm\frac{2\hat\kappa^2M\sqrt{3}}{\pi}\log(\eta) +h(l,\eta)\;,
\end{equation}
where $h(l,\eta)=\cO(\eta^0)$. The independence of $l$ in the first term of $c_2$ shows that it is valid to estimate $F$ by matching $F^{\text{in}}$ and $F^{\text{out}}$. Ignoring $h(l,\eta)$ since it is finite as $\eta\rightarrow0^+$, we reconstruct the leading order of $F(r,\rho-\eta)$ when $r\gg1$ by multiplying our solution for $F^0$ by the zero mode $\xi_0(\rho)$ to find
\begin{equation}
    F(r,\rho -\eta) = \frac{12\hat\kappa^2M}{\pi^2}\log\tanh\rho \log(\eta)\frac{1}{r} +\cO(\eta^0) +\cO\left(\frac{1}{r^2}\right)\;,
\end{equation}
which agrees with our preliminary analysis \eqref{homogeneouslarger} in the $\eta\rightarrow0^+$ limit.
\subsection{Type III and Mass Gaps}
In order for $F^0$ to produce a $\tfrac{1}{r}$ behaviour for large $r$, it was crucial that the leading behaviour of $G^N$ change from a $\tfrac{1}{R^3}$ structure near the origin to a $\tfrac{\exp(-gr)}{r^2}$ structure for large $r$. If the leading behaviour of $G^N$ stayed like $\tfrac{1}{R^3}$ everywhere, then $F^0$ would have asymptotic structure $\tfrac{\log r}{r}$ for large $r$, as seen from \eqref{inside}. The reason that $G^N$ is exponentially suppressed for large $r$, as we saw in the previous section, is because the scattering states of the transverse operator $\Delta$ have eigenvalues starting from $\omega > 1$ rather than from $\omega =0$. There is a mass gap $\delta\omega = 1$, and it is this mass gap that allows for the emergence of lower-dimensional physics as $r\to\infty$. The immediate question now is whether a mass gap is necessary for all types of transverse geometries. Suppose we have a system with an $\mb{R}^{1,3}$ worldvolume and a non-compact transverse space whose geometry factorises into $\mb{R}^b\times\{\text{compact}\}$ near the worldvolume. For us, the SS--CGP solution has $b = 2$, recalling that the geometry of Eguchi--Hanson space is $\mb{R}^2\times S^2$ near $\rho=0$. Then, near the $(r,\rho)=(0,0)$ origin, the corresponding asymptotic equation for the zero-mode of $F$ is 
\begin{equation}
\left(\partial_r^2 + \frac{2}{r}\partial_r\right)F^0 = \frac{1}{(r^2+\eta^2)^{\frac{b+1}{2}}}\ +\hbox{subleading}  \,,
\end{equation}
with the right-hand-side being the leading behaviour of the $G^N$ function for this system near the origin. Let's now assume that the leading behaviour of $G^N$ is unchanged as we move towards $r=\infty$. This is the case when there is no mass gap. Then the large $r$ behaviour of $F^0$ is given by 
\begin{equation}
F^0 \sim \begin{cases} \log r \,,\quad b=1 \\ \ -\frac{\log r}{r} \,,\quad b=2 \\ - \frac{\sqrt{\pi}\Gamma\left(\frac{b-2}{2}\right)}{4\eta^{b-2}\Gamma\left(\frac{b+1}{2}\right)}\frac{1}{r} \,,\quad b\geq3\,.\end{cases}
\end{equation}
For $b\geq3$, the leading behaviour is $F^0 \propto 1/r$. This suggests that for such geometries, even in the case where there is no mass gap, as long as there is a normalisable zero mode, there is a possibility of localisation to lower-dimensional physics at large $r$. On the other hand, for $b\leq2$, it is clear that the leading behaviour of $G^N$ must be modified at large $r$ in order for there to be gravity localisation. In our case, $G^N$ is modified by a mass gap so that it becomes exponentially suppressed.

\subsection{Mass Gaps and the Randall--Sundrum Model}
We will now relate the ideas we have developed so far to the Randall--Sundrum model \cite{Randall:1999vf}, whose underlying geometry is an orbifolded $AdS_5$, 
\begin{equation}
ds^2_{\text{RS}} = \frac{1}{\Lambda^2z^2}(\eta_{\mu\nu}dx^\mu dx^\nu + dz^2) \,,
\end{equation}
where $z\in[1/k,\infty)$, $k>0$, $z=1/k$ is the orbifold point,\footnote{It is also common to change to the coordinate $z = e^{k|y|}/k$ with $y\in[0,\infty)$.} and the Ricci tensor is normalised to $R_{MN} = -4\Lambda^2g_{MN}$. The relevant perturbation equation for a transverse, traceless perturbation $\eta_{\mu\nu} \mapsto \eta_{\mu\nu} + H_{\mu\nu}$ is
\begin{equation}
    \left(\Box_4+\partial^2_z -\frac{3}{z}\partial_z\right) H_{\mu\nu}(x,z) = 0\;,
\end{equation}
where $\Box_4$ is the d'Alembertian on $\mb{R}^{1,3}$. We are therefore interested in time independent Green functions associated with this differential operator: 
\begin{equation}\label{rseqn}
    \left({\partial^2_r}+\frac{2}{r}\partial_r +{\partial^2_z} -\frac{3}{z}\partial_z\right) G(r,z) = \frac{\Lambda^3z^3 k^3\delta(r)\delta(z-1/k)}{4 \pi  r^2}\;,
\end{equation}
where $\mu(z)=1/z^3$ is the appropriate measure for integration over $z$, as can be seen from the $H^2_{\mu\nu}$ terms in the perturbative action, and we have chosen $\Lambda>0$ for convenience.\footnote{If $\Lambda<0$, the factor on the RHS of \eqref{rseqn} will just be $|\Lambda|^3$, as it comes from the square-root of the determinant of the metric.} As stated in \cite{Giddings:2000mu}, the solutions of interest are ones that obey the Neumann condition at $z=1/k$, 
\begin{equation}
\partial_zG(r,z)\Big|_{z=1/k} = 0 \,.
\end{equation}
The Neumann Green function $G^N(r,z)$ will be built out of eigenmodes of the transverse operator, which in this case is
\begin{equation}
\Delta_{\rm RS} = \partial^2_z -\frac{3}{z}\partial_z\,.
\end{equation}
The zero mode of $\Delta_{\rm RS}$ is given by
\begin{equation}
   \Delta_{\rm RS}  f(z) = 0 \implies f(z) = A + B z^4 =\zeta_0(z)+ \xi_0(z)\;.
\end{equation}
Applying the Neumann boundary condition, only the $\zeta_0(z) = A$ mode survives. Unlike the SS--CGP case, this constant mode is normalisable due to the fact that the transverse space of the Randall--Sundrum geometry has finite volume; the orbifolding of $AdS_5$ has essentially \say{compactified} the transverse space. In particular, we choose $A$ such that 
\begin{equation}
    \int_0^\infty \mu(z) \zeta_0(z)^2 dz = \int_{1/k}^\infty \frac{A^2}{z^3} dz = \frac{k^2}{2}A^2=1\;.\label{gneumannRS}
\end{equation}
We will choose the positive root, so $A=\sqrt{2}/k$. The Green function is then 
\begin{equation}\label{RSneumann}
G^N(r,z) =-\frac{\Lambda^3 k}{2\pi r} - \frac{\Lambda^2 k^3}{4\pi r}\int e^{-\omega r}\zeta_\omega(z) d\omega \,,
\end{equation} 
where the integral is over the eigenvalues of the scattering states of $\Delta_{\rm RS}$. From \eqref{RSneumann}, it is clear that the Green function exhibits lower-dimensional behaviour at large $r$.

We can use the long distance mirrors framework to understand the Randall--Sundrum model, giving a varied perspective on the treatment of Reference \cite{Giddings:2000mu}. As in the previous section, we write
\begin{equation}
    G^N(r, z) = G_{AdS}(r,z) + F(r, z) \;,
\end{equation}
where
\begin{equation}
G_{AdS}(r,z) = \Lambda^3 k^3\int_{-\infty}^{\infty}d\tau K_{AdS}(\tau,r,z) \,,\quad K_{AdS} = \frac{3}{\pi^2}\left(\frac{\xi}{2}\right)^4{}_2F_1\left(2,\frac{5}{2},3,\xi^2\right) \,.
\end{equation}
with $K_{AdS}$ the Green function on Euclidean $AdS_5$ with unit value of $\Lambda$ localised at $(\tau,r,z) = (0,0,1/k)$, and 
\begin{equation}
\xi = \frac{1}{k}\frac{2z}{z^2+k^{-2}+\tau^2+r^2} \,.
\end{equation}
The function $G_{AdS}$ is the standard time-independent Green function on $AdS_5$ which solves \eqref{rseqn}, with a source at $(r,z) = (0,1/k)$. Near this source,
\begin{equation}
G_{AdS} \sim -\frac{\Lambda^3}{2\pi^2}\frac{1}{r^2+(z-1/k)^2} \,,\quad (r,z) \to (0,1/k) \,,
\end{equation}
while as $r\to\infty$ with $z\to 1/k$, one has
\begin{equation}
G_{AdS} \sim -\frac{15\Lambda^3}{4\pi k^7}\frac{1}{r^7} \,, \quad (r,z) \to (\infty,1/k) \,.
\end{equation}
The function $F(r,z)$ solves the homogeneous equation
\begin{equation}
    \left({\partial^2_r}+\frac{2}{r}\partial_r +{\partial^2_z} -\frac{3}{z}\partial_z\right) F(r,z) = 0\;.
\end{equation}
As in the previous section, we can expand $F(r,z)$ in the $\{\zeta_\omega\}$ basis, and we find that the zero-mode projection $F^0(r)$ satisfies the equation
\begin{equation}\label{zeromodeRS}
    \left(\partial^2_r+\frac{2}{r}\partial_r\right) F^0(r) = \frac{1}{z^3} \zeta_0(z)\partial_z G_{AdS}(r,z)\Big|_{z=1/k}=\sqrt{2}k^2\partial_z G_{AdS}(r,z)\Big|_{z=1/k}\;.
\end{equation}
The closed-form solution to \eqref{zeromodeRS} is not known, but we can identify its asymptotic behaviour. For convenience, it is easier to work with a regularised limit on the right-hand-side of \eqref{zeromodeRS}, $z\to1/k + \eta$, where $\eta \ll 1$, as we did for SS--CGP. Then, in the $r\to0$ limit, we have
\begin{equation}\label{RSorigin}
 \left(\partial^2_r+\frac{2}{r}\partial_r\right) F^0(r) = \frac{\sqrt{2}\Lambda^3k^2\eta}{\pi^2(r^2+\eta^2)^2} \,.
\end{equation}
The solution to \eqref{RSorigin} that is regular at $r=0$, which we will denote by $F^{\rm in}$, is
\begin{equation}
F^{\rm in}(r) = -\frac{\Lambda^3k^2}{\sqrt{2}
\pi^2}\frac{\tan^{-1}\left(\frac{r}{\eta}\right)}{r} + c_1 \,.
\end{equation}
In the large $r$ limit, the leading part of \eqref{zeromodeRS} becomes 
\begin{equation}
 \left(\partial^2_r+\frac{2}{r}\partial_r\right) F^0(r) = -\frac{15\Lambda^3}{2\sqrt{2}k^2\pi r^7} \,,
\end{equation}
and the solution that vanishes at infinity, which we will denote by $F^{\rm out}$, is 
\begin{equation}
F^{\rm out} = \frac{c_2}{r}-\frac{3\Lambda^3}{8\sqrt{2}k^2\pi r^5} \,.
\end{equation}
Using the same junction technique as in the previous section, we consider a point $r=l$ where $F^{\rm in}$ and $F^{\rm out}$ and their first derivatives match. This fixes the constants $c_1$ and $c_2$ to be 
\begin{equation}
c_1 = \frac{\Lambda^3}{2\sqrt{2}\pi^2}\left(\frac{3\pi}{l^5}+\frac{2k^4\eta}{l^2+\eta^2}\right)\,,\quad c_2 = \frac{\Lambda^3}{8\sqrt{2}k^2\pi^2}\left(\frac{15\pi}{l^4}+\frac{8k^4l\eta}{l^2+\eta^2}-8k^4\tan^{-1}\left(\frac{l}{\eta}\right)\right) \,.
\end{equation}
The constant $c_1$ is irrelevant for the large $r$ behaviour, while $c_2$ encodes the lower-dimensional behaviour at large $r$. For $\eta\ll1$, we find
\begin{equation}
c_2 = -\frac{\Lambda^3k^2}{2\sqrt{2}\pi} + h(l,\eta) \,,
\end{equation}
where $h(l,\eta)=\mc{O}(\eta^0)$, so the leading behaviour of $F(r,z)$ as $r\to\infty$ is given by 
\begin{equation}
F(r,z) \sim \frac{\sqrt{2}}{k}F^0(r) \sim -\frac{\Lambda^3k}{2\pi r} \,,
\end{equation}
in agreement with \eqref{RSneumann}.

\subsection{A Type III Example Without Localisation}

An example of a Type III situation without localisation is a BPS magnetic $p$-brane in $d$-dimensions. The metric is 
\begin{equation}
ds^2_d = {\cal H}^{\frac{-d_m}{d-2}}\big(ds^2(\mb{R}^{1,p}) + {\cal H}\,ds^2(\mb{R}^{d_m+2})\big) \,, \label{eq:magbrane}
\end{equation} 
where $d_m = d-p-3$, and $\cal H$ is the radially symmetric harmonic function on $\mb{R}^{d_m+2}$, not to be confused with the trace of a gravitational perturbation. For simplicity, we will consider $d_m\geq1$, and work in spherical polar coordinates on the transverse space, so 
\begin{equation}
ds^2(\mb{R}^{d_m+2}) = dr^2 + r^2ds^2(S^{d_m+1}) \,,\quad {\cal H} = 1 + \frac{k}{r^{d_m}} \,,
\end{equation} 
where $ds^2(S^{d_m+1})$ is the metric on the unit $(d_m+1)$-sphere, and $k > 0$. The antisymmetric tensor that sources the brane has the field strength 
\begin{equation}
F_{(d_m+1)} = d_mk\vol(S^{d_m+1}) \,.
\end{equation} 
For theories with $d\leq10$, there is also a dilaton $\phi$ that is given by 
\begin{equation}
e^{\phi} = {\cal H}^{-\frac{a}{2}} \,,
\end{equation} 
where $a$ is the dilaton coupling constant. Considering a pure $H_{MN}$ gravitational perturbation as in previous sections, we find that the equation of motion for $H_{00}$ is given by
\begin{equation}
\partial^2H_{00} + \frac{1}{\mc{H}}\Big(\partial^2_r + \frac{d_m+1}{r}\partial_r\Big)H_{00} = 0 \,,
\end{equation} 
where $\partial^2=\eta^{\mu\nu}\partial_\mu\partial_\nu$. The relevant transverse eigenvalue equation for the transverse operator $\Delta=\partial^2_r + \frac{d_m+1}{r}\partial_r$ is then
\begin{equation}\label{eq:evaluemag}
\Big(\partial^2_r + \frac{d_m+1}{r}\partial_r\Big)\xi_{(m)}=-m^2{\cal H}\xi_{(m)}\, .
\end{equation}
The kernel of this equation is simply the Laplace equation on $\mb{R}^{d_m+2}$. The measure is given by 
\begin{equation}
\mu(r) = {\cal H}r^{d_m+1} \,,
\end{equation}
from which we observe that the zero eigenfunction of \eqref{eq:evaluemag} is non-normalisable. Let $\xi_{(m)} = r^{-\frac{d_m+1}{2}}\psi_{(m)}$. Then \eqref{eq:evaluemag} becomes the Schr\"odinger equation
\begin{equation}
-\frac{d^2\psi}{dr^2} + \frac{d_m^2-1}{4r^2}\psi = m^2{\cal H} \psi \,, \label{eq:tise}
\end{equation} 
with Schr\"odinger potential 
\begin{equation}
V(r) = \frac{d_m^2-1}{4r^2} \,.
\end{equation} 
\textbf{Case 1}: $d_m = 1$. $V(r) = 0$, and the Schr\"odinger equation reduces to $\psi'' = - m^2{\cal H} \psi$, where the primes indicate $r$ derivatives. The spectrum of the eigenvalue equation is then a continuous spectrum of scattering states. 

\noindent\textbf{Case 2}: $d_m \geq 2$. At large $r$, the Schr\"odinger equation becomes $\psi'' \approx -m^2 \psi$, as both $V(r)$ and the non-constant term in $\cal H$ are suppressed by inverse powers of $r$. As with the case of $d_m=1$, there are no bound state solutions. 

The analysis above shows that for $d_m\geq1$, the spectrum of the Schr\"odinger equation in \eqref{eq:tise} is continuous with no bound states. Consequently, there is no localisation of gravity to the worldvolume in such magnetic brane cases.

\section{The Worldvolume Newton Constant}\label{se: Newton effective}
Now that we have the effective Newton potentials for the Type I to Type III cases, we want to understand their physics. In particular, we are interested in studying whether these potentials have a lower-dimensional (four-dimensional) behaviour, and if they do, what is the effective, four-dimensional Newton constant. Let's begin by analysing the Type I and II cases. Type I solutions (black spokes), as we recall, correspond to worldvolume Ricci-flat solutions. Although these solutions are clearly four-dimensional in nature and actually solve a full nonlinear self-interacting equation, they do not correspond to a specific four-dimensional Newton constant. This is because of the worldvolume `trombone' symmetry that is inherent in the Ricci-flat family of solutions - any rescaling of the worldvolume metric by any positive constant remains a solution. More details on this are presented in Appendix \ref{AppendixA}. Due to the existence of this symmetry, there is no well-defined four-dimensional Newton's constant, as its value can always be scaled to a different value by a trombone transformation. The trombone symmetry, however, can be broken when we couple to external sources. In such cases, since our transverse space is non-compact, the usual argument of Reference \cite{Hull:1988jw} states that the four-dimensional Newton's constant vanishes. What this really means is that, in contrast to the Ricci-flat self interactions, coupling of the black spokes to external sources would be inherently higher-dimensional instead of four-dimensional.

As for Type II solutions, they clearly do not exhibit four-dimensional behaviour. From \eqref{expsup}, we found that the large-distance behaviour of the Type II potential is of a form corresponding to massive gravity in 5 dimensions. We will not be examining this further, but will move on to the Type III solutions, which as we can see from \eqref{typeIIIlargedist}, do indeed have four-dimensional behaviour at large $r$ worldvolume distance. However, it is not immediately obvious what the four-dimensional Newton constant should be, as it appears to depend on the non-compact transverse coordinate $\rho$. In the following, we offer three different approaches to identifying an appropriate four-dimensional Newton constant. 

\subsection{Newton's Constant or the Gravitational Coupling \texorpdfstring{$\kappa$}{kappa} in Type III}
\indent All three interpretive approaches centre on the geodesic equations derived in Section \ref{se: Newton's Constant}, which we will reproduce here in radial coordinates on the worldvolume,
\begin{align}
    R''(t) - \frac{l_W^2}{R(t)^3} &= -\frac{6g\hat\kappa^2M}{ \pi^3 R(t)^2} \log \tanh\left(P(t)\right)\log\tanh(\eta) + \cO\left(\frac{1}{R(t)^3}\right)\;,\label{eq:worldvolumegeo}\\
    P''(t) + \frac{g^2}{2}P(t) &= \frac{12g^3\hat\kappa^2M}{ \pi^3 R(t)} \frac{\log \tanh(\eta)}{\sinh(2 P(t))}+ \cO\left(\frac{1}{R(t)^2}\right)\;,\label{eq:transversegeo}
\end{align}
where $0<\eta\ll1$ is the transverse coordinate of the mass $M$ source, $R^2(t) = X^i(t)X^i(t)$, and $l_W^2$ is the worldvolume angular momentum. We note that although there is a sign difference between the two equations, the potential is attractive in both the worldvolume and transverse coordinates since for all $x>0$, $\tanh x\in(0,1)$, so the logarithmic terms are negative definite. 
\subsubsection*{Method 1: Fixed Points}
Our first approach is to find fixed points of the geodesic equation where $P(t)=\text{constant}$. We can then infer the four-dimensional Newton constant by substituting this fixed point into \eqref{eq:worldvolumegeo}. At first glance, however, we find that there are no fixed points for $P(t)$. In order to generate one, recall that our 5-dimensional system can be embedded in 10 dimensions, where the $\rho$ coordinate is paired with the angular coordinate $\chi$, forming an $\mb{R}^2$. So we may suppose that by restoring nontrivial $\chi$ dependence, there will be an additional angular momentum term in \eqref{eq:transversegeo}. More precisely, we have
\begin{equation}
    P''(t) + \frac{g^2}{2}P(t) - \frac{l_T^2}{P(t)^3} = \frac {6g^3 \hat\kappa^2 M}{\pi^3  R(t)}\frac{\log(\eta)}{P(t)} + \cO\left(\frac{1}{R(t)^2}\right)\;,
\end{equation}
where $l_T$ is the transverse angular momentum.
If we ignore the higher order corrections involving the radius $R(t)$ and take $P(t) = P$ to be constant, this equation simplifies to
\begin{equation}\label{eq:quartic P fixed point}
    \frac{g^2}{2}P^4 -\frac {6g^3 \hat\kappa^2 M}{\pi^3  R(t)}\log(\eta)P^2 - l_T^2 = 0\;.
\end{equation}
The only positive solution for this is 
\begin{equation}
    P = 2^{\frac{1}{4}}\sqrt{\frac{l_T}{ g}} + \frac{3 g^\frac{3}{2} \hat\kappa ^2 M \log (\eta )}{2^{1/4}\pi ^3 R(t) \sqrt{l_T}}+ \cO\left(\frac{1}{R(t)^2}\right)\;.
\end{equation}
\indent Of course, we can find the leading order of this expression by simply suppressing the quadratic term in $P$ in equation \eqref{eq:quartic P fixed point}. This reflects the structure of the background: since there is an attractive potential, there is a stable circular orbit where
\begin{equation}\label{stableorbit}
    P = 2^\frac{1}{4}\sqrt{\frac{l_T}{ g}}\;.
\end{equation}
The additional attractive potential from the mass $M$ source `squeezes' this orbit, but at large world volume radius this squeezing fades out. If we suppose there is some minimum non-zero transverse angular momentum $l_T$, as in the Bohr--Sommerfeld quantisation condition, then we may suppose that $P$ takes this value. One may make a similar interpretation for the value of the mass $M$ source transverse coordinate $\eta$.

Substituting \eqref{stableorbit} into \eqref{eq:worldvolumegeo}, we then find that the $R(t)$ equation becomes 
\begin{align}
    R''(t) - \frac{l_W^2}{R(t)^3} &= -\frac{6g\hat\kappa^2M}{\pi^3 R(t)^2} \left(\log \tanh\left(2^\frac{1}{4}\sqrt{\frac{l_T}{ g}}\right)\right)^2 + \cO\left(\frac{1}{R(t)^3}\right)\;,\\
    &\approx - \frac{6g \hat \kappa^2 \log\left(\sqrt 2\frac{ g}{l_T}\right)^2 M}{ 4 \pi^3 R(t)^2}\;.
\end{align}
If we compare this to the usual radial geodesic equation in 4 dimensions
\begin{equation}
    r''(t) - \frac{l_W^2}{r(t)^3} = -\frac{\kappa^2 M}{4 \pi r(t)^2}\;,
\end{equation}
we find a value for the effective four-dimensional gravitational coupling
\begin{equation}
    \kappa = \frac{\sqrt{6 g}}{\pi}\Big|\log\left(\frac{\sqrt2l}{g}\right)\Big|\hat \kappa\;,
\end{equation}
where we recall that $\hat \kappa$ is the five-dimensional gravitational coupling constant.

\subsubsection*{Method 2: Quantum Localisation}
We can go beyond the above semiclassical picture, if we want to consider that our geodesic equation becomes nonsingular due to quantum effects. Specifically we may ask what the instantaneous worldvolume radial force is on a purely quantum test particle, defined by a separable wavefunction
\begin{equation}
    \Psi(x^i,\rho) = \psi(x^i) \phi(\rho)\;,
\end{equation}
where we assume that the worldvolume wavefunction $\psi$ is some Gaussian wave packet with a negligible width compared to the worldvolume radius $r$ of its centroid. To apply a quantum mechanical analysis, we note that our geodesic equations can be obtained from the following Lagrangian:
\begin{equation}
    L=\frac{1}{2}\left(\frac{d}{dt}X^i(t)\right)^2+\frac{1}{2g^2}\left(\frac{d}{dt}P(t)\right)^2- \frac{1}{2} P(t)^2- \frac{\mu}{R(t)}\log\tanh(P(t))\;,
\end{equation}
where $\mu= \tfrac{12 g\hat \kappa^2 M}{\pi^3} \log \tanh(\eta)$. If we assume that the $X^i$ are effectively constant, the associated Hamiltonian is
\begin{equation}
    H = \frac{1}{2}\Pi(t)^2 + \frac{g^2}{2} P(t)^2+ \frac{\mu}{r} \log \tanh(P(t))\;.
\end{equation}
Therefore, we may study functions $\phi_{E}(\rho)$ that solve the associated time independent Schr\"odinger equation (TISE),
\begin{equation}\label{TISE}
    E\phi_{E}(\rho) = \left(-\frac{\hbar^2}{2} \frac{d^2}{d\rho^2}+\frac{g^2}{2}\rho^2+ \frac{\mu}{r} \log \tanh(\rho)\right) \phi_{E}(\rho)\;,
\end{equation}
and we will focus on the ground state $\phi_{0} =\phi$, as we are interested in small (low-energy) quantum excitations. \\
\indent The TISE \eqref{TISE} was derived with the assumption that the worldvolume motions $X^i$ are effectively constant. This is a good approximation when $r^2=X^iX^i\gg1$. Now, if we assume that $\rho$ is finite, then along with the assumption $r^2\gg1$, the TISE asymptotes to the equation describing a quantum harmonic oscillator, with its well-known solutions. The ground state, in particular, is \footnote{Since the domain is the positive real line, we are a factor of $\sqrt{2}$ different from the standard normalisation of the ground state of the quantum harmonic oscillator.}
\begin{equation}\label{phiharmonic}
    \phi(\rho) = \sqrt{2}\left(\frac{g}{\pi \hbar}\right)^\frac{1}{4} \exp\left(-\frac{g\rho^2}{2\hbar}\right) + \cO\left(\frac{\mu}{r}\right)\;.
\end{equation}
For $\rho\ll1$ on the other hand, the logarithmic term in the TISE is no longer negligible even in the large $r$ approximation, and the equation for the ground state at fixed $r$ approximates instead to
\begin{equation}
    \left(-\frac{\hbar^2}{2} \frac{d^2}{d\rho^2}+ \frac{\mu}{r}\log(\rho)\right) \phi(\rho)=0\;.
\end{equation}
To our knowledge, the exact solution to this differential equation is unknown. However, if we make a WKB approximation (noting that the $f_{n}(\rho)$ will in general be complex)
\begin{equation}
    \phi(\rho) = \exp\left(\frac{1}{\hbar} f_{-1}(\rho)+ f_0(\rho)+ \hbar f_1(\rho)+\cO(\hbar^2)\right)\;,
\end{equation}
then, to leading order in $\hbar$, we find,
\begin{equation}
    \left(\frac{d}{d\rho} f_{-1}(\rho)\right)^2 = \frac{2 \mu}{r} \log(\rho)\;,
\end{equation}
which has solutions
\begin{equation}
    f^{\pm}_{-1} = k^{\pm} \pm i\sqrt{\frac{2\mu}{r}} \left(\rho\sqrt{-\log\rho}-\frac{\sqrt{\pi}}{2}\text{Erf}\left(\sqrt{-\log \rho}\right)\right)\;,
\end{equation}
where $k^{\pm}$ are integration constants, and $\text{Erf}$ is the error function. Therefore in the $\rho\ll1$ regime, $\phi$ is given by a superposition
\begin{equation}\begin{split}
    \phi(\rho) &= A \exp\left(\frac{i}{\hbar}\sqrt{\frac{2\mu}{r}} \left(\rho\sqrt{-\log\rho}-\frac{\sqrt{\pi}}{2}\text{Erf}\left(\sqrt{-\log \rho}\right)\right)
   \right)\\
   &\quad+ B\exp\left(-\frac{i}{\hbar}\sqrt{\frac{2\mu}{r}}\left(\rho\sqrt{-\log\rho}-\frac{\sqrt{\pi}}{2}\text{Erf}\left(\sqrt{-\log \rho}\right)\right)
   \right)\;.\label{phiWKB}
\end{split}\end{equation}
Since $\phi$ is a ground-state quantum wavefunction, we will require that it obey the special Neumann boundary condition at $\rho=0$:  
\begin{equation}\label{radneumann}
\partial_\rho\phi\big\rvert_{\rho=0} = 0 \,.
\end{equation}
This is the same condition as that obeyed by the quantum harmonic oscillator ground state \eqref{phiharmonic}, as is appropriate for an S-wave ground state when one recalls that Equation \eqref{TISE} is the radial part of a transverse two-dimensional Schr\"odinger problem in $(\rho,\chi)$. Condition \eqref{radneumann} relates the coefficients $A$ and $B$:
\begin{equation}
A = B\exp\left(i\sqrt{\frac{2\pi\mu}{r}}\right) \,.
\end{equation}
\indent We can now determine the remaining coefficient $B$ by matching the large $r$ limit of \eqref{phiWKB} with the harmonic oscillator ground state \eqref{phiharmonic} at $\rho=0$, which gives
\begin{equation}
    B = \sqrt{\frac{g}{\pi\hbar}}\;.
\end{equation}
In particular, since we can match \eqref{phiWKB} with \eqref{phiharmonic}, we can, up to corrections of order $\mc{O}(\mu/r)$, compute expectation values using just the harmonic oscillator. The expectation value we are interested in is the transverse-space dependent part of the right-hand-side of \eqref{eq:worldvolumegeo}. This allows us to deduce the four-dimensional effective Newton constant. Explicitly, we find
\begin{equation}
    \kappa^2 = \frac{24 g\hat \kappa^2}{\pi^2}\left\langle \log\tanh(\rho)\right\rangle\log\tanh(\eta)\;,
\end{equation}
with the expectation value for an operator $f(\rho)$ defined as
\begin{equation}
    \left\langle f(\rho)\right\rangle = \int_0^\infty 2 \sqrt{\frac{g}{\pi \hbar}}\exp\left(- \frac{g \rho^2}{\hbar}\right) f(\rho)d \rho+\cO\left(\frac{\mu}{r}\right)\;.
\end{equation}
We may similarly choose to consider both the test particle and the source to be governed by the same transverse quantum Schr\"odinger problem. Given that, we find the effective four-dimensional gravitational coupling at large $r$ distance
\begin{equation}
    \kappa =- \sqrt{24g}\frac{ \hat \kappa}{\pi}\left\langle \log\tanh(\rho)\right\rangle\;.
\end{equation}
We are unable to compute such expectation values analytically. But, if we set $\hbar/g=1$, we can give a numerical approximation:
\begin{equation}\label{eq:qhokappa}
    \kappa=\sqrt{g} \left(1.73338\ldots\right)\hat\kappa\;.
\end{equation}
\subsubsection*{Method 3: Smeared Transverse Expectation Values}
Of course, calculating expectation values given some transverse profile function does not require a fully quantum treatment. We can instead imagine measuring the instantaneous acceleration of a particle whose transverse position is drawn from a smeared distribution of possible positions in the transverse direction.\\
\indent We may suppose that the test particles have $P(0)=P$ and suppose that the probability $\cP$ of $P$ taking a given value between $0<a<b<\infty$ is
\begin{equation}
    \cP(a < P < b) = \int_a^b f_P(\rho) d\rho\;,
\end{equation}
where we define our random variable, $P$, by its probability density function $f_P$. The average instantaneous acceleration we measure for a test particle drawn from this distribution is
\begin{equation}
    \left\langle R''(0)\right\rangle - \left\langle\frac{l_W^2}{R(0)^3}\right\rangle = -\left\langle \frac{6 \hat \kappa^2 M}{\pi^3 g R(0)^2}\log\tanh(P(0))\log\tanh(\eta)\right\rangle + \cO\left(\frac{1}{R(0)^3}\right)\;.
\end{equation}
Assuming $R(0)$ and $P$ are independent variables and our probability density function is correctly normalised, \ie $\left\langle1\right\rangle = 1$, then
\begin{equation}
    R''(0)-\frac{l_W^2}{R(0)^3}=-\frac{6g \hat \kappa^2 M}{\pi^3 R(0)^2}\left\langle\log\tanh(P)\right\rangle \log\tanh(\eta) + \cO\left(\frac{1}{R(0)^3}\right)\;.
\end{equation}
Here
\begin{equation}\label{eq: PDF expectation}
    \left\langle\log\tanh(P)\right\rangle = \int_0^\infty f_P(\rho) \log\tanh(\rho)d\rho\;.
\end{equation}
\indent We might choose to study any number of random distributions, but, given the suggestive form of the right-hand-side of equation \eqref{eq: PDF expectation} we will take
\begin{equation}
    f_P(\rho) = \mu(\rho) \xi_0(\rho)^2 = \frac{12}{\pi^2} \sinh(2\rho) \left(\log\tanh(\rho)\right)^2\;.
\end{equation}
Given this,
\begin{equation}
    \left\langle\log\tanh(P)\right\rangle = \frac{9 \zeta(3)}{\pi^2}\;,
\end{equation}
with $\zeta(x)$ the Riemann zeta function. This determines the four-dimensional $\kappa$ to be
\begin{equation}
    \kappa^2 = -\frac{216 \zeta(3)g\hat \kappa^2}{\pi^4} \log\tanh(\eta)\;.
\end{equation}
We can similarly average to get an expected value for $\log\tanh(\eta)$, to find
\begin{equation}\label{eq:smearkappa}
    \kappa = \sqrt{6 g}\frac{18 \zeta(3)}{\pi^3}\hat \kappa\;.
\end{equation}
We may compare this with the numerical value of $\kappa$ given by the quantum treatment of the geodesic equation  \eqref{eq:qhokappa}, finding here
\begin{equation}
    \kappa = \sqrt{g}\left(1.70932\ldots\right)\hat \kappa\;,
\end{equation}
and observe that these two approaches calculations agree to $3$ parts in $100$. Although the numerical result from the quantum treatment required setting $\hbar/g=1$, the result will not change significantly if $\hbar/g$ is set to another finite constant. This is because the quantum expectation value is dominated by behaviour of $\exp(-g\rho^2/\hbar)\log\tanh\rho$ near the origin, which only deviates very slowly as a function of the ratio $\hbar/g$. The result of \eqref{eq:smearkappa} also agrees precisely with the value found in Reference \cite{Crampton:2014hia} for the four-dimensional graviton self-coupling $\kappa$, up to corrections arising from the compactification of higher transverse dimensions other than $\rho$.
\subsubsection*{The Force Lines of the Newtonian Potential}
In order to help visualising the effect of the source near the origin on a test particle at some distance $r$ away, and specifically to show how the resulting near field evolves into the far field, we have made approximate illustrations for Type II and Type III potentials.\\
\indent These images were created by taking the leading orders of the potential in the near ($R\ll1$) and far ($r\sim1$) field limits and interpolating. The change brought about by the source perturbation needs to be considered in comparison to the effect of the unperturbed SS--CGP background. The effect of the background is a uniform attraction to $\rho=0$ proportional to $\rho$. At small values of $\rho$, or for relatively massive sources, then this background effect may be neglected. There is one additional scale of relevance, which  is the ratio of $g$, the SS--CGP background parameter, to $\eta$, the height above the $\rho=0$ plane at which the source is placed. In our illustrations we have chosen $\frac{\eta}{g}=0.1\,.$ We did not take any obvious limits, such as $\frac{\eta}{g}\rightarrow0$ or $\infty$, because the Type III solution becomes infinite or vanishes in those limits respectively.\\
\begin{figure}[t]
    \centering
    \includegraphics[width=\textwidth]{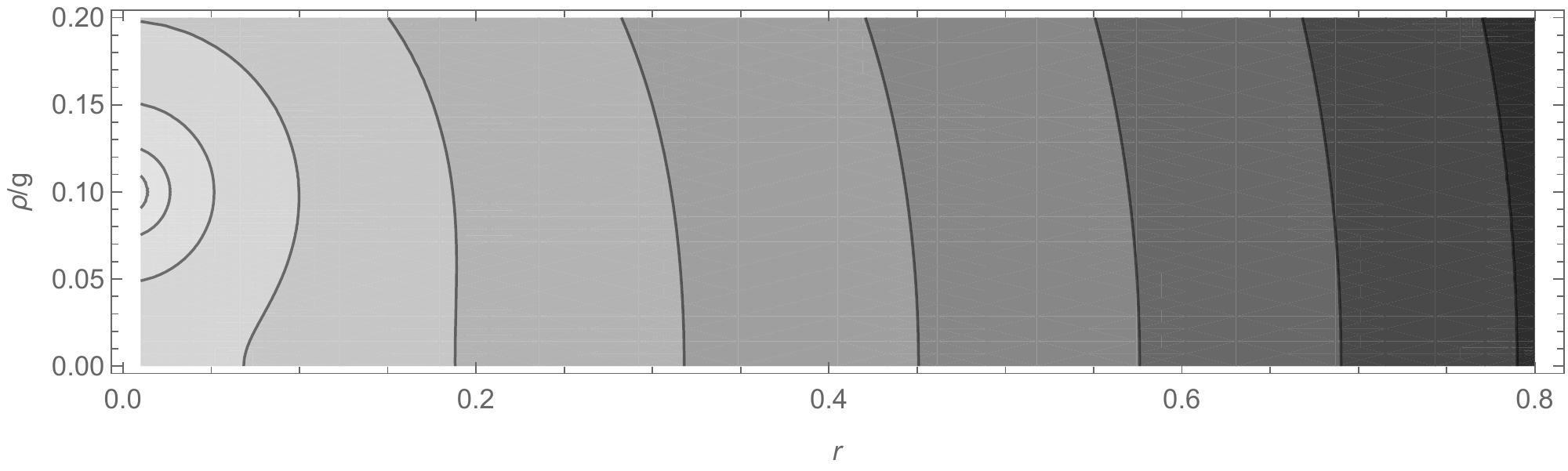}
    \caption{Equipotential surfaces of a Type II potential}
    \label{fig:taxonomy2B}
\end{figure}
\begin{figure}
    \centering
    \includegraphics[width=\textwidth]{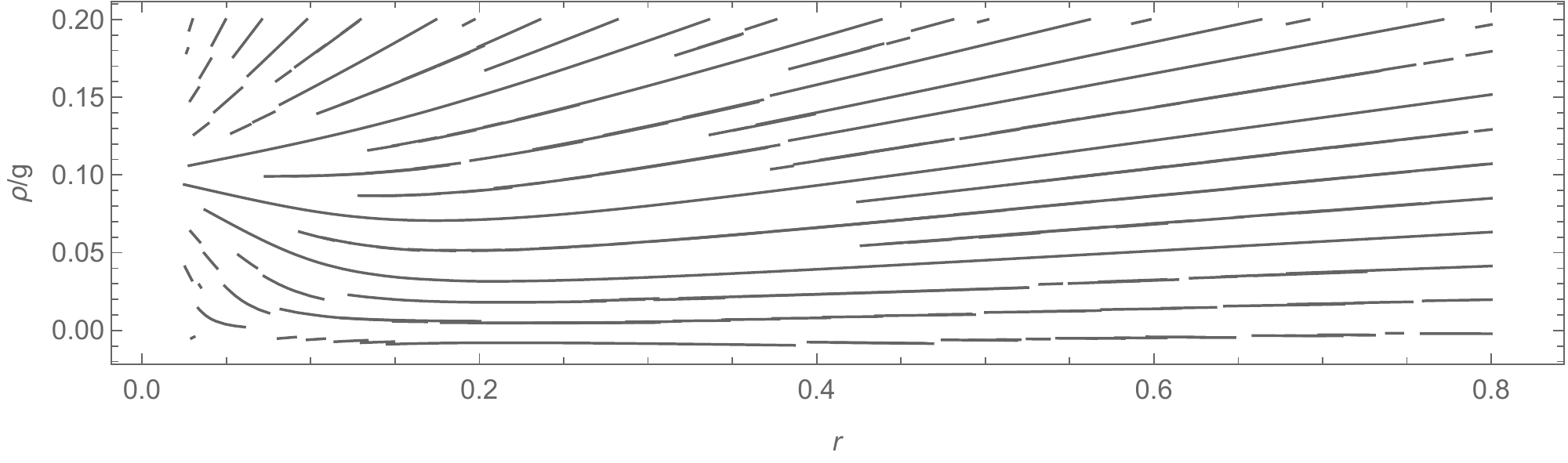}
    \caption{Force lines (gradient flows) of a Type II potential}
    \label{fig:taxonomy2A}
\end{figure}
\indent We can see from Figures \eqref{fig:taxonomy2B} and \eqref{fig:taxonomy2A} that, near to the source, the Type II potential asymptotes to a spherically symmetric potential $\left(\frac{1}{R^3}\right)$. Note that the lines in the two figures are orthogonal to each other. Arbitrarily far away, the equipotential surface shapes asymptote to an oblate spheroid which has twice the radius in the $\rho$ direction as in the $r$ direction. It is not seen from the illustration that the Type II solution is exponentially decaying at large $r$. Overall, the particle is drawn towards the source with relative disregard (in comparison to Type III) for its $\rho$ position.\\
\begin{figure}[t]
    \centering
    \includegraphics[width=\textwidth]{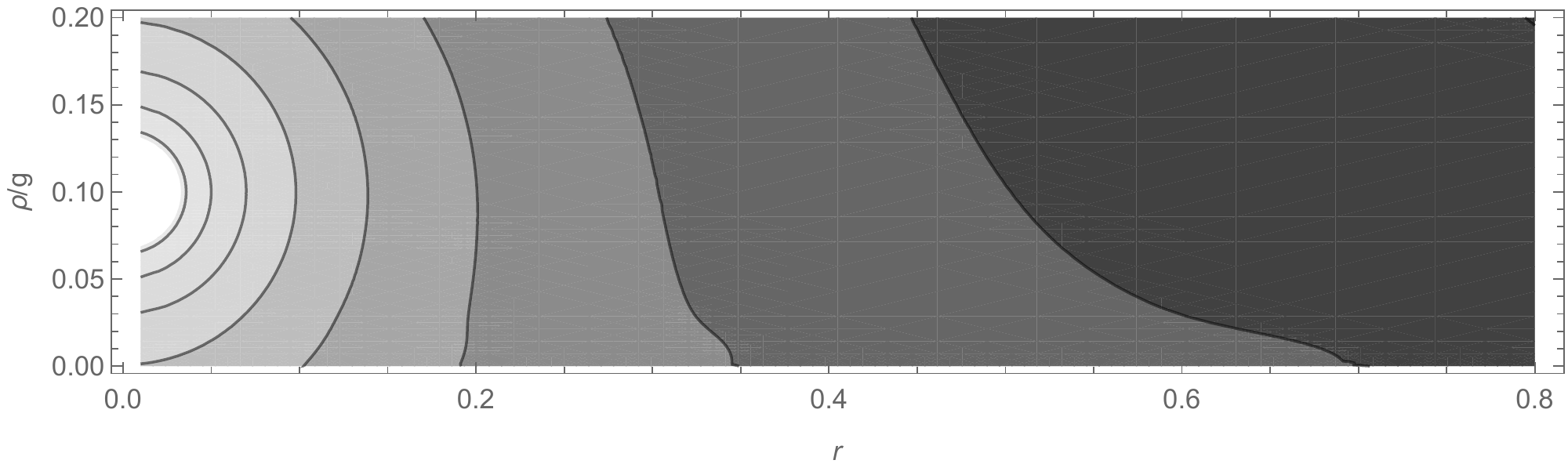}
    \caption{Equipotential surfaces of a Type III potential}
    \label{fig:taxonomy3B}
\end{figure}
\begin{figure}
    \centering
    \includegraphics[width=\textwidth]{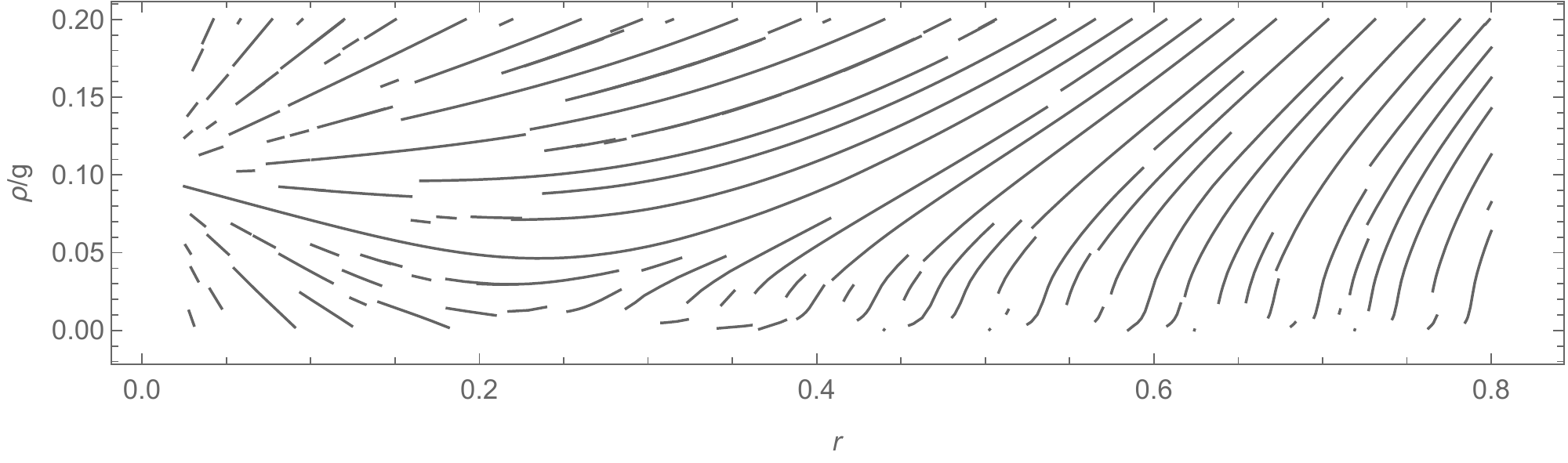}
    \caption{Force lines (gradient flows) of a Type III potential}
    \label{fig:taxonomy3A}
\end{figure}
\indent Now contrast this with the Type III potential shown in figures \eqref{fig:taxonomy3B} and \eqref{fig:taxonomy3A}. Near to the source on the SS--CGP background in the Type III situation, the potential behaves asymptotically in a similar fashion as in the Type II situation. The difference occurs for large $r$.\\
\indent For the sake of clarity, we have regularized the $\xi_0\propto \log\tanh(\rho)$ transverse wavefunction. The equation for the regularised $\tilde\xi_0$ is
\begin{equation}\label{eq: xi0reg}
    \left({\partial^2_\rho}+2 \coth(2 \rho)\right)\tilde\xi_0(\rho) = \frac{1}{\epsilon}\left(\tanh\left(\alpha\left(\epsilon-\rho\right)\right)-1\right)\;.
\end{equation}
We have chosen to regularise $\xi_0$ in this way so that all force lines in the illustration end on the perturbative source at the displace point $r=0$, $\rho=\eta$. When $\alpha\gg1\,,$ the right-hand side of equation \eqref{eq: xi0reg} approximates a step function, normalised so that it integrates to one over the half open integral. In our illustration we have chosen $\alpha = 100$ and $\epsilon = 0.02$.\\
\indent One can see in Figures \eqref{fig:taxonomy3B} and \eqref{fig:taxonomy3A} that the Type III force lines concentrate as one approaches $\rho=0$ or, alternately, that the equipotential surfaces spread out with increasing $r$ along the $\rho=0$ subsurface. The fact that the potential at large $r$ is proportional to the $\xi_0$ transverse wavefunction is due to the Type III boundary condition \eqref{robinbcforG}, or, equivalently, to the presence of a boundary term placed at $\rho=0$ in order to enforce the boundary condition. The $\xi_0\to\tilde\xi_0$ regularisation is equivalent to the smearing of that condition/source.\\
\indent Due to the $\tilde\xi_0$ smearing/regularising, the effect of the boundary term can be seen near to $\rho=0$, as opposed to at $\rho=0$ in our illustration. Specifically one sees the force lines on the far right travel downwards towards $\rho=0$, in response to the presence of the boundary term. Close to $\rho=0$, the force lines bend left as the $r$ dependence of the boundary term draws them towards the source. Then as they approach the origin they then bend back upwards towards the source at $(r,\rho)=(0,\eta)$. If one removes the $\tilde\xi_0$ regularisation, almost all force lines concentrate within the $\rho=0$ subplane. That does not largely effect the long-range potential, but the regularised $\tilde\xi_0$ helps the visualisation.\\
\indent Due to the boundary condition/term at $\rho=0$, the force in the Type III situation falls more slowly at large $r$ than in Type II, \ie it does not decay exponentially when $r\gg1$. Instead, the potential has an $1/r$ falloff as we found in the Type III Green function \eqref{typeIIIlargedist}. The total effect is similar to the RSII `brane bending' as described by Giddings, Katz, and Randall \cite{Giddings:2000mu}.
\subsection{A Comment on Type III and Corresponding Effective Field Theories}
The solutions that we have presented above provide hints towards the structure of the corresponding four-dimensional effective field theories that can be constructed about the SS--CGP background. The Type III solution, in particular, shows that its corresponding effective field theory contains, at the linearised level, a massless graviton. This massless graviton is supported by a non-constant zero-mode, $\xi_0(\rho) \propto \log\tanh\rho$, on the non-compact transverse space. A study of the dynamics of effective theories describing massless fields supported by non-constant zero-modes, focusing on Maxwell theory and scalar QED, was initiated in \cite{Erickson:2020oda}. There, it was found that the existence of a non-constant zero-mode leads to a non-linear realisation of the underlying $U(1)$ gauge symmetry by fields arising from the transverse component of the higher-dimensional Maxwell gauge potential. A consequence of this is that there is an \textit{apparent} symmetry breaking -- called covert symmetry breaking -- appearing in the quartic vector-scalar interaction terms, where the coefficient of the quartic interaction is not the square of the cubic interaction, as is required by a scalar QED theory in which the $U(1)$ symmetry is linearly realised. This suggests that the effective field theory about the SS--CGP background associated to the Type III solution will similarly display covert symmetry breaking as a consequence of the underlying gauge symmetry, four-dimensional diffeomorphisms, being non-linearly realised.

\section{Conclusion; Review of the Taxonomy}
%
Expanding the different contexts for the embedding of lower-dimensional effective-theory physics into a higher-dimensional theory is highly relevant for the exploitation of string and supergravity theories. Reductions to subsurfaces of non-compact higher-dimensional spacetimes have been of interest since early work on supergravity dimensional reductions that generate non-compact symmetries in reduced theories \cite{Hull:1988jw}. From the perspective of the present work, such reductions correspond to the Type I scenarios, which work when there happens to be a mathematically consistent truncation to the lower dimension. In general, however, there is a problem that was already outlined in Reference \cite{Hull:1988jw}, that coupling to matter not contained within the framework of a given consistent reduction (\ie to matter not purely contained within a consistently reduced supergravity) is not of the structure expected for a lower-dimensional theory. This has been summarised by saying that the corresponding lower-dimensional Newton constant vanishes for such non-compact reductions. A characteristic feature of such situations is the existence of a continuous higher-mode spectrum that extends right down to zero mass eigenvalue.
 
Mathematically consistent reductions are highly interesting in themselves, but they are not the only approach to developing lower-dimensional physics within a higher-dimensional theory. More generally, one may look for embeddings that generate a lower-dimensional effective theory that may be subject to corrections arising from higher-dimensional modes that are, however, suppressed on appropriate scales. This happens already in more traditional Kaluza--Klein reductions on compact spaces that do not admit strictly mathematically consistent truncations to a lower-dimensional theory. There can be situations, such as for Calabi--Yau  reductions of pure Type II theories, where the effects of integrating out the higher modes produce only higher-derivative corrections, suppressed by powers of a large mass arising from the compactification \cite{Duff:1989cr}.

Another approach to developing lower-dimensional effective theories within a higher-dimensional origin is to look for situations involving non-compact reduction spaces where the corrections arising from integrating out higher-dimensional modes are nonetheless suppressed on appropriate scales. An example of a construction that went partway in this direction was the RSII construction \cite{Randall:1999vf}. \say{Partway} because by folding AdS spacetime and consequently excluding an infinite volume of it, it has been debated whether that construction was a genuinely non-compact reduction. 

A genuinely non-compact reduction 
\cite{Cvetic:2003xr} involving $H(2,2)$ hyperbolic space has been developed, in which the emergence of a lower-dimensional effective theory occurs because the higher-mode spectrum has a single zero mode, generating a massless effective theory in the lower dimension, which is however separated from the continuous part of the higher-mode spectrum by a mass gap \cite{Crampton:2014hia}. To date, this is the only example that we know of such a mass-gap-protected construction, but most likely it is not the only one -- searching for other analogous examples is clearly a promising open area for research.

In the present paper, the aim has been to explore further the nature of lower-dimensional effective theories with non-compact reduction spaces through analysis, at the linearised level, of the response to the inclusion of an additional mass source that is genuinely localised within the full higher-dimensional space. What we have found is the key r\^ole played by boundary conditions that are imposed as one approaches the lower-dimensional \say{worldvolume}. This is analogous to the r\^ole of boundary conditions imposed in the RSII scenario, as laid out more clearly in Reference \cite{Giddings:2000mu}. Lower-dimensional gravitational behaviour emerges only when boundary conditions are applied that permit the inclusion of the relevant transverse-space zero mode.

From our analysis, we have categorised effective theory reductions into three distinct groups:
\subsection*{Type I}
Type I Ricci-flat reductions \cite{Brecher:1999xf,Chamblin:1999by} (generalisable to surviving supergravity reductions as in Reference \cite{Lu:2000xc}), involve solutions to which may be viewed as \say{black spokes}. These couple to sources that are actually extended into the transverse space via the reduction ansatz. They are solutions to a consistent truncation of the higher-dimensional theory and they are consequently known to full nonlinear order. Owing to the surviving `trombone' scaling symmetry for the Ricci-flat condition, such constructions do not have a well-defined lower-dimensional Newton constant, even though they are self-interacting solutions. Inclusion of a separate matter source that is genuinely localised in the higher-dimensional spacetime will excite the erstwhile-truncated higher modes, rendering the truncation to a lower-dimensional system mathematically inconsistent. Consequently, such couplings to external localised higher-dimensional sources will produce higher-dimensional gravitational behaviour instead of lower-dimensional behaviour.
\subsection*{Type II}
Type II localisations correspond to solutions that are \say{native} in the higher-dimensional theory. Technically, at linear order, they may couple to a localised source in the higher dimension and then give rise to a gravitational potential that is proportional to the canonical Green function for the spacetime when the transverse space is replaced with its universal cover (such as $\bR^1$ for $\cS^1$). No special boundary conditions aside from regularity away from sources and falloff at spatial infinity are imposed. The leading component of the linearised perturbative response to a higher-dimensional massive source is, in all Type II cases which we have seen, correspondingly higher-dimensional. Such constructions, accordingly, do not produce a lower-dimensional Newton constant.
\subsection*{Type III}
Type III localisations modify the Green functions of Type II localisations by the imposition of boundary conditions (given in \eqref{robinbcforG} in the present case) on a worldvolume-spanning submanifold that are compatible with the existence of a normalisable transverse-space zero mode. This modifies the behaviour of coupling to higher-dimensional matter at large distances on the lower-dimensional worldvolume. This is the only one of the three reduction scenarios that genuinely permits the emergence of an effective theory of lower-dimensional gravity at large worldvolume distance scales. In this case, finite Newton constant values can be defined for various notions of source and test particle localisation near the worldvolume.

\section*{Acknowledgments}

We are grateful to Carl Bender, Chris Hull, Chris Pope and Toby Wiseman for helpful discussions. The work of KSS was supported in part by the STFC under Consolidated Grants ST/P000762/1 and ST/T000791/1 and the work of CWE was supported by the United States Department of Veterans Affairs under the Post 9/11 GI Bill.
\addtocontents{toc}{\protect\newpage}
\addcontentsline{toc}{section}{Appendices}
\section*{Appendices}
\appendix
\section{Doubly-Ricci-flat Branes}\label{AppendixA}

In this appendix, we will review the doubly-Ricci-flat brane solutions in a single scalar supergravity model. The Lagrangian in $d$ dimensions is
\begin{equation}
\mc{L}=R{\ast} 1 - \frac{1}{2}d\phi\wedge{\ast} d\phi - \frac{1}{2}e^{a\phi}F_{(p+2)}\wedge{\ast} F_{(p+2)} \,, \label{eq:toy}
\end{equation} 
where $F_{(p+2)} = dA_{(p+1)}$. The equations of motion are 
\begin{align}
&\Box\phi = \frac{a}{2(p+2)!}e^{a\phi}\big(F_{(p+2)}\big)^2 \,, \label{eq:e1} \\
&\partial_M\big(\sqrt{-g}e^{a\phi}F^{MN_1\cdots N_{p+1}}\big) = 0 \,, \label{eq:e2} \\
&R_{MN} = \frac{1}{2}\partial_M\phi\partial_N\phi + \frac{1}{2(p+1)!}e^{a\phi}\Big(F^{}_{M\cdots}F_{N}^{\ph{N}\cdots}- \frac{p+1}{(p+2)(d-2)}\big(F_{(p+2)}\big)^2g_{MN}\Big) \label{eq:e3} \,.
\end{align} 
This system can be obtained as a consistent truncation of maximal supergravity in $d$-dimensions, or equivalently, from M-theory reduced on $T^{11-d}$.  Working within this system, we will derive a set of doubly Ricci-flat brane solutions. These are solutions whose worldvolume and transverse space are arbitrary Ricci-flat manifolds. 

\subsection{Electric Branes}

Consider a warped product manifold $M_d = M_{p+1}\times B_{d-p-1}$ with metric  
\begin{equation}
ds^2_d = e^{2A(y)}\ov g_{\mu\nu}(x)dx^\mu dx^\nu + e^{2B(y)}\wt g_{ij}(y)dy^idy^j \,, \label{eq:elmet}
\end{equation} 
where $\ov g_{\mu\nu}$ is the Lorentzian metric on $M_{p+1}$, and $\wt g_{ij}$ is the Riemannian metric on $B_{d-p-1}$. Defining the constants $d_e = p+1$ and $d_m = d-p-3$, the non-zero components of the Ricci tensor are 
\begin{equation}
\begin{split}
R_{\mu\nu} &= \ov R_{\mu\nu} - e^{2(A-B)}\Big(\wt\nabla^2A + \wt g^{ij}\partial_iA\big(d_e\partial_j A+d_m\partial_j B\big)\Big)\ov g_{\mu\nu} \,, \\
R_{ij} &= \wt R_{ij} -d_e\wt\nabla_i\wt\nabla_jA - d_m\wt\nabla_i\wt\nabla_jB + d_m\partial_iB\partial_jB - d_e\partial_iA\partial_jA  \\
&\quad+ 2d_e\partial_{(i}A\partial_{j)}B- \Big(\wt\nabla^2B + \wt g^{ij}\partial_i B\big(d_e\partial_j A+d_m\partial_j B\big)\Big)\wt g_{ij} \,, \label{eq:ricci2}
\end{split}
\end{equation} 
where $\ov R_{\mu\nu}$ and $\wt R_{ij}$ are the Ricci tensors of $\ov g_{\mu\nu}$ and $\wt g_{ij}$ respectively, $\wt\nabla_i$ is the covariant derivative with respect to $\wt g_{ij}$, and $\wt\nabla^2 = \wt g^{ij}\wt\nabla_i\wt\nabla_j$. For electric branes, we take $M_{p+1}$ and $B_{d-p-1}$ to be Ricci-flat, and the gauge field and scalar to be 
\begin{equation}
A_{(p+1)} = \pm e^{C(y)}\vol(M_{p+1}) \,,\quad \phi = \phi(y) \,. \label{eq:elecanz}
\end{equation} 
By imposing the linear relations 
\begin{equation}
d_eA + d_mB = 0 \,, \quad A = -\frac{d_m}{a(d-2)}\phi\,,\quad C = \log\frac{2}{\sqrt{\Delta}} - \frac{\Delta}{2a}\phi  \,, \label{eq:lin}
\end{equation} 
where the constant $\Delta$ is defined as 
\begin{equation}
\Delta = a^2 + \frac{2d_ed_m}{d-2} \,, \label{eq:deltaconst}
\end{equation} 
the equations of motion \eqref{eq:e1}-\eqref{eq:e3} reduce to one equation given by 
\begin{equation}
\wt\nabla^2 \phi + \frac{\Delta}{2a} \wt g^{ij}\partial_i\phi\partial_j\phi = 0 \,. \label{eq:sca}
\end{equation} 
Writing $H = e^{\frac{\Delta}{2a}\phi}$, \eqref{eq:sca} becomes 
\begin{equation}
\wt\nabla^2H = 0 \,, \label{eq:harm}
\end{equation} 
meaning that $H$ is the harmonic function on $B_{d-p-1}$. To avoid confusion with the constant $\Delta$ in \eqref{eq:deltaconst}, we will not use $\Delta_B$ to denote the Laplacian on $B_{d-p-1}$. If $B_{d-p-1}$ is compact and closed, then $H$ is a constant. In that case, the metric becomes Ricci-flat, and $A_{(p+1)}$ is gauge equivalent to zero, which is not particularly interesting. Consequently, we will now assume that $B_{d-p-1}$ is non-compact. 

When $a=0$, we can consistently set $\phi=0$. In this case, instead of having $A$ and $C$ being related to $\phi$ as in \eqref{eq:lin}, we have 
\begin{equation}
C= \log\frac{2}{\sqrt{\Delta}} + \frac{\Delta(d-2)}{2d_m}A \,,
\end{equation} 
and $H=e^{-C}$ satisfies \eqref{eq:harm}. In the case where $d_m=0$, we can consistently set $A=0$, and we then define $C$ as 
\begin{equation}
C= \log\frac{2}{\sqrt{\Delta}} - \frac{\Delta(d-2)}{2d_e}B \,,
\end{equation} 
with $H=e^{-C}$ again satisfying \eqref{eq:harm}. 

In full, the electric brane solution for generic values of $a$ and $d_m$ is 
\begin{equation}
\begin{split}
&ds^2_d = H^{-\frac{4d_m}{\Delta(d-2)}}ds^2(M_{p+1}) + H^{\frac{4d_e}{\Delta(d-2)}}ds^2(B_{d-p-1}) \,,  \\
&A_{(p+1)} =\pm\frac{2}{\sqrt{\Delta}}H^{-1}\vol(M_{p+1}) \,,\quad e^{\phi} = H^{\frac{2a}{\Delta}} \,,\quad \wt\nabla^2 H = 0\,. \label{eq:elecsol}
\end{split}
\end{equation} 
This solution is a generalisation of the static brane solutions, in which both the worldvolume and the transverse space are arbitrary Ricci-flat manifolds. In particular, this solution allows for a consistent worldvolume truncation to general relativity in $p+1$ dimensions. 

For later convenience, we will define the conserved electric charge of \eqref{eq:elecsol}. The equation of motion for the gauge field is given by 
\begin{equation}
d(e^{a\phi}{\ast} F_{(p+2)}) = 0 \,,
\end{equation} 
This defines a constant of motion, the electric charge $Q_{\text{el}}$, given by 
\begin{equation}
Q_{\text{el}} = \int_{\partial B}e^{a\phi}{\ast} F_{(p+2)} \,, \label{eq:eleccharge}
\end{equation} 
where we removed the subscript from $B_{d-p-1}$ for convenience. 

\subsection{Magnetic Branes}

To derive the magnetic analogue of the above electric brane solution, we have to dualise the theory. This is done using the standard Poincar\'e duality technique, and the theory in the dualised variables is given by 
\begin{equation}
\mc{L}=R{\ast} 1 - \frac{1}{2}d\phi\wedge{\ast} d\phi - \frac{1}{2}e^{-a\phi}G_{(d-p-2)}\wedge{\ast} G_{(d-p-2)} \,,
\end{equation} 
where $G_{(d-p-2)} = dK_{(d-p-3)}$, and is related to the original field strength $F_{(p+2)}$ by 
\begin{equation}
F_{(p+2)} = e^{-a\phi}{\ast} G_{(d-p-2)} \,.
\end{equation} 
The dualisation procedure replaces $A_{(p+1)}$ with $K_{(d-p-3)}$, both of which have the same number of degrees of freedom, and flips the sign of the dilaton. Using our results in the previous subsection, the theory admits the solution 
\begin{equation}
\begin{split}
&ds^2_d = H^{-\frac{4d_e}{\Delta(d-2)}}ds^2(M_{d-p-3}) + H^{\frac{4d_m}{\Delta(d-2)}}ds^2(B_{p+3}) \,,  \\
&K_{(d-p-3)} =\pm\frac{2}{\sqrt{\Delta}}H^{-1}\vol(M_{d-p-3}) \,,\quad e^{\phi} = H^{-\frac{2a}{\Delta}} \,,\quad \wt\nabla^2H = 0 \,,
\end{split}
\end{equation} 
where $M_{d-p-3}$ and $B_{p+3}$ are Ricci-flat. Again, we choose $B_{p+3}$ to be non-compact so that $H$ is not constant, and $K_{(d-p-3)}$ is not gauge equivalent to zero. In the original variables, this is 
\begin{equation}\label{magrf}
F_{(p+2)} = \pm\frac{2}{\sqrt{\Delta}}\frac{\sqrt{\wt g}}{(p+2)!}\wt g^{ij}\epsilon_{ik_1\cdots k_{p+2}}\partial_j H dy^{k_1}\wedge\cdots\wedge dy^{k_{p+2}} \,.
\end{equation} 
Taking the exterior derivative, we have 
\begin{equation}
dF_{(p+2)} = \pm\frac{2}{\sqrt{\Delta}}\wt\nabla^2 H\vol(B_{p+3}) = 0 \,,
\end{equation} 
which follows from the fact that $H$ is harmonic. This Bianchi identity gives rise to a conserved magnetic charge, defined by 
\begin{equation}
Q_{\text{mag}} = \int_{\partial B}F_{(p+2)} \,, \label{eq:magcharge}
\end{equation} 
where we again removed the subscript from $B_{p+3}$ for convenience. 

\subsection{Dyonic Branes}

Dyonic solutions are charged under a self-dual or anti-self-dual field strength. These only occur in even dimensions with $p+2 = d/2$, for odd $p$. In this case, the scalar field decouples, as it is sourced by $F_{(p+2)}\wedge\ast F_{(p+2)} = \pm F_{(p+2)}\wedge F_{(p+2)} = 0$. So, effectively, we can set $a=0$, and the resulting solution is
\begin{equation}
\begin{split}
&ds^2_d = H^{-\frac{2}{\Delta}}ds^2(M_{p+1}) + H^{\frac{2}{\Delta}}ds^2(B_{p+3}) \,, \nonumber \\
&F^{\pm}_{(p+2)} = L_{(p+2)} \pm \ast L_{(p+2)} \,,\quad L_{(p+2)} = \mp \frac{\sqrt{2(d-2)}}{d_D}H^{-2}dH\wedge\vol(M_{p+1}) \,,\quad \wt\nabla^2 H = 0\,,
\end{split}
\end{equation} 
where the $\pm$ indicates whether the field strength is self-dual or anti-self-dual. Again, $M_{p+1}$ and $B_{p+3}$ are Ricci-flat, and $H$ is a harmonic function on $B_{p+3}$. 

The conserved charge associated with the dyonic brane is both electric and magnetic. 

\subsection{Trombone Symmetries}

The doubly-Ricci-flat brane solutions presented above have three global symmetries, which have been called `trombone' symmetries \cite{Cremmer:1997xj}. The first trombone symmetry is a rescaling of the $d$-dimensional fields
\begin{equation}\label{largetrombone}
    g_{MN}\mapsto k_H^2g_{MN}\,,\quad \phi\mapsto \phi\,,\quad  F_{(p+2)}\mapsto k_H^{p+1}F_{(p+2)} \,.
\end{equation}
The next two trombone symmetries are individual rescalings of the worldvolume and transverse metrics,
\begin{equation}\label{littletrombones}
    \ov g_{\mu\nu}\mapsto k_W^2\ov g_{\mu\nu}\,,\quad \tilde g_{ij}\mapsto k_T^2\tilde g_{ij}\,,
\end{equation}
with no active rescaling on $\phi$ and $F_{(p+2)}$, though $F_{(p+2)}$ will, in general, be scaled through the rescalings of $\ov g_{\mu\nu}$ and $\tilde g_{ij}$. For example, the magnetic flux \eqref{magrf} scales as $k_T^{p+1}$. We stress that $k_H$, $k_W$, and $k_T$ are three independent parameters. 

To see that \eqref{littletrombones} are symmetries, we note that the doubly-Ricci-flat brane solutions are defined by three equations,
\begin{equation}
    \ov R_{\mu\nu}=0\,,\quad \tilde R_{ij}=0\,,\quad \tilde\nabla^2H = 0 \,.
\end{equation}
Rescaling the worldvolume and transverse metrics as in \eqref{littletrombones} do not rescale their Ricci tensors, so they remain Ricci-flat. For the third equation, recall that 
\begin{equation}
    \tilde\nabla^2H=\frac{1}{\sqrt{\tilde g}}\partial_i\left(\sqrt{\tilde g}\tilde g^{ij}\partial_j H\right) \mapsto k_T^{-2}\tilde\nabla^2H \,.
\end{equation}
Thus, if $H$ is harmonic with respect to $\tilde g_{ij}$, it also harmonic with respect to $k_T^2\tilde g_{ij}$. The first trombone symmetry \eqref{largetrombone} is a symmetry that is present irrespective of the particular solution. On the other hand, the second and third trombone symmetries \eqref{littletrombones} only occur within our specific doubly-Ricci-flat ansatz. 

\section{A Consistent Truncation about the SS--CGP Background}\label{se:consistenttrunc}
We will start with the bosonic sector of ten-dimensional type I supergravity, 
\begin{equation}
\mc{\hat{L}}_{10} = \hat R\hat {\ast}1-\frac{1}{2}d\hat \phi\wedge\hat {\ast} d\hat \phi - \frac{1}{2}e^{-\hat \phi}\hat H_{(3)}\wedge\hat {\ast} \hat H_{(3)} \,,
\end{equation} 
where $\hat H_{(3)} = d\hat B_{(2)}$. The solution of interest is the lifted Salam--Sezgin vacuum with an NS5-brane inclusion. This was derived in Reference \cite{Crampton:2014hia}, and we reproduce it here:
\begin{align}
&d\hat s^2_{10} = W(\rho)^{-\frac{1}{4}}\Big(\eta_{\mu\nu}dx^\mu dx^\nu + dy^2 + \frac{1}{4g^2}\Big(d\psi + \sech2\rho\big(d\chi+\cos\theta\,d\varphi\big)\Big)^2 + \frac{1}{g^2}W(\rho)\,ds^2_{EH}\Big) \,, \nonumber \\
&e^{2\hat \phi} = W(\rho) \,, \quad \hat B_{(2)} = \frac{1}{4g^2}\big((1+k)d\chi + \sech2\rho\,d\psi\big)\wedge\big(d\chi + \cos\theta\,d\varphi\big) \,, \label{eq:ss10}
\end{align} 
where $W(\rho) = \sech2\rho-k\log\tanh\rho$, and $ds^2_{EH}$ is the Eguchi-Hanson metric 
\begin{equation}
ds^2_{EH} = \cosh2\rho\Big(d\rho^2 + \frac{1}{4}(\tanh2\rho)^2\big(d\chi+\cos\theta\,d\varphi\big)^2 + \frac{1}{4}\big(d\theta^2 + \sin^2\theta\,d\varphi^2\big)\Big) \,, \label{eq:eguchi}
\end{equation} 
The coordinates $\psi$, $y$, and $\chi$ are $S^1$ coordinates, $(\theta,\varphi)$ parametrises an $S^2$, $\rho\in[0,\infty)$ is the non-compact radius, and $k$ is a positive constant. The 6-dimensional worldvolume of the NS5-brane is parametrised by $(x^\mu, \psi, y)$. Our goal is to reduce the type I theory on $T^3 \ni (y, \psi, \chi)$ via the usual Kaluza--Klein methods, and $S^2$ on the background given in \eqref{eq:ss10} to obtain a 5-dimensional theory. 

\subsection{\texorpdfstring{$10\to9$}{10 to 9}: Reduce on \texorpdfstring{$y$}{y}} 

The background metric in \eqref{eq:ss10} does not have a fibre over the circle parametrised by $y$. So, the appropriate Kaluza--Klein ansatz is 
\begin{equation}
d\hat s^2_{10} = e^{-\frac{1}{2\sqrt{7}}\phi_2}ds^2_9 + e^{\frac{\sqrt{7}}{2}\phi_2}dy^2 \,,\quad \hat B_{(2)} = B_{(2)} \,,\quad \hat \phi = \phi_1 \,,
\end{equation} 
where the un-hatted fields are 8-dimensional fields. The resulting equations of motion are encoded in the action 
\begin{equation}
\mc{L}_9 = R{\ast} 1- \frac{1}{2}d\phi_i\wedge{\ast} d\phi_i - \frac{1}{2}e^{-\phi_1+\frac{1}{\sqrt{7}}\phi_2}H_{(3)}\wedge{\ast} H_{(3)} \,, \label{eq:9daction}
\end{equation} 
where $H_{(3)}=dB_{(2)}$, and $i\in\{1,2\}$. The background solution \eqref{eq:ss10} reduces to 
\begin{align}
&ds^2_9 = W^{-\frac{2}{7}}\Big(\eta_{\mu\nu}dx^\mu dx^\nu + dy^2 + \frac{1}{4g^2}\Big(d\psi + \sech2\rho\big(d\chi+\cos\theta\,d\varphi\big)\Big)^2 + \frac{1}{g^2}W\,ds^2_{EH}\Big) \,, \nonumber \\
&e^{2\phi_1} = e^{-2\sqrt{7}\phi_2} = W \,,\quad B_{(2)} = \frac{1}{4g^2}\big((1+k)d\chi + \sech2\rho\,d\psi\big)\wedge\big(d\chi + \cos\theta\,d\varphi\big) \,. \label{eq:ss9} 
\end{align} 
We observe that in the background solution, the combination $\phi_1+\sqrt{7}\phi_2=0$. This suggests a field redefinition,
\begin{equation}\label{10to9matrix}
\begin{pmatrix} \Phi_2 \\ \Phi_1 \end{pmatrix} = \frac{1}{2\sqrt{2}}\begin{pmatrix} 1 & \sqrt{7} \\ -\sqrt{7} & 1 \end{pmatrix}\begin{pmatrix}\phi_1 \\ \phi_2 \end{pmatrix} \,.
\end{equation} 
Substituting this into \eqref{eq:9daction}, and noting that the transformation matrix \eqref{10to9matrix} is orthogonal, we have 
\begin{equation}
\mc{L}_9 = R{\ast} 1- \frac{1}{2}d\Phi_i\wedge{\ast} d\Phi_i - \frac{1}{2}e^{2\sqrt{\frac{2}{7}}\Phi_1}H_{(3)}\wedge{\ast} H_{(3)} \,, \label{eq:9daction2}
\end{equation} 
and the background solution is 
\begin{equation}
\Phi_2 = 0 \,,\quad e^{-\sqrt{\frac{7}{2}}\Phi_1} = W \,.
\end{equation} 
From \eqref{eq:9daction2}, we find that $\Phi_2$ is decoupled, so there is a consistent truncation of the 9-dimensional theory given by $\Phi_2 = 0$, $\Phi_1 = \phi$. For completeness, the truncated theory is given by 
\begin{equation}
\mc{\hat L}_9 = \hat R\hat {\ast} 1- \frac{1}{2}d\hat \phi\wedge\hat{\ast} d\hat \phi - \frac{1}{2}e^{2\sqrt{\frac{2}{7}}\hat \phi}\hat H_{(3)}\wedge\hat {\ast} \hat H_{(3)} \,, \label{eq:9daction3}
\end{equation} 
where we have reintroduced hats on all 9-dimensional fields. We will take \eqref{eq:9daction3} as the starting point for the next reduction step. 

\subsection{\texorpdfstring{$9\to8$}{9 to 8}: Reduce on \texorpdfstring{$\psi$}{Psi}}

The background metric in \eqref{eq:ss9} is fibred over the $\wt\psi = \psi/2g$ coordinate. This suggests that the appropriate Kaluza--Klein ansatz is 
\begin{equation}
d\hat s^2_9 = e^{-\frac{1}{\sqrt{21}}\phi_2}ds^2_8 + e^{\frac{6}{\sqrt{21}}\phi_2}\big(d\wt\psi + \mc{A}_{(1)}\big)^2 \,,\quad \hat B_{(2)} = B_{(2)} + A_{(1)}\wedge d\wt\psi \,,\quad \hat\phi = \phi_1 \,.
\end{equation} 
The resulting equations of motion are encoded in the action 
\begin{align}
\mc{L}_8 &= R{\ast} 1 - \frac{1}{2}d\phi_i\wedge{\ast} d\phi_i - \frac{1}{2}e^{\sqrt{\frac{7}{3}}\phi_2}\mc{F}_{(2)}\wedge{\ast}\mc{F}_{(2)} \nonumber \\
&-\frac{1}{2}e^{2\sqrt{\frac{2}{7}}\phi_1-\frac{5}{\sqrt{21}}\phi_2}F_{(2)}\wedge{\ast} F_{(2)} - \frac{1}{2}e^{2\sqrt{\frac{2}{7}}\phi_1+\frac{2}{\sqrt{21}}\phi_2}H_{(3)}\wedge{\ast} H_{(3)} \,, \label{eq:8daction}
\end{align} 
where $\mc{F}_{(2)}=d\mc{A}_{(1)}$, $F_{(2)}=dA_{(1)}$, $H_{(3)} = dB_{(2)}-dA_{(1)}\wedge\mc{A}_{(1)}$, and $i\in\{1,2\}$. The background solution in \eqref{eq:ss9} reduces to 
\begin{align}
&ds^2_8 = W^{-\frac{1}{3}}\Big(\eta_{\mu\nu}dx^\mu dx^\nu + \frac{W}{g^2}ds^2_{EH}\Big) \,,\quad B_{(2)} = \frac{1+k}{4g^2}\cos\theta\,d\chi\wedge d\varphi \,, \nonumber \\ 
&\mc{A}_{(1)} = - A_{(1)} = \frac{1}{2g}\sech2\rho\,(d\chi + \cos\theta\,d\varphi)\,,\quad e^{-\sqrt{\frac{7}{2}}\phi_1} = e^{-\sqrt{21}\phi_2} = W \,. \label{eq:ss8}
\end{align} 
From this, we observe that $\phi_1 -\sqrt{6}\phi_2 = 0$. This suggests the field redefinition 
\begin{equation}\label{9to8matrix}
\begin{pmatrix} \Phi_2 \\ \Phi_1 \end{pmatrix} = \frac{1}{\sqrt{7}}\begin{pmatrix} 1 & -\sqrt{6} \\ \sqrt{6} & 1 \end{pmatrix}\begin{pmatrix}\phi_1 \\ \phi_2 \end{pmatrix} \,.
\end{equation} 
For the background solution, we have 
\begin{equation}
\Phi_2 = 0 \,,\quad e^{-\sqrt{3}\Phi_1} = W \,.
\end{equation} 
Now, substituting the field redefinition into \eqref{eq:8daction}, and noting that the transformation matrix \eqref{9to8matrix} is orthogonal, we obtain
\begin{align}
\mc{L}_8 &= R{\ast} 1 - \frac{1}{2}d\Phi_i\wedge{\ast} d\Phi_i - \frac{1}{2}e^{-\sqrt{2}\Phi_2+\frac{1}{\sqrt{3}}\Phi_1}\mc{F}_{(2)}\wedge{\ast}\mc{F}_{(2)} \nonumber \\
&-\frac{1}{2}e^{\sqrt{2}\Phi_2+\frac{1}{\sqrt{3}}\Phi_1}F_{(2)}\wedge{\ast} F_{(2)} - \frac{1}{2}e^{\frac{2}{\sqrt{3}}\Phi_1}H_{(3)}\wedge{\ast} H_{(3)} \,. \label{eq:8daction2}
\end{align} 
Let us now examine the equations of motion of $\mc{A}_{(1)}$, $A_{(1)}$, and $\Phi_2$: 
\begin{align}
\mc{A}_{(1)}: \quad &d\big(e^{-\sqrt{2}\Phi_2+\frac{1}{\sqrt{3}}\Phi_1}{\ast} d\mc{A}_{(1)}\big) - e^{\frac{2}{\sqrt{3}}\Phi_1}dA_{(1)}\wedge{\ast} H_{(3)} = 0 \,, \nonumber \\
A_{(1)}: \quad &d\big(e^{\sqrt{2}\Phi_2+\frac{1}{\sqrt{3}}\Phi_1}{\ast} d{A}_{(1)}\big) - e^{\frac{2}{\sqrt{3}}\Phi_1}d\mc{A}_{(1)}\wedge{\ast} H_{(3)} = 0 \,, \nonumber \\
\Phi_2: \quad & d{\ast} d\Phi_2 = \frac{1}{2\sqrt{2}}e^{\frac{1}{\sqrt{3}}\Phi_1}\big(e^{\sqrt{2}\Phi_2}F_{(2)}\wedge{\ast} F_{(2)} - e^{-\sqrt{2}\Phi_2}\mc{F}_{(2)}\wedge{\ast}\mc{F}_{(2)}\big) \,.
\end{align} 
These equations admit the solution 
\begin{equation}
\mc{A}_{(1)} = \pm A_{(1)} \,,\quad \Phi_2 = 0 \,. 
\end{equation} 
For our background solution, we have $\mc{A}_{(1)} = -A_{(1)}$. Using this simplifying ansatz, we find that the rest of the equations of motion are encoded in the action 
\begin{equation}
\mc{L}_8 = R{\ast} 1 - \frac{1}{2}d\phi\wedge{\ast} d\phi - e^{\frac{1}{\sqrt{3}}\phi}F_{(2)}\wedge{\ast} F_{(2)} - \frac{1}{2}e^{\frac{2}{\sqrt{3}}\phi}H_{(3)}\wedge{\ast} H_{(3)} \,,
\end{equation} 
where $F_{(2)}=dA_{(1)}$, $H_{(3)} = dB_{(2)} + dA_{(1)}\wedge A_{(1)}$, and we relabelled $\Phi_1 = \phi$. To put the action in canonical form, we have to rescale $A_{(1)}$ by a factor of $1/\sqrt{2}$. The final 8-dimensional theory is 
\begin{equation}
\mc{\hat L}_8 = \hat R\hat {\ast} 1 - \frac{1}{2}d\hat \phi\wedge\hat {\ast} d\hat \phi - \frac{1}{2}e^{\frac{1}{\sqrt{3}}\hat \phi}\hat F_{(2)}\wedge\hat {\ast} \hat F_{(2)} - \frac{1}{2}e^{\frac{2}{\sqrt{3}}\hat \phi}\hat H_{(3)}\wedge\hat {\ast} \hat H_{(3)} \,,
\end{equation} 
where $\hat F_{(2)}=d\hat A_{(1)}$, $\hat H_{(3)} = d\hat B_{(2)} + \tfrac{1}{2}d\hat A_{(1)}\wedge \hat A_{(1)}$, and we have restored the hats for all 8-dimensional fields. 

\subsection{\texorpdfstring{$8\to7$}{8 to 7}: Reduce on \texorpdfstring{$\chi$}{Chi}}

The background solution \eqref{eq:ss8} is fibred over $\wt\chi = \chi/2g$. The appropriate Kaluza--Klein ansatz is then 
\begin{align}
&d\hat s^2_8 = e^{-\frac{1}{\sqrt{15}}\phi_2}ds^2_7 + e^{\sqrt{\frac{5}{3}}\phi_2}\big(d\wt\chi + \tilde{\mc{A}}_{(1)}\big)^2 \,, \nonumber \\
&\hat B_{(2)} = B_{(2)} + B_{(1)}\wedge d\wt\chi \,,\quad \hat A_{(1)} = A_{(1)} + \sigma\,d\wt\chi \,,\quad \hat\phi=\phi_1 \,.
\end{align} 
The resulting equations of motion are encoded in the action 
\begin{align*}
\mc{L}_7 &= R{\ast} 1 -\frac{1}{2}d\phi_i\wedge{\ast} d\phi_i - \frac{1}{2}e^{\frac{1}{\sqrt{3}}\phi_1-\sqrt{\frac{5}{3}}\phi_2}d\sigma\wedge{\ast} d\sigma - \frac{1}{2}e^{2\sqrt{\frac{3}{5}}\phi_2}\mc{F}_{(2)}\wedge{\ast}\mc{F}_{(2)} \numberthis  \label{eq:7daction} \\
&- \frac{1}{2}e^{\frac{1}{\sqrt{3}}\phi_1+\frac{1}{\sqrt{15}}\phi_2}F_{(2)}\wedge{\ast} F_{(2)} - \frac{1}{2}e^{\frac{2}{\sqrt{3}}\phi_1-\frac{4}{\sqrt{15}}\phi_2}H_{(2)}\wedge{\ast} H_{(2)} - \frac{1}{2}e^{\frac{2}{\sqrt{3}}\phi_1+\frac{2}{\sqrt{15}}\phi_2}H_{(3)}\wedge{\ast} H_{(3)} \,,
\end{align*} 
where $\mc{F}_{(2)} = d\tilde{\mc{A}}_{(1)}$, $F_{(2)} = dA_{(1)}-d\sigma\wedge\tilde{\mc{A}}_{(1)}$, $H_{(2)} = dB_{(1)} + \tfrac{1}{2}(\sigma dA_{(1)}-d\sigma\wedge A_{(1)})$, $H_{(3)} = dB_{(2)} + \tfrac{1}{2}dA_{(1)}\wedge A_{(1)} - H_{(2)}\wedge\tilde{\mc{A}}_{(1)}$, and $i\in\{1,2\}$. The background solution \eqref{eq:ss8} reduces to 
\begin{align}
&ds^2_7 = \frac{(\sinh2\rho)^{\frac{2}{5}}}{(W\cosh2\rho)^{\frac{1}{5}}}\Big(\eta_{\mu\nu}dx^\mu dx^\nu + \frac{W\cosh2\rho}{g^2}d\rho^2 + \frac{W\cosh2\rho}{4g^2}\big(d\theta^2 + \sin^2\theta\,d\varphi^2\big)\Big) \,, \nonumber \\
&\sigma = \sqrt{2}\sech2\rho \,, \quad \tilde{\mc{A}}_{(1)} = \frac{1}{2g}\cos\theta\,d\varphi \,, \quad A_{(1)} = \sigma\tilde{\mc{A}}_{(1)} \,,\quad B_{(1)} = -(1+k)\tilde{\mc{A}}_{(1)} \,,\nonumber \\
&e^{-\sqrt{3}\phi_1}= W \,,\quad e^{\sqrt{\frac{5}{3}}\phi_2}=W^{\frac{2}{3}}(\sinh2\rho)^2\sech2\rho \,,\quad B_{(2)} = 0 \,. \label{eq:ss7}
\end{align} 
It is convenient to perform the field redefinition 
\begin{equation}\label{8to7matrix}
\begin{pmatrix} \Phi_1 \\ \Phi_2 \end{pmatrix} = \frac{1}{\sqrt{6}}\begin{pmatrix} 1 & -\sqrt{5} \\ \sqrt{5} & 1 \end{pmatrix}\begin{pmatrix}\phi_1 \\ \phi_2 \end{pmatrix} \,.
\end{equation} 
The background solution is then 
\begin{equation}
e^{-\sqrt{2}\Phi_1} = W(\sinh2\rho)^2\sech2\rho \,,\quad e^{\sqrt{10}\Phi_2} = W^{-1}(\sinh2\rho)^2\sech2\rho \,.
\end{equation} 
Substituting the field redefinition into \eqref{eq:7daction}, and noting that the transformation matrix \eqref{8to7matrix} is orthogonal, we find that 
\begin{align}
\mc{\hat L}_7 &= \hat R\hat {\ast} 1 -\frac{1}{2}d\hat \Phi_i\wedge\hat {\ast} d\hat \Phi_i - \frac{1}{2}e^{\sqrt{2}\hat\Phi_1}d\hat \sigma\wedge\hat {\ast} d\hat \sigma - \frac{1}{2}e^{-\sqrt{2}\hat\Phi_1 + \sqrt{\frac{2}{5}}\hat\Phi_2}\hat{\mc{F}}_{(2)}\wedge\hat {\ast}\hat{\mc{F}}_{(2)} \nonumber \\
&- \frac{1}{2}e^{\sqrt{\frac{2}{5}}\hat\Phi_2}\hat F_{(2)}\wedge\hat {\ast} \hat F_{(2)} - \frac{1}{2}e^{\sqrt{2}\hat\Phi_1 + \sqrt{\frac{2}{5}}\hat\Phi_2}\hat H_{(2)}\wedge\hat {\ast} \hat H_{(2)} - \frac{1}{2}e^{2\sqrt{\frac{2}{5}}\hat\Phi_2}\hat H_{(3)}\wedge\hat {\ast} \hat H_{(3)} \,, \label{eq:7daction2}
\end{align} 
where we have restored the hats to the 7-dimensional fields.

\subsection{\texorpdfstring{$7\to5$}{7 to 5}: Reduce on \texorpdfstring{$S^2$}{S2}}

The reduction ansatz that is consistent with the 7-dimensional background solution \eqref{eq:ss7} is 
\begin{align}
&d\hat s^2_7 = e^{-\frac{2}{\sqrt{15}}\Phi_3}ds^2_5 + \frac{1}{4g^2}e^{\sqrt{\frac{3}{5}}\Phi_3}ds^2(S^2) \,, \quad \hat\Phi_{1,2} = \Phi_{1,2} \,,\quad \hat\sigma = \sigma \,, \nonumber \\
&\hat{\mc{F}}_{(2)} = -\frac{1}{2g}\vol(S^2) \,,\quad \hat F_{(2)} = -\frac{\sigma}{2g}\vol(S^2) \,,\quad \hat{H}_{(2)} = -\frac{\sigma^2+m}{4g}\vol(S^2) \,,\quad \hat{H}_{(3)} = 0 \,, \label{eq:ansatz}
\end{align} 
where $ds^2(S^2)$ and $\vol(S^2)$ are the metric and volume form on the unit 2-sphere respectively, $m$ is a constant, and all un-hatted fields are 5-dimensional fields. The ansatz for the field strengths is consistent with the 7-dimensional Bianchi identities: 
\begin{equation}
d\hat{\mc{F}}_{(2)} = 0 \,,\quad d\hat F_{(2)} = d\hat\sigma\wedge\hat{\mc{F}}_{(2)} \,,\quad d\hat H_{(2)} = d\hat\sigma\wedge\hat F_{(2)} \,,\quad d\hat H_{(3)} = \frac{1}{2}\hat F_{(2)}\wedge\hat F_{(2)} - \hat H_{(2)}\wedge\hat{\mc{F}}_{(2)} \,.
\end{equation} 
Let us first look at the 7-dimensional gauge field equations of motion: 
\begin{align}
\hat B_{(2)}: \quad &d\big(e^{2\sqrt{\frac{2}{5}}\hat\Phi_2}\hat{\ast} \hat H_{(3)}\big) = 0  \,, \\
\hat B_{(1)}: \quad &d\big(e^{\sqrt{2}\hat\Phi_1 + \sqrt{\frac{2}{5}}\hat\Phi_2}\hat{\ast}\hat H_{(2)}\big) - e^{2\sqrt{\frac{2}{5}}\hat\Phi_2}\hat{\mc{F}}_{(2)}\wedge\hat{\ast} \hat H_{(3)} = 0 \,, \\
\hat A_{(1)}: \quad &d\big(e^{\sqrt{\frac{2}{5}}\hat\Phi_2}\hat{\ast}\hat F_{(2)}\big) + e^{\sqrt{2}\hat\Phi_1 + \sqrt{\frac{2}{5}}\hat\Phi_2}d\hat\sigma\wedge\hat{\ast}\hat H_{(2)} + e^{2\sqrt{\frac{2}{5}}\hat\Phi_2}\hat F_{(2)}\wedge\hat{\ast}\hat H_{(3)} = 0 \,, \\
\hat{\mc{A}}_{(1)}: \quad &d\big(e^{-\sqrt{2}\hat\Phi_1 + \sqrt{\frac{2}{5}}\hat\Phi_2}\hat{\ast}\hat{\mc{F}}_{(2)}\big) + e^{\sqrt{\frac{2}{5}}\hat\Phi_2}d\hat\sigma\wedge\hat{\ast}\hat F_{(2)} - e^{2\sqrt{\frac{2}{5}}\hat\Phi_2}\hat H_{(2)}\wedge\hat{\ast}\hat H_{(3)} = 0 \,, \\
\hat\sigma: \quad &d\big(e^{\sqrt{2}\hat\Phi_1}\hat{\ast}\hat d\hat\sigma\big) - e^{\sqrt{\frac{2}{5}}\hat\Phi_2}\hat{\mc{F}}_{(2)}\wedge\hat{\ast}\hat F_{(2)} - e^{\sqrt{2}\hat\Phi_1 + \sqrt{\frac{2}{5}}\hat\Phi_2}\hat F_{(2)}\wedge\hat{\ast}\hat H_{(2)} = 0 \,.
\end{align} 
We note that 
\begin{equation}
\hat{\ast}\vol(S^2) = 4g^2e^{-\frac{8}{\sqrt{15}}\Phi_3}\vol(M_5) \,,\quad \hat{\ast} d\hat\sigma = \frac{1}{4g^2}({\ast} d\sigma)\wedge\vol(S^2) \,, \label{eq:hodge}
\end{equation} 
where $\vol(M_5)$ and ${\ast}$ are the volume form and Hodge star defined with respect to the 5-dimensional metric $ds^2_5$ in \eqref{eq:ansatz} respectively. From this, we find that $\hat{\ast}\hat{\mc{F}}_{(2)}$, $\hat{\ast}\hat F_{(2)}$, and $\hat{\ast}\hat H_{(2)}$ are all proportional to $\vol(M_5)$, which is a top-form on $M_5$. This means that $d(e^S\hat{\ast}\hat{\mc{F}}_{(2)}) = d(e^S\hat{\ast}\hat{F}_{(2)}) = d(e^S\hat{\ast}\hat{H}_{(2)}) = 0$ for any field $S\in C^{\infty}(M_5)$, and $d\hat\sigma\wedge\hat{\ast}\hat F_{(2)} = d\hat\sigma\wedge\hat{\ast}\hat H_{(2)} = 0$. Therefore, the only non-trivial equation from the above is the $\hat\sigma$ equation. After some algebra, we find that the $\hat\sigma$ equation reads 
\begin{equation}
d\big(e^{\sqrt{2}\Phi_1}{\ast} d\sigma\big) = 2g^2e^{\sqrt{\frac{2}{5}}\Phi_2 - \frac{8}{\sqrt{15}}\Phi_3}\big(2 + e^{\sqrt{2}\Phi_1}(\sigma^2+m)\big)\sigma{\ast}1 \,, \label{eq:sig}
\end{equation} 
where we used the identity ${\ast} 1 =\vol(M_5)$. 

\indent Next, we have the 7-dimensional dilaton equations, 
\begin{align}
&d\hat{\ast} d\hat\Phi_1 + \frac{1}{\sqrt{2}}\big(e^{-\sqrt{2}\hat\Phi_1 + \sqrt{\frac{2}{5}}\hat\Phi_2}\hat{\mc{F}}_{(2)}\wedge\hat{\ast}\hat{\mc{F}}_{(2)} - e^{\sqrt{2}\hat\Phi_1}d\hat\sigma\wedge\hat{\ast} d\hat\sigma - e^{\sqrt{2}\hat\Phi_1 + \sqrt{\frac{2}{5}}\hat\Phi_2}\hat H_{(2)}\wedge\hat{\ast}\hat H_{(2)}\big) = 0 \,, \\
&d\hat{\ast} d\hat\Phi_2 - \frac{1}{\sqrt{10}}e^{\sqrt{\frac{2}{5}}\hat\Phi_2}\big(e^{-\sqrt{2}\hat\Phi_1}\hat{\mc{F}}_{(2)}\wedge\hat{\ast}\hat{\mc{F}}_{(2)} + \hat F_{(2)}\wedge\hat{\ast} \hat F_{(2)} + e^{\sqrt{2}\hat\Phi_1}\hat H_{(2)}\wedge\hat{\ast}\hat H_{(2)}\big) = 0 \,,
\end{align} 
where we have substituted the ansatz $\hat{H}_{(3)} = 0$. Using \eqref{eq:hodge}, we find that these equations become 
\begin{align}
&d{\ast} d\Phi_1 = \frac{1}{\sqrt{2}}e^{\sqrt{2}\Phi_1}d\sigma\wedge{\ast} d\sigma + \frac{g^2}{\sqrt{2}}e^{\sqrt{\frac{2}{5}}\Phi_2-\frac{8}{\sqrt{15}}\Phi_3}\big(e^{\sqrt{2}\Phi_1}(\sigma^2+m)^2 - 4e^{-\sqrt{2}\Phi_1}\big){\ast}1\,, \label{eq:p1} \\ 
&d{\ast} d\Phi_2 = \frac{1}{\sqrt{10}}g^2e^{\sqrt{\frac{2}{5}}\Phi_2-\frac{8}{\sqrt{15}}\Phi_3}\big(e^{\sqrt{2}\Phi_1}(\sigma^2+m)^2 + 4e^{-\sqrt{2}\Phi_1} + 4\sigma^2\big){\ast}1 \,. \label{eq:Bp2} 
\end{align} 
Finally, we have the 7-dimensional Einstein equation 
\begin{align*}
\hat R_{MN} &= \frac{1}{2}\partial_M\hat\Phi_i\partial_N\hat\Phi_i + \frac{1}{2}e^{\sqrt{2}\hat\Phi_1}\partial_M\hat\sigma\partial_N\hat\sigma \\
&\quad+ \frac{1}{2}e^{-\sqrt{2}\hat\Phi_1 + \sqrt{\frac{2}{5}}\hat\Phi_2}\Big(\hat{\mc{F}}^{}_{MP}\hat{\mc{F}}_{N}^{\ph{N}P} - \frac{1}{10}\big(\hat{\mc{F}}_{(2)}\big)^2\hat g_{MN}\Big) \\
&\quad+ \frac{1}{2}e^{\sqrt{\frac{2}{5}}\hat\Phi_2}\Big(\hat{{F}}^{}_{MP}\hat{{F}}_{N}^{\ph{N}P} - \frac{1}{10}\big(\hat{{F}}_{(2)}\big)^2\hat g_{MN}\Big) \\
&\quad+ \frac{1}{2}e^{\sqrt{2}\hat\Phi_1 + \sqrt{\frac{2}{5}}\hat\Phi_2}\Big(\hat{{H}}^{}_{MP}\hat{{H}}_{N}^{\ph{N}P} - \frac{1}{10}\big(\hat{{H}}_{(2)}\big)^2\hat g_{MN}\Big) \numberthis \,, 
\end{align*} 
where $i\in\{1,2\}$, and we have substituted the $\hat H_{(3)} = 0$ ansatz. We have to consider the equations where the indices $M, N$ lie in the 5-dimensional directions and the $S^2$ directions independently. Let $A, B, \dots$ be the 5-dimensional indices, and $m, n, \dots$ be the $S^2$ indices. The $\hat R_{mn}$ equations give 
\begin{equation}
d{\ast} d\Phi_3 = -\frac{4g^2}{\sqrt{15}}e^{\sqrt{\frac{2}{5}}\Phi_2-\frac{8}{\sqrt{15}}\Phi_3}\big(e^{\sqrt{2}\Phi_1}(\sigma^2+m)^2 + 4\sigma^2 + 4e^{-\sqrt{2}\Phi_1} - 10 e^{-\sqrt{\frac{2}{5}}\Phi_2+\sqrt{\frac{3}{5}}\Phi_3}\big){\ast}1 \,. \label{eq:Sig}
\end{equation} 
The $\hat R_{Am}$ equations give a $0=0$ identity, and the remaining $\hat R_{AB}$ equations read 
\begin{align}
R_{AB} &= \frac{1}{2}\partial_A\Phi_i\partial_B\Phi_i + \frac{1}{2}e^{\sqrt{2}\Phi_1}\partial_A\sigma\partial_B\sigma \nonumber \\
&+ \frac{2g^2}{3}e^{\sqrt{\frac{2}{5}}\Phi_2-\frac{8}{\sqrt{15}}\Phi_3}\Big(e^{-\sqrt{2}\Phi_1}+\sigma^2 + \frac{1}{4}e^{\sqrt{2}\Phi_1}(\sigma^2+m)^2 - 4e^{-\sqrt{\frac{2}{5}}\Phi_2+\sqrt{\frac{3}{5}}\Phi_3}\Big)g_{AB} \,, \label{eq:Beinstein}
\end{align} 
where $i\in\{1,2,3\}$. The 5-dimensional equations \eqref{eq:sig}, \eqref{eq:p1}, \eqref{eq:Bp2}, \eqref{eq:Sig}, and \eqref{eq:Beinstein} can be obtained from the action 
\begin{equation}
\mc{L}_5 = R{\ast} 1 - \frac{1}{2}d\Phi_i\wedge{\ast} d\Phi_i - \frac{1}{2}e^{\sqrt{2}\Phi_1}d\sigma\wedge{\ast} d\sigma - V{\ast} 1\,,
\end{equation} 
where $V$ is the scalar potential given by 
\begin{equation}
V = 2g^2e^{\sqrt{\frac{2}{5}}\Phi_2-\frac{8}{\sqrt{15}}\Phi_3}\Big(e^{-\sqrt{2}\Phi_1}+\sigma^2 + \frac{1}{4}e^{\sqrt{2}\Phi_1}(\sigma^2+m)^2 - 4e^{-\sqrt{\frac{2}{5}}\Phi_2+\sqrt{\frac{3}{5}}\Phi_3}\Big) \,.
\end{equation} 
The 5-dimensional Newton constant is related to the 10-dimensional one by
\begin{equation}
\hat\kappa^2 = \frac{g^4\hat\kappa_{10}^2}{2\pi^3l_y}\,.
\end{equation}
The 7-dimensional background solution \eqref{eq:ss7} is now reduced to 
\begin{align}
&ds^2_5 = (W\cosh2\rho)^{\frac{1}{3}}(\sinh2\rho)^{\frac{2}{3}}\Big(\eta_{\mu\nu}dx^\mu dx^\nu + \frac{W\cosh2\rho}{g^2}d\rho^2\Big) \,, \quad e^{-\sqrt{2}\Phi_1} = W(\sinh2\rho)^2\sech2\rho \,,\nonumber \\
&e^{\sqrt{10}\Phi_2} = W^{-1}(\sinh2\rho)^2\sech2\rho \,,\quad e^{\sqrt{15}\Phi_3} = (W\cosh2\rho)^4(\sinh2\rho)^2 \,,\quad \sigma = \sqrt{2}\sech2\rho \,. \label{eq:Bss5}
\end{align} 
The NS5-brane charge $k$ is related to the parameter $m$ by $k = -1 - m/2$. Since $k\geq0$, we find that $m\leq-2$. For the purposes of the present paper, we set $k=0$, so $m=-2$.
\section{Special Functions and Green Functions}\label{se:spegre}
\subsection{Orthonormalised Transverse Wavefunctions}
To eek out any higher precision than the expression given in equation \eqref{expsup}, first note that our separated solutions
\begin{equation}\label{eq: sourced Laplacian}
    \left({\partial^2_r}+\frac{2}{r}\partial_r +g^2\left({\partial^2_\rho}+2 \coth(2 \rho)\partial_\rho\right)\right)f^\omega(r)\zeta_\omega(\rho) = 0\;,
\end{equation}
have transverse factors $\zeta_\omega(\rho)$, which, after changing variables to $y = \cosh(2 \rho)$ and $\zeta_\omega(\rho) = \psi_\omega(\cosh(2\rho))= \psi_\omega(y)$, solve
\begin{equation}\label{transverseq}
    \left(4 \partial_y\left(y^2 -1 \right)\partial_y+\omega^2\right) \psi_\omega(y) = 0\;.
\end{equation}
This is a known version of Legendre's differential equation with the general solution given by Legendre functions (since the order is in general complex).
\begin{equation}
    \psi_\omega(y) = a_\omega \cP_{-\frac{1}{2}+ \frac{\sqrt{1- \omega^2}}{2}}(y) + b_\omega \cQ_{-\frac{1}{2}+ \frac{\sqrt{1- \omega^2}}{2}}(y)\;.
\end{equation}
The Legendre functions of the second type ($\cQ$) have a logarithmic divergence as $y\rightarrow1$. For the moment we want to consider only solutions that are regular when $r\neq 0$ and $\rho\rightarrow0$ ($y\rightarrow1$), so we consider solutions involving only the Legendre function of the first type ($\cP$).\\
\indent Returning to the $\rho$ variables, we now investigate orthonormality. We require 
\begin{equation}
    \int_0^\infty\sinh(2 \rho)\zeta_\omega( \rho) \zeta_\upsilon( \rho)  d\rho = \delta( \omega - \upsilon)\;.
\end{equation}
Applying our transverse operator and integrating by parts, we find that this integral may be given purely in terms of contact terms at infinity. We recall the identity\footnote{We ignore momentarily the numerical factors of the form $\sqrt{2}$, $\sqrt{\pi}$, etc.}
\begin{equation}
    \lim_{R\rightarrow \infty} \frac{1}{\omega^2-\upsilon^2}\left(\omega \sin(\omega \rho)\cos(\upsilon \rho)-\upsilon \cos(\omega \rho)\sin(\upsilon \rho)\right)\bigg|_{\rho=R} \propto \delta (\omega - \upsilon)\;.\end{equation}
Our solutions do not asymptote to sinusoidal functions with frequency $\omega$. Instead, they asymptote with frequency $\sigma = \sqrt{\omega^2-1}$ as can be seen both from the asymptotic form of Equation \eqref{transverseq}, and via the properties of Legendre functions. Specifically, the large $y$ asymptote is
\begin{equation}
    \cP_{\nu}(y) \sim B\left(\nu + \frac{1}{2},\frac{1}{2}\right)^{-1}\left(2 y\right)^{-\nu-1}\;,
\end{equation}
where $B$ is the Euler beta function. There are actually two asymptotic regimes that we need to consider: when $Re(\nu)> -\frac{1}{2}$ and when $Re(\nu)< -\frac{1}{2}$ (although they actually agree in the present case). Furthermore, we will need the connection formula for Legendre functions, the definition of the Euler beta function and the reflection formula for Euler gamma functions:
\begin{equation}
    \cP_\nu(y) = \cP_{-1-\nu}(y)\;,\qquad B(x,y) = \frac{\Gamma(x)\Gamma(y)}{\Gamma(x+y)}\;,\qquad \Gamma(z)\Gamma(1-z) = \pi \csc(\pi z)\;.
\end{equation}
These allow us to derive the necessary normalisation $a_\omega$ so that the amplitude of our solutions as $\rho\rightarrow\infty$ is $\omega$ independent. That is, given
\begin{equation}
    \zeta_\omega(\rho) \propto \sqrt{\pi \sigma \tanh\left(\frac{\pi \sigma}{2}\right)} \cP_{-\frac{1}{2}+\frac{i \sigma}{2}}(\cosh(2 \rho))\;,
\end{equation}
we have $\sqrt{\sinh(2\rho)}\zeta_\omega(\rho)\sim 2 \sin( \sigma \rho + \delta)$. The shift $\delta$ is irrelevant for orthonormalisation. These almost satisfy the equation that we require. We require one additional normalisation, since the asymptotic frequency is given by a function of the separation constant, rather than the separation constant we na\"{\i}vely expected. That is, since $\sigma=\sqrt{\omega^2-1}$, we have
\begin{align*}
    &\int_0^\infty \sinh(2 \rho) \left(\sqrt{\pi \sigma \tanh\left(\frac{\pi \sigma}{2}\right)} \cP_{-\frac{1}{2}+\frac{i \sigma}{2}}(\cosh(2 \rho))\right)\left(\sqrt{\pi \tau \tanh\left(\frac{\pi \tau}{2}\right)} \cP_{-\frac{1}{2}+\frac{i \tau}{2}}(\cosh(2 \rho))\right) d\rho\\
    &= \delta\left( \sigma - \tau\right)\;.\numberthis
\end{align*}
To build our Green functions we require this integral to generate a delta function distribution with respect to $\omega$, not $\sigma$. We use the well-known following property of delta function distributions,
\begin{equation}
    \delta(f(\omega)-f(\tau)) = \frac{\delta(\omega - \tau)}{f'(\omega)}\;,
\end{equation}
then divide by the derivative of the function of the asymptotic frequency with respect to the separation constant, to find the correctly normalised transverse wavefunctions. At the end, they are
\begin{equation}
    \zeta_\omega (\rho) = \sqrt{\frac{\pi \left(\sigma^2+1\right)}{\sigma} \tanh\left(\frac{\pi \sigma}{2}\right)} \cP_{-\frac{1}{2}+ i \frac{\sigma}{2}}\left(\cosh(2 \rho)\right)\;,
\end{equation}
written in terms of $\sigma$ and
\begin{equation}
    \zeta_\omega (\rho) = \sqrt{\frac{\pi \omega^2}{\sqrt{\omega^2-1}} \tanh\left(\frac{\pi}{2} \sqrt{\omega^2-1}\right)} \cP_{-\frac{1}{2}+ \frac{\sqrt{1-\omega^2}}{2}}\left(\cosh(2 \rho)\right)\;,
\end{equation}
when written in terms of $\omega$. These now, by construction, obey the identity
\begin{equation}
    \int_1^\infty \zeta_\omega(\rho)\zeta_\omega(\eta) d \omega = \frac{\delta(\rho-\eta)}{\sinh(2 \rho)}\;,
\end{equation}
We set $\eta=0$ for ease since $\cP_\nu(0) = 1$ for all $\nu$. \\

As for the worldvolume factors $f^\omega(\eta)$, we know the fundamental solution to the corresponding worldvolume differential equation:
\begin{equation}
    \left({\partial_r}^2 +\frac{2}{r}\partial_r - g \omega^2 \right) \frac{\exp\left(-g^2 \omega r\right)}{ 4 \pi r} = \frac{\delta(r)}{4 \pi r^2}\;.
\end{equation}
We may then write the fundamental solution to the total Laplacian
\begin{equation}\label{eq: explicit F}
    G(r,\rho)= \int_1^\infty\frac{\exp \left(- g\omega r\right)}{4 \pi r} \left(\frac{\pi \omega^2}{\sqrt{\omega^2-1}} \tanh\left(\frac{\pi}{2} \sqrt{\omega^2-1}\right)\right) \cP_{-\frac{1}{2} +  \frac{\sqrt{1-\omega^2}}{2}}\left(\cosh (2 \rho)\right) d \omega\;.
\end{equation}
Alternately, we may state the integral in terms of $\sigma$
\begin{equation}\label{integralforray}
    \int_0^\infty\frac{\exp\left(-g\sqrt{\sigma^2+1}r\right)}{4 \pi r}\pi \sqrt{\sigma^2+1} \tanh\left(\frac{\pi \sigma}{2}\right) \cP_{-\frac{1}{2}+i \frac{\sigma}{2}} (\cosh(2 \rho)) d \sigma\;.
\end{equation}
\subsection{The Ray Trick}
No general form of the integral \eqref{integralforray} is known, as it involves an integral with respect to the order of a Legendre function. However, we can find some limits of this integral. Let us introduce the \say{ray trick}. If we want to consider the limit of some integral, say
\begin{equation}
  I(r,\rho) =  \int_0^\infty \frac{\exp(-\omega r)}{4 \pi r} \cos(\omega \rho) d \omega\;,
\end{equation}
we can take explicit ratios of $r = x t$ and $\rho = t$ as $t \rightarrow 0^+$. Then our integral becomes
\begin{equation}
  I(x,t) = \int_0^\infty \frac{\exp (- \omega x t)}{ 4 \pi x t} \cos( \omega t) d\omega\;.
\end{equation}
Multiplying this integral by $xt^2$, and taking the limit $t\rightarrow 0^+$, we define
\begin{equation}
  J(x) =  \lim_{t\rightarrow 0^+} x t^2 \int_0^\infty \frac{\exp (- \omega x t)}{ 4 \pi x t} \cos( \omega t) d\omega\;.
\end{equation}
This can be rewritten as
\begin{equation}
 J(x) =   \lim_{t\rightarrow 0^+} \int_0^\infty t f_x(\omega t) d \omega\;,
\end{equation}
which after a variable redefinition $y = \omega t$ becomes
\begin{equation}
 J(x)   \lim_{t\rightarrow 0^+} \int_0^\infty f_x(y) t\; t^{-1} d y = \lim_{t\rightarrow 0^+} \int_0^\infty f_x(y) d y\;,
\end{equation}
where, crucially, the integrand is $t$ \textit{independent}. Thus,
\begin{equation}
 J(x)=   \lim_{t\rightarrow 0^+} x t^2 \int_0^\omega \frac{\exp (- \omega x t)}{4 \pi x t} \cos( \omega t) d\omega = \frac{1}{4 \pi}\int_0^\infty \exp(- x y) \cos(y) dy = \frac{1}{4 \pi}\frac{x}{1+x^2}\;.
\end{equation}
We can now divide by the factor that we used to get the equation into the $t$ independent form and we find
\begin{equation}
  I(x,t) =  \int_0^\infty \frac{\exp (- \omega x t)}{4 \pi x t} \cos( \omega t) d\omega \sim \frac{1}{4 \pi}\frac{1}{t^2 x}\frac{x}{1 + x^2} =\frac{1}{4 \pi} \frac{1}{(1+x^2)t^2} =\frac{1}{4 \pi} \frac{1}{r^2+ \rho^2}\;.
\end{equation}
This gives us the expected value of the integral in the $r\sim\rho\sim0$ region.
\subsection{The \texorpdfstring{$R\ll1$}{R<<1} Expansion}
Let us verify that the solution \eqref{eq: explicit F} is the same (or at least proportional to) the solution \eqref{eq:Rtom3}.  We begin by multiplying the total function by the SS--CGP parameter $g$, redefining $\tilde r = g r$, then dropping the $\tilde\;$ tilde. That is, by rescaling $r$ and $G^N$ by $g$ we may find the solution in terms of the integral when $g=1$.
\begin{equation}\begin{split}\label{eq: denaturing g}
    G^N\left(r,\rho\right) &= g \int_1^\infty\frac{\exp \left(- \omega \tilde r\right)}{2 \pi \tilde r} \left(\frac{\pi \omega^2}{\sqrt{\omega^2-1}} \tanh\left(\frac{\pi}{2} \sqrt{\omega^2-1}\right)\right) \cP_{-\frac{1}{2} +  \frac{\sqrt{1-\omega^2}}{2}}\left(\cosh (2 \rho)\right) d \omega\\&=g \tilde G\left(g r,\rho\right).
\end{split}\end{equation}
where $\tilde G=G^N\big|_{g=1}$. We then break the dual space into low frequency and high frequency contributions separated at a value $\Lambda$, writing the integrand as $E(\omega,r,\rho)$ 
\begin{equation}
    \tilde G(r,\rho) = \int_1^\Lambda E(\omega, r,\rho)d \omega + \int_\Lambda^\infty E(\omega, r, \rho)d\omega\;.
\end{equation}
Let us now focus on the large frequency integral. When $\omega \gg 1$ most of the terms of the integral simplify. We use the following asymptotic forms for square root, hyperbolic tangent, and Legendre functions
\begin{equation}\begin{aligned}
\sqrt{\omega^2 -1} &\sim \omega - \frac{1}{2 \omega} + \cO(\omega^{-3})\;,\\
\tanh(X) &\sim 1 - 2 \exp(- 2 X) + \cO\left(\exp(-4 X)\right)\;,\\
\cP_{-\frac{1}{2}+i \frac{\omega}{2}}(\cosh(2 \rho))\sqrt{\sinh{2 \rho}} & \sim \sqrt{2 \rho}  J_0( \omega \rho) + \text{subleading}\;.
\end{aligned}\end{equation}
Given these expansions and a sufficiently large cut off, we may now write the high frequency integral in terms of a new simpler integrand $\cE$, plus subleading corrections
\begin{equation}\begin{split}
    \int_\Lambda^\infty E(\omega,r,\rho) d \omega &= \pi\int_\Lambda^\infty \frac{\exp ( - \omega r)}{4 \pi r} \omega J_0(\omega \rho)\frac{\sqrt{ 2 \rho}}{\sqrt{\sinh(2 \rho)}} + \cO\left(\frac{1}{\omega}\right) d \omega\\
    &=\int_\Lambda^\infty\cE(\omega,r,\rho)+\cO\left(\frac{1}{\omega}\right)d\omega\;.
\end{split}\end{equation}
This integral is still unknown. However, using the fundamental theorem of calculus, we may approximate it in the small $\Lambda$ limit:
\begin{equation}\begin{gathered}
    \int_\Lambda^\infty\cE(\omega,r,\rho)d\omega = \int_0^\infty \cE(\omega,r,\rho)d\omega-\int_0^\Lambda \cE(\omega,r,\rho)d \omega\\
    = \frac{\sqrt{ 2 \rho}}{4 \sqrt{\sinh(2 \rho)}}\left(\frac{1}{R^3}+\frac{ \Lambda^2}{2 r} -\frac{\Lambda^3}{3}+\frac{\left(2 r^2-\rho^2\right) \Lambda^4}{r}+\cO\left(\Lambda^5\right)\right)\;,
\end{gathered}\end{equation}
where we recall that $R^2 = g^2r^2+\rho^2$. We will address the validity of a the small $\Lambda$ limit momentarily. The low frequency contribution may be done using different approximations of these functions. First we shift $\omega = \tilde \omega + 1$ so that our integral is from $\tilde \omega =0$ to $\tilde\Lambda = \Lambda - 1$. Our integrand becomes
\begin{equation}\begin{split}
    \int_1^\Lambda E(\omega, r,\rho) d\omega =\exp(-r) \int_0^{\tilde \Lambda}\frac{\exp(-\tilde \omega r)}{4 \pi r} &\left(\frac{\pi ^2}{2}+\left(\pi ^2-\frac{\pi ^4}{12}\right) \tilde{\omega }+\cO\left(\tilde{\omega
   }^{2}\right)\right)\\
   \times&\left(1+\frac{1}{4} \rho ^2 \left(-\tilde{\omega }^2-2 \tilde{\omega }-1\right)+\cO\left(\rho
   ^4\right)\right)d\tilde\omega\;.
\end{split}\end{equation}
We can expand this in the small $\tilde\Lambda$ and small $\rho$ limit to find
\begin{equation}
    \int_1^\Lambda E(\omega, r,\rho) d\omega =\frac{\pi \tilde\Lambda  e^{-r}}{8 r}-\frac{\pi \tilde\Lambda  \rho ^2 e^{-r}}{32 r}+\cO\left(\tilde\lambda^2\right)+\cO\left(\rho^4\right)\;.
\end{equation}
Therefore as $R\rightarrow0$, $G\rightarrow\frac{1}{R^3}$. We confirm that this is (proportional to) the solution given above. All terms that contain factors of the cutoff accurately represent the forms of corrections. However, since the cutoff is arbitrary the exact function will, of course, be independent of the cutoff, but the actual coefficients of these corrections remain unknown.
\subsection{The \texorpdfstring{$\rho\ll1$}{Rho<<1} Expansion}
First we set $\rho=0$ when $\cP_\nu(1)=1$. We shift our integrand as before to find
\begin{equation}
    \tilde G(r,0) = \exp(-r)\int_0^\infty \frac{\exp(-\omega r)}{4 \pi r} \left(\frac{ \pi \left(\omega+1\right)^2}{\sqrt{(\omega+1)^2-1}}\tanh\left(\frac{\pi}{2}\sqrt{(\omega + 1)^2-1}\right)\right) d\omega\;.
\end{equation}
This integral still escapes the domain of known integrals giving named functions, but we can expand the integrand excluding the $\exp(-\omega r)$ term in the small $\omega$ limit. This is valid when $r$ becomes large as all large $\omega$ terms become exponentially suppressed. This gives us the following series
\begin{equation}
    \tilde G(r,0) =\frac{\exp(-r)}{4 \pi}\left(\frac{\pi ^2}{r^2}-\frac{2 \pi ^2 \left(\pi ^2-3\right)}{3 r^3}+\frac{2 \pi ^2 \left(15-25 \pi ^2+8 \pi ^4\right)}{15 r^4}+\cO\left(\frac{1}{r^6}\right)\right)\;.
\end{equation}
If we expand our transverse functions at small $\rho$, we find the following series
\begin{equation}
    \cP_{-\frac{1}{2}+\frac{\sqrt{1-(\omega+1)^2}}{2}}\left(\cosh(2 \rho)\right)=
    1+\left(-\frac{1}{4}-\frac{\omega}{2}-\frac{\omega^2}{4}\right)\rho^2+\left(\frac{11}{192}+\frac{7 \omega }{48}+\frac{13 \omega ^2}{96}+\frac{\omega ^3}{16}+\frac{\omega ^4}{64}\right)\rho^4+\cO\left(\rho^6\right).
\end{equation}
Using these two series we can find $\exp(r)\tilde G(r,\rho)$ to arbitrary order in $\frac{1}{r}$ and $\rho$. Furthermore, we may find the exact coefficient of the leading term in the expansion in $\frac{\exp(-r)}{r}$ by first substituting $\omega=0$ into our transverse wavefunction. We find
\begin{equation}
    \cP_{-\frac{1}{2}}\left(\cosh(2 \rho)\right) = \frac{2}{\pi} K\left(-\sinh^2(\rho)\right)\;,
\end{equation}
where $K$ is the complete elliptic integral of the first kind. The best estimate we have for $\tilde G$ is therefore
\begin{equation}\begin{split}
    \tilde G(r,\rho) =  &\frac{\exp(-r)}{4 r^2} K\left(-\sinh^2(\rho)\right)\\
    &+\frac{\exp(-r)}{4 \pi r^3}\left(-\left(\frac{\pi ^4}{12}+\pi ^2\right)+\left(\frac{\pi ^4}{48}- \frac{\pi ^2}{2}\right) \rho ^2+\left(\frac{25 \pi ^2}{192}-\frac{11 \pi ^4}{2304}\right) \rho ^4\right)\;,
\end{split}\end{equation}
up to corrections of order $\cO\left(\frac{\exp(-r)}{r^4}\right)$ or $\cO\left(\rho^5\right)$. We find a similar solution when we assume $\tilde G$ is given in an expansion in $\exp(-r)r^{-n}f_n(\rho)$ with minimum $n=2$ and solve equation \eqref{eq: sourced Laplacian} in the large $r$ limit order by order, using the same technique as for finding the large $R$ expansion. Unfortunately the first sourced order ($f_3$, the coefficient of $\exp(-r)r^{-3}$) cannot be solved analytically except for the case when $\rho\ll1$.
\subsection{The \texorpdfstring{$\rho \gg 1$}{Rho>>1} Expansion}
When $\rho \gg 1$ we may approximate our differential equation as
\begin{equation}
    \left({\partial^2_r}+\frac{2}{r}\partial_r +{\partial^2_\rho} + 2 \partial_\rho\right)\tilde G(r,\rho) = 0\;.
\end{equation}
To simplify, we change variables to 
\begin{equation}
    \tilde G(r,\rho) = \frac{\exp\left(-r -\rho\right)}{r}U(r,\rho)\;.
\end{equation}
We can further simplify our differential equation by multiplying by $r \exp(r + \rho)$. We find $f$ satisfies
\begin{equation}
    \left({\partial^2_r}+ {\partial^2_\rho} - 1\right) U(r,\rho)=0\;.
\end{equation}
Unfortunately we cannot translate our boundary conditions onto any condition on this $U$, other than that it must not grow exponentially fast as $\rho\rightarrow\infty$ of $r\rightarrow\infty$. We may, however, suppose the ansatz that it has a Laurent series in $r$ starting with $\frac{1}{r}$. Given that choice and using the same technique as for small $R$ we find
\begin{equation}
    U(r,\rho)=\frac{a_1\rho+b_1}{r}+\frac{-\frac{a_1}{3}\rho^3-b_1\rho^2+a_2 \rho+b_2}{r^2}+\cO\left(\frac{1}{r^3}\right)\;.
\end{equation}
Returning to the actual solution given in integral form we may approximate the Legendre function when $\rho \gg 1$ as above:
\begin{equation}
    \cP_{\nu}\left(\cosh(2\rho)\right)\sim\frac{1}{2}\left(B\left(\nu+\frac{1}{2},\frac{1}{2}\right)^{-1}{\exp(2 \rho)}^{-1-\nu}+B\left(\mu+\frac{1}{2},\frac{1}{2}\right)^{-1}{\exp(2 \rho)}^{-1-\mu}\right)\;,
\end{equation}
where $\nu$ and $\mu$ are complex conjugates given that the real part of $\nu =-\frac{1}{2}$. Using the mirror symmetry of gamma functions ($\Gamma(z^*)=\Gamma(z)^*$). We may identify $B(\nu+\frac{1}{2},\frac{1}{2})^{-1}= m \exp(i \delta)$ for some real variables $m$ and $\delta$. Given our $\nu=-\frac{1}{2} + i \frac{\sigma}{2}$, we have
\begin{equation}
    \cP_{\nu}\left(\cosh(2\rho)\right)\sim \exp\left(-\rho\right) m \frac{1}{2}\left(\exp(i \delta)\exp(-i \sigma \rho)+\exp(-i \delta)\exp(i \sigma \rho)\right)\;.
\end{equation}
Simplifying, we find
\begin{equation}\label{eq: asymptotic Legendre}
    \cP_{-\frac{1}{2}+i\frac{ \sigma}{2}}\left(\cosh(2\rho)\right)\sim \exp\left(-\rho\right) m \cos\left(\rho\sigma - \delta\left(\sigma\right)\right)\;.
\end{equation}
Since we require an expansion of this quantity in $\omega$ we may no longer ignore the frequency shift, $\delta$. The formulae for $m$ and $\delta$ are
\begin{equation}
    m=\sqrt{\frac{\Gamma\left(\frac{1}{2}\right)^2\Gamma \left(-\frac{i \sigma }{2}\right) \Gamma \left(\frac{i \sigma }{2}\right)}{\Gamma \left(\frac{1}{2}-\frac{i \sigma }{2}\right) \Gamma \left(\frac{i \sigma }{2}+\frac{1}{2}\right)}}\;,\qquad \delta = \arctan\left(\frac{\text{Im}\left\{B(-i\frac{\sigma}{2},\frac{1}{2})\right\}}{\text{Re}\left\{B(-i\frac{\sigma}{2},\frac{1}{2})\right\}}\right)\;,
\end{equation}
where we may use the reflection formula for gamma functions to find an exact value for the first and a Taylor expansion for the second:
\begin{equation}\label{eq: explicit mag and delta}
    m = \sqrt{\frac{2 \pi}{\sigma \tanh\left(\frac{\pi}{2}\sigma\right)}}\;,\qquad \delta = \frac{\pi}{2}-\frac{1}{2}\left(\psi_0(1)-\psi_0\left(\frac{1}{2}\right)\right)\sigma+\frac{1}{48}\left(\psi_2(1)-\psi_2\left(\frac{1}{2}\right)\right)+\cO(\sigma^3)\;.
\end{equation}
Inserting \eqref{eq: explicit mag and delta} into \eqref{eq: asymptotic Legendre} then into \eqref{eq: explicit F}, we must then change coordinates from $\sigma = \sqrt{\omega^2-1}$ to $\omega$, then $\omega=\tilde\omega - 1$ to $\tilde \omega$. We may then expand the integrand (save $\exp(-\tilde \omega r)$) as a Taylor series in $\tilde \omega$. This becomes
\begin{equation}\begin{gathered}
    \tilde G (r, \rho) = \frac{\exp(- r - \rho)}{4 \pi r}\int_0^\infty \exp(-\tilde \omega r) \Bigg( \frac{\pi^2 \left( 4 \rho+\log(16)\right)}{4 \sqrt{2}}\\
    -\frac{\pi ^2 \left((\rho +\log (2)) \left(4 \left(\rho  (\rho +\log (4))-6+\log ^2(2)\right)+\pi ^2\right)+6 \zeta (3)\right)}{12 \sqrt{2}}\tilde \omega+ \cO\left(\tilde \omega^2\right)\Bigg)d \tilde \omega \;.
\end{gathered}\end{equation}
From this we approximate
\begin{equation}\begin{split}
    \tilde G(r,\rho) &= \frac{\exp(-r-\rho)}{4 \pi r}\Bigg(\frac{\pi ^2 (\rho +\log (2))}{\sqrt{2} r}\\&
    -\frac{\pi ^2 \left((\rho +\log (2)) \left(4 \left(\rho  (\rho +\log (4))-6+\log ^2(2)\right)+\pi ^2\right)+6 \zeta (3)\right)}{12 \sqrt{2} r^2}+\cO\left(\frac{1}{r^3}\right)\Bigg)\;,
\end{split}\end{equation}
which we see obeys the expansion that we derived previously as the most general solution.
\section{Product Space Green Functions}\label{se:psgf}
In this appendix, we will derive several useful formulae for calculating Green functions. We begin with the resolution of the identity. First, we consider the eigenvalues of some second-order ordinary operator differential operator $\Delta_\rho$
\begin{equation}
    \Delta_\rho \xi_\omega= \left(\frac{1}{\mu(\rho)}\partial_\rho \nu(\rho)\partial_\rho\right)\omega= -\omega^2 \xi_\omega\;.
\end{equation}
These lie within a self-adjoint domain of $\Delta_\rho$, given they are orthonormal with respect to the transverse inner product on their domain $\cI$
\begin{equation}
    \int_\cI \mu \xi_\omega \xi_\tau d\rho = \frac{\nu}{\omega^2-\tau^2}\left(\xi_\omega \partial_\rho\xi_\tau-\xi_\tau\partial_\rho\xi_\omega\right)\Big|_{\partial\cI}=0\;.
\end{equation}
This will be true for some solutions to our ODE which obey some boundary condition that causes the right-hand-side of this expression to vanish. A complete set of such functions with eigenvalues $\omega\in\cM_\omega$ (with $\cM_\omega^+$ explicitly excluding $\omega=0$) forms a basis for $L_2(\cI,\mu)$. This can be summarised by the identity
\begin{equation}\label{eq: rhospacedelta}
    \int_{\cM_\omega} \xi_\omega(\rho)\xi_\omega(\eta) d\omega = \frac{\delta(\rho-\eta)}{\mu(\rho)}\;.
\end{equation}
From \eqref{eq: rhospacedelta} we can derive that the integral over \say{massive} modes divided by their corresponding eigenvalue is closely related to the Green function
\begin{equation}
    G(\rho-\eta) = -\int_{\cM_\omega^+}\frac{\xi_\omega(\rho)\xi_\omega(\eta)}{\omega^2} d \omega + K(\rho-\eta) \;.
\end{equation}
Here $G$ is the transverse Green function
\begin{equation}
    \Delta_\rho G(\rho-\eta) = \frac{\delta(\rho-\eta)}{\mu(\rho)}\;,
\end{equation}
and $K$ is the solution to the differential equation
\begin{equation}
    \Delta_\rho K(\rho- \eta) = \xi_0(\rho)\xi_0(\eta)\;,
\end{equation}
which obeys our desired boundary condition. Note that $K$ is zero when the spectrum omits a zero mode.\\
\indent There is a slight generalisation of this identity which we will require to identify Green functions on the total space. Note that this now includes the zero mode in contrast to the above discussion. It is
\begin{equation}\label{eq: greenfrommodes}
    G_{-\tau}(\rho-\eta) = -\int_{\cM_\omega}\frac{ \xi_\omega(\rho)\xi_\omega(\eta)}{\omega^2+\tau^2} d \omega\;.
\end{equation}
Where $G_{-\tau}$ is the Green function for the modified differential operator\footnote{Strictly taking $\tau\rightarrow-\tau$ would not change the differential operator as it depends only on $\tau^2$. The minus sign therefore denotes the change of sign of the total eigenvalue.}
\begin{equation}
    \left(\Delta_\rho-\tau^2\right)G_{-\tau}(\rho-\eta) = \frac{\delta(\rho-\eta)}{\mu(\rho)}\;.
\end{equation}
\indent This generalises even further. For instance, we may consider a separable partial differential operator
\begin{equation}
    \Delta = \Delta_r + \Delta_\rho = \frac{1}{\mu_r(r)}\partial_r \nu_r(r)\partial_r+\frac{1}{\mu_\rho(\rho)}\partial_\rho \nu_\rho(\rho)\partial_\rho\;,
\end{equation}
for which we define a total space Green function
\begin{equation}
    \left(\Delta_r + \Delta_\rho\right)G(r-s,\rho-\eta) = \frac{\delta(r-s)\delta(\rho-\eta)}{\mu_r(r)\mu_\rho(\rho)}\;.
\end{equation}
Then we find
\begin{equation}\label{eq: productspacegreenf}
    G(r-s,\rho-\eta)= -\int_{\cM_\tau(r)}\int_{\cM_\omega(\rho)} \frac{f_\tau(r)f_\tau(s)\xi_\omega(\rho)\xi_\omega(\eta)}{\tau^2 + \omega^2} d \tau d\omega\;.
\end{equation}
Here $f_\tau(r)$, $\cM_\tau(r)$, $\xi_\omega(\rho)$, and $\cM_\omega(\rho)$ are the orthonormalised eigenfunctions and spectra, respectively of $\Delta_r$ and $\Delta_\rho$, respectively. Note that unlike the case for non-product-space Green functions, this integral \textit{does not} exclude the zero modes. \\
\indent \eqref{eq: productspacegreenf} gives us two\footnote{In general $n!$ for $n$ separations.} paths for evaluating the total Green function $G$. We may either evaluate the external rather than internal integrals or the internal rather than external integrals.
\begin{gather}
    -\int_{\cM_c(r)}\int_{\cM_c(\rho)} \frac{f_\tau(r)f_\tau(s)\xi_\omega(\rho)\xi_\omega(\eta)}{\tau^2 + \omega^2} d \tau d\omega\nonumber\\
    \swarrow \hspace{3 cm} \searrow\nonumber\\
    =\int_{\cM_c(\rho)} G_{-\omega}(r-s) \xi_\omega(\rho)\xi_\omega(\eta) d\omega \hspace{3 cm}
=\int_{\cM_c(r)} f_\tau(r)f_\tau(s) G_{-\tau}(\rho-\eta)d\tau\\
    \searrow \hspace{3 cm} \swarrow\nonumber\\
    = G (r-s,\rho-\eta)\nonumber
\end{gather}
\indent The explicit evaluation of these integrals is impossible in all but the simplest cases. However, it is generally straightforward to find the Green function. This in turn helps us find actual values for some novel integrals of special functions.

\end{document}